\documentclass[iop]{emulateapj}

\usepackage{amsmath,amssymb,latexsym,graphics,epsfig,exscale,graphicx}
\usepackage{pstricks}
\usepackage{multirow}
\usepackage{graphicx}
\usepackage{subfigure}

\newcommand{\gad}{{\sc Gadget-2}}

\newcommand{\hmpc}{h^{-1}{\rm Mpc}}
\newcommand{\lcdm}{$\Lambda$CDM}
\newcommand{\Msun}{$M_{\odot}$}

\newcommand{\figsz}{0.46}

\newcommand{\pathI}{figures}

\begin{document}

\title{Cosmological Zoom Simulations of $\lowercase{z} = 2$ Galaxies: The Impact of Galactic Outflows}
\shorttitle{The Impact of Galactic Outflows on $\lowercase{z} = 2$ Galaxies}

\shortauthors{D. Angl{\'e}s-Alc{\'a}zar, R. Dav\'e, F. {\"O}zel \& B. Oppenheimer}
\author{
Daniel Angl{\'e}s-Alc{\'a}zar\altaffilmark{1},
Romeel Dav\'e\altaffilmark{2,3,4,5},
Feryal {\"O}zel\altaffilmark{2,6},
Benjamin D. Oppenheimer\altaffilmark{7}}
\altaffiltext{1}{Department of Physics, University of Arizona, Tucson, AZ 85721, USA;
{\rm anglesd@email.arizona.edu}}
\altaffiltext{2}{Department of Astronomy, University of Arizona, Tucson, AZ 85721, USA}
\altaffiltext{3}{University of the Western Cape, Bellville, Cape Town 7535, South Africa}
\altaffiltext{4}{South African Astronomical Observatories, Observatory, Cape Town 7525, South Africa}
\altaffiltext{5}{African Institute for Mathematical Sciences, Muizenberg, Cape Town 7545, South Africa}
\altaffiltext{6}{Radcliffe Institute for Advanced Study, Harvard University, 
Cambridge, MA 02138, USA}
\altaffiltext{7}{Leiden Observatory, Leiden University, PO Box 9513, 2300 RA Leiden, The Netherlands}

 \begin{abstract}

We use high-resolution cosmological zoom simulations with $\sim
200$~pc resolution at $z=2$ and various prescriptions for galactic
outflows in order to explore the impact of winds on the morphological,
dynamical, and structural properties of eight individual galaxies 
with halo masses $\sim 10^{11}$--$2 \times 10^{12}$\,\Msun~at $z = 2$.  
We present a detailed comparison to spatially and
spectrally resolved $H{\alpha}$ and other observations of
$z\approx 2$ galaxies.  We find that simulations without winds
produce massive, compact galaxies with low gas fractions, super-solar
metallicities, high bulge fractions, and much of the star formation
concentrated within the inner kpc.  Strong winds are required to
maintain high gas fractions, redistribute star-forming gas over
larger scales, and increase the velocity dispersion of simulated
galaxies, more in agreement with the large, extended, turbulent
disks typical of high-redshift star-forming galaxies.  Winds also suppress
early star formation to produce high-redshift cosmic star formation
efficiencies in better agreement with observations.  Sizes, rotation
velocities, and velocity dispersions all scale with stellar mass in accord
with observations.  
Our simulations produce a diversity of morphological characteristics---among our three most massive galaxies, we find a quiescent
grand-design spiral, a very compact star-forming galaxy, and a clumpy disk undergoing a minor merger;
the clumps are evident in H$\alpha$ but not in the stars.  
Rotation curves are generally slowly rising, particularly when
calculated using azimuthal velocities rather than enclosed mass.
Our results are broadly resolution-converged.  These results show
that cosmological simulations including outflows can
produce disk galaxies similar to those observed during the peak
epoch of cosmic galaxy growth.

\end{abstract}

\section{Introduction}

The epoch around redshift $z\sim 2$ is the most active period of
cosmic star formation~\citep{mad96,hopk06}, and appears to be the period when
the familiar {\it Hubble} sequence first began to emerge.  It is thus a
critical epoch for understanding how galaxies form, grow, and evolve
into the populations we see today.  The advent of near-infrared
integral field spectrometers on 8--10\,m class ground-based telescopes
has enabled spatially and spectrally resolved observations of an
increasing number of galaxies at this epoch
\citep[e.g.,][]{for06,cres09,for09,law09,wri09,ala12,swi12b}. These studies
are complemented by space-based observations with {\it Hubble},
enabling a detailed study of the distributions of star formation
and stellar populations \citep[e.g.,][]{elmegreen04,elmegreen09,wuy12}, 
and {\it Spitzer} and {\it Herschel},
providing infrared data that better constrains the total stellar
mass and bolometric emission of these galaxies
\citep[e.g.,][]{nordon10,rodighiero10,wuy11}.  Such data are
providing a comprehensive view of the structure, kinematics, and
star formation properties of galaxies at early stages of their
evolution.

These observations provide new and unexplored avenues with which
to constrain models of galaxy formation and evolution.  In particular,
the resolved (both spatially and spectrally) information that is
now available provides detailed constraints on the assembly of
galaxies at $z\sim 2$.  The galaxy population at this epoch displays
a remarkable level of diversity, with many properties unlike anything
seen locally, from extremely compact and (relatively) quiescent
ellipticals \citep{vandok08,barro13}, to extended turbulent and clumpy disks \citep{for09,genz11}, to galaxies
forming stars at hundreds to thousands of solar masses per year \citep{chap10,ala12,tar13}.
Capturing this diversity, both qualitatively and quantitatively,
presents an enormous challenge for galaxy formation models, one
that is only now beginning to be addressed using the latest
computational tools.

Cosmological hydrodynamic simulations have now matured to the point
that they yield galaxy populations broadly in agreement with
observations across a range of redshifts \citep[e.g.,][]{dav11a,dav11b,mccarthy12,torr13,kannan13}.
While there are still many discrepancies with even the latest
models~\citep[e.g.,][]{weinmann12}, overall it appears that the
galaxy population is in general agreement with expectations from a
$\Lambda$ cold dark matter ($\Lambda$CDM) cosmology, combined with numerous feedback
mechanisms from star formation, photo-ionizing radiation, and active
galactic nuclei that help establish the properties of galaxies
assembling within hierarchically-growing halos.  The emerging
paradigm is that galaxy evolution at this epoch is governed by a
balance between inflows from the intergalactic medium (IGM) and
powerful, ubiquitous outflows that intimately connect galaxies and
their surroundings in a ``baryon cycle" of exchanging mass, energy,
and metals \citep{dav12,lilly13}.

Resolved studies of distant galaxies present a new challenge to
such models.  In order to meet this challenge, numericists have
begun to employ the ``zoom" technique to expand the dynamic range
sufficiently to model such observations.  In zoom simulations, a
sub-volume extracted from a larger volume is re-simulated at
significantly higher resolution, providing a substantial increase
in resolution at a manageable computational cost.  Such simulations
can simultaneously capture the crucial baryon cycle on larger scales,
while still achieving sufficiently high enough resolution to robustly
model the internal structure and dynamics of galaxies.

While inflows are generally well-predicted within the $\Lambda$CDM
paradigm (albeit difficult to detect), galactic outflows remain a
highly uncertain and poorly constrained ingredient in galaxy formation
models.  It is now evident that powerful galactic outflows are
ubiquitous in high-redshift galaxies, as indicated by the high
frequency of blue-shifted rest-frame UV absorption lines
\citep[e.g.,][]{wei09,ste10,kornei12,mar12} and broad H$\alpha$ emission-line
profiles \citep[e.g.,][]{genz11,new12}.  These winds carry out
masses comparable to the star formation rates~\citep[SFRs;][]{ste10,genz11},
though these estimates can be uncertain by an order of magnitude.
Hence outflows are likely to play a central role in the evolution
of galaxies and the IGM.

Theoretically, outflows are the primary candidate for regulating
the baryon and metal content of galaxies, while concurrently
explaining the enrichment of the IGM.  The physical origin and
launching mechanism of such outflows is still in debate, and it is
usually attributed to energy and/or momentum input from supernovae
(SNe) and/or radiation pressure from massive stars \citep{mur05,mur10,kru13}.
Galaxy scale and cosmological simulations by \citet{spr03a,spr03b}
showed that the injection of energy from SNe in the form of kinetic
outflows with constant velocity can be effective in removing gas
from galaxies, potentially solving the overcooling problem by
regulating SFRs and enriching the IGM.

However, observations of dwarf starbursts and low-redshift luminous
infrared galaxies \citep{mar05,rup05} as well as higher redshift
galaxies \citep{wei09} suggest that the properties of galactic
outflows scale with galaxy properties---galaxies with higher masses
and SFRs drive faster and more energetic winds---in broad agreement
with predictions of momentum-driven models \citep{mur05}.  In
the momentum-driven wind scenario, radiation from massive stars is
absorbed by dust that collisionally couples to the gas, resulting
in galactic outflows for which the velocity and mass loading factor
(i.e., the mass loss rate relative to the SFR) scale linearly
and inversely, respectively, with the circular velocity of galaxies.
These scalings are also favored by recent high-resolution galaxy-scale simulations
including explicit stellar feedback models \citep{hop12}, 
though there are concerns regarding the efficiency of radiative momentum coupling required to drive sufficiently strong outflows \citep{dek13,kru13}.

\citet{opp06,opp08} implemented a variety of outflow models into
cosmological hydrodynamic simulations of galaxy formation, and found
that the scalings arising for momentum-driven winds yielded a
significant improvement over the original \citet{spr03b} ``constant
wind" models toward matching a wide range of observables, including
(1) the chemical enrichment of the IGM at $z > 2$
\citep{opp06,opp08}, (2) the luminosity function of high-redshift
galaxies \citep{dav06,fin07,dav11a}, and (3) the galaxy
mass--metallicity relation \citep{fin08,dav11b}.  However, with
spatial resolution of typically several kpc, these simulations were
unable to examine the internal structural properties of galaxies
on sub-kpc scales.

In this work, we use cosmological hydrodynamic zoom simulations
including galactic outflows using the exact same prescriptions as
in \citet{dav11a} to make predictions for internal structure and
dynamics of eight re-simulated galaxies \citep{ang13}.  Our modeling is similar to
that used in \citet{gen12b}, who focused on studying the properties
of star-forming clumps~\citep{genz11,for11} in models with and without
winds, finding that a similar wind prescription provided both
realistic suppression of star formation while disrupting clumps on
short timescales.  Here, we examine the impact of galactic outflows
on the morphological, dynamical, and star formation properties in
a sample of eight re-simulated central galaxies in two re-simulated
regions, and compare these properties to observations from various
surveys, focusing particularly on the 
Spectroscopic Imaging survey in the Near-infrared with SINFONI
(SINS) Survey of $z\sim 2$ galaxies \citep{for09}.

We begin by describing our simulations in Section~\ref{sec:sim} and
present and overview of the sample of simulated galaxies in
Section~\ref{sec:samp}.  We analyze the impact of galactic outflows
on the time evolution and radial structure of three example galaxies
in Section~\ref{sec:eff}, where we also evaluate the effects of
different wind models on global properties of galaxies
such as halo baryonic fractions and cosmic star formation efficiencies, and
their rotation curves.  In Section~\ref{sec:obs} we compare the
properties of our simulated galaxies to available spatially and
spectrally resolved observations of $z \sim 2$ galaxies.  We present resolution convergence tests of our key results in Section~\ref{sec:num} and we
summarize our results in Section~\ref{sec:con}.

\section{Simulations}\label{sec:sim}

\subsection{Simulation Code}

Our simulations were run with an extended version of the $N$-body + smoothed 
particle hydrodynamics (SPH) cosmological galaxy formation code
\gad~\citep{spr05}.  \gad~combines an entropy-conserving formulation of SPH \citep{spr02}
along with a tree-particle-mesh algorithm for computing gravitational forces.
Additions to the public version of the code have been described in \citet{opp08}
and include models for gas cooling, star formation, chemical enrichment, 
and galactic outflows.

We include photoionization heating starting at $z = 9$ via a spatially
uniform, optically thin UV background \citep{haa01}.
We account for radiative cooling from primordial gas assuming ionization
equilibrium as in \citet{kat96} and metal-line cooling 
using the collisional ionization equilibrium tables of \citet{sut93}.  
We track the production of four metal species (C, O, Si, and
Fe) from Type II SNe, Type Ia SNe, and asymptotic giant branch (AGB) stars 
using metallicity-dependent yields as described in \citet{opp08}.
We also account for energy feedback from Type II and Type Ia SNe, and 
mass loss from AGB stars.

Star formation is modeled using the sub-grid prescription of \citet{spr03a}.  
Gas particles that are sufficiently dense to become
Jeans unstable ($n>0.13$~cm$^{-3}$) are treated as a two-phase 
interstellar medium (ISM) consisting of hot gas that condenses into 
cold star-forming clouds via a thermal instability \citep{mck77}.  
Stars form from the cold phase and thermal feedback
from Type II SNe causes the evaporation of cold clumps into the hot medium.
Star formation is implemented probabilistically, such that at any
given step, a sufficiently dense gas particle can spawn a star based
on a \citet{schmidt59} law.  The star particle's mass is half the
original gas particle mass.  The resulting SFRs are
tuned to be in accord with the observed \citet{ken98a} relation.  

Our simulations include a galactic outflow mechanism
that imparts kinetic energy to gas particles which is identical to that used in larger-scale simulations
of, e.g., \citet{dav11a}.  
The outflow is described by two parameters, the wind speed $v_w$ and the mass
loading factor $\eta$, which is the mass outflow rate in units of
the SFR.  If a particle is eligible to form stars,
it is likewise eligible to be kicked into an outflow, with a
probability given by $\eta$ times the star formation probability.
If selected, the gas particle is kicked with a velocity $v_w$ in
the direction $v \times a$, where $v$ and $a$ are
the particle's instantaneous velocity and acceleration.  
Hydrodynamic forces are turned off until the particle reaches a density of
0.013~cm$^{-3}$, or a time of 20~kpc/$v_w$ has passed \citep{opp08}.
Decoupling from hydrodynamics allows outflowing gas to escape from the galactic ISM and travel to large distances as observed in $z \approx 2$ galaxies \citep{ste10}, a process not properly captured at the resolution of our simulations, and also yields results that are less sensitive to numerical resolution \citep{spr03b}.  Similar implementations of galactic outflows have been used in recent zoom simulations \citep{gen12b} as well as full cosmological simulations \citep{barai13,puch13}.

Choices for $v_w$ and $\eta$ define our ``wind model".  Throughout
this paper we compare the following wind models:

\begin{itemize}
\item {\it No wind model (nw)}:  We turn-off galactic winds ($\eta=0$).

\item {\it Constant wind model (cw)}:  $v_{\rm w} = 680$\,km\,s$^{-1}$ and
$\eta = 2$ for all galaxies.  

\item {\it Momentum-driven wind model (vzw)}:  The kick velocity
scales with galactic velocity dispersion, $\sigma$, and the mass
loading factor scales as $1/\sigma$, as in the momentum-conserving
case \citep{mur05}.  Following \citet{opp08} we take $v_{\rm w} =
3\sigma \, \sqrt{f_{\rm L} -1}$ and $\eta = \sigma_0/\sigma$, where
$f_{\rm L}$ is the (metallicity dependent) critical luminosity
necessary to expel gas from the galaxy and $\sigma_{0} =
150$\,km\,s$^{-1}$.  We run an on-the-fly galaxy finder to calculate
galaxy masses that are then converted to $\sigma$ using standard
relations \citep{mo98}.
\end{itemize}

Throughout this paper we assume a
\citet{cha03} initial mass function (IMF) and a 
\lcdm~concordance cosmology with parameters $\Omega_{\rm \Lambda} = 0.72$,
$\Omega_{\rm M} = 0.28$, $\Omega_{\rm b} = 0.046$, $h = 0.7$,
$\sigma_{8} = 0.82$, and $n = 0.96$, consistent with 
five-year {\it Wilkinson Microwave Anisotropy Probe} ({\it WMAP}) data combined with baryon acoustic oscillations and SNe constraints
 \citep{kom09} as well as the final nine-year {\it WMAP} data \citep{hin13}.

\subsection{Simulation Runs and Analysis}

Current high-resolution near-infrared observations being carried
out by adaptive optics-assisted 10\,m class telescopes as well as
the {\it Hubble Space Telescope} are able to resolve $\sim 1$\,kpc scales
at $z \sim 2$ \citep[e.g.,][]{for11,kar12}.  In order to get full
advantage of these observations, we used the ``zoom-in" technique
\cite[e.g.,][]{nav94} to carry out cosmological simulations that
follow the evolution of galaxies down to $z \sim 2$ with sub-kiloparsec
spatial resolution.  The simulations presented here have been used
in \citet{ang13} in a different context, where we constrain the
growth of massive black holes at the centers of galaxies on
cosmological time scales and discuss the implications of the observed
black hole--galaxy correlations.  

We selected two $\sim [5\,\hmpc]^3$ regions for re-simulation from an intermediate-resolution 
full cosmological simulation surrounding two central galaxies characterized
by similar stellar masses ($\sim 3 \times 10^{10}$\,\Msun) 
but different morphologies and merger histories.
The parent simulation had $2 \times 256^{3}$ gas+dark matter particles in a $[24\,\hmpc]^3$ box
and was run including momentum-driven winds.
As described in \citet{ang13}, zoom initial conditions were generated
by identifying all dark matter particles within the virial
radius of each selected $z = 2$ galaxy, tracing their positions back to their locations on the initial grid.  Refinement regions were initially defined by the grid cells containing these particles and, subsequently, significantly enlarged by an iterative cleaning procedure incorporating an increasing number of neighboring cells.  High-resolution regions were populated with a large number of lower mass particles (yielding $\times 64$ mass resolution increase for our highest resolution simulations) and small-scale power spectrum fluctuations were applied according to the spatial resolution of the refined grid.  Additionally, two nested concentric layers of progressively lower resolution were defined surrounding the high-resolution regions in order to reduce numerical artifacts due to the difference in particle masses and to ensure that the large scale gravitational torques acting on re-simulated halos are accurately represented.

\begin{figure*}
\begin{center}
\includegraphics[scale=0.45]{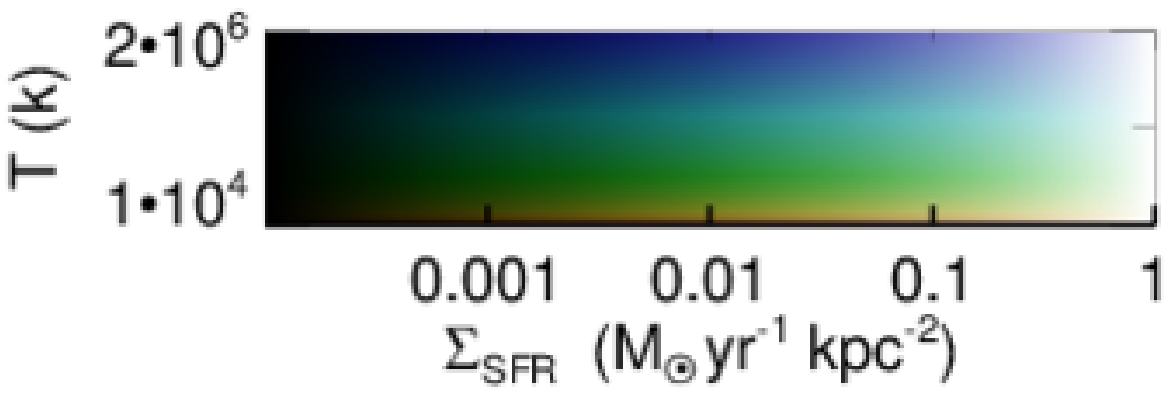}

\includegraphics[scale=\figsz]{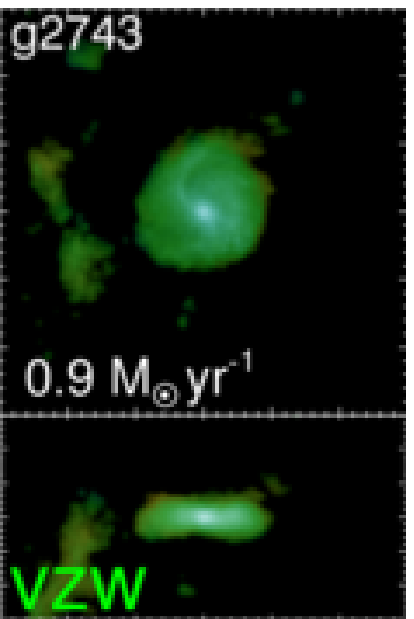}
\includegraphics[scale=\figsz]{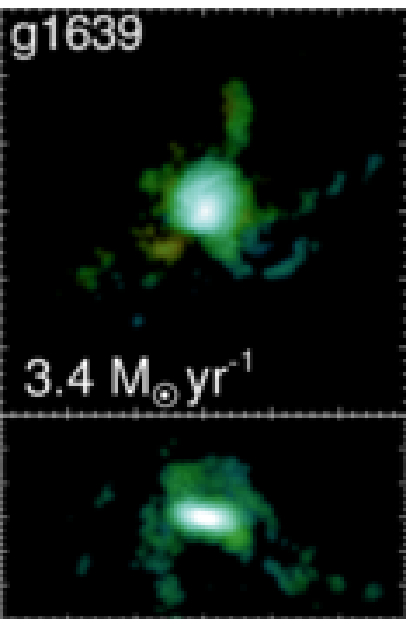}
\includegraphics[scale=\figsz]{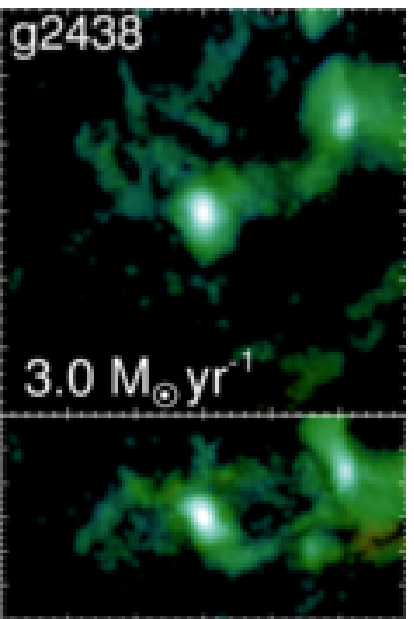}
\includegraphics[scale=\figsz]{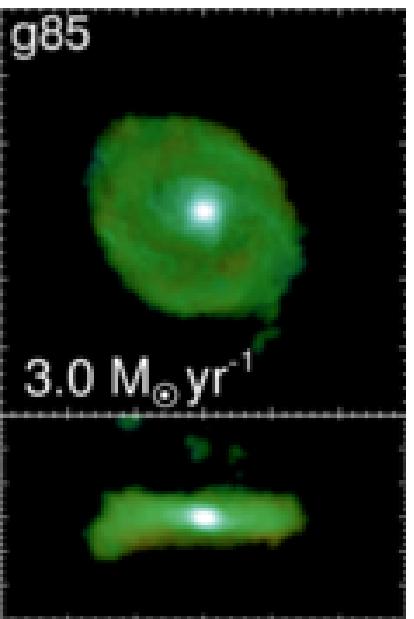}
\includegraphics[scale=\figsz]{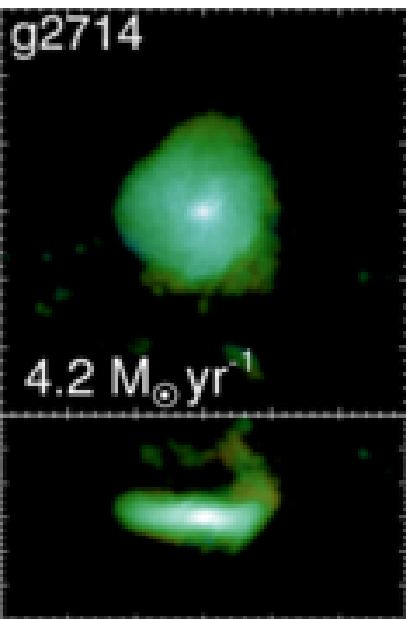}
\includegraphics[scale=\figsz]{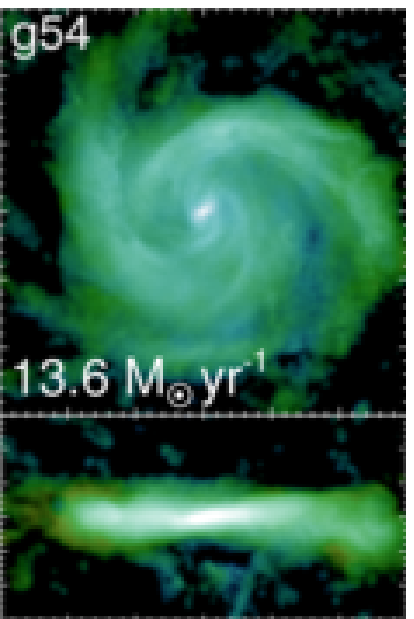}
\includegraphics[scale=\figsz]{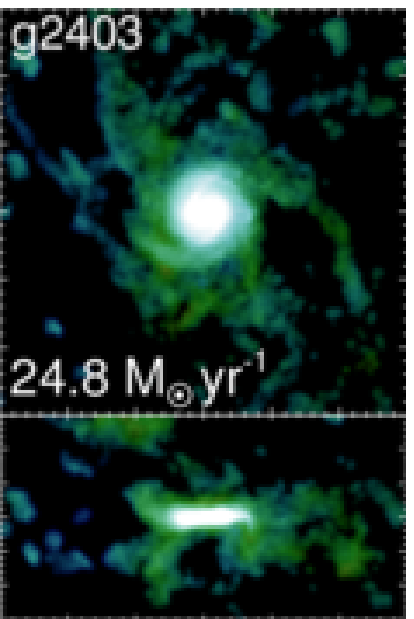}
\includegraphics[scale=\figsz]{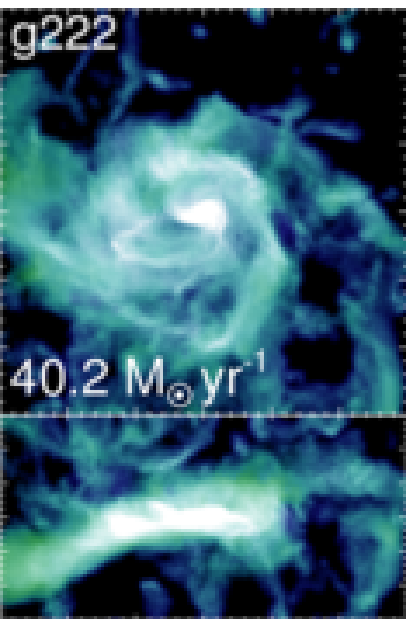}

 \smallskip

\includegraphics[scale=\figsz]{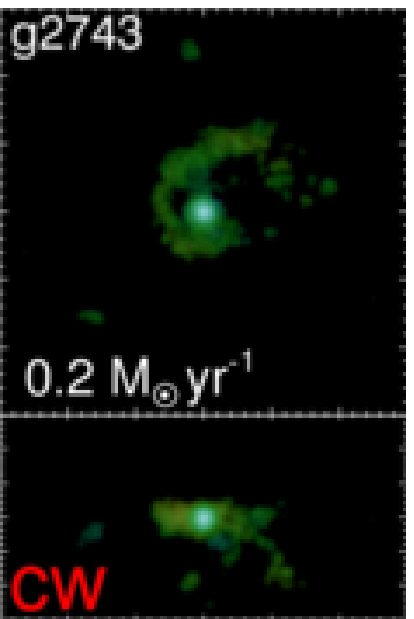}
\includegraphics[scale=\figsz]{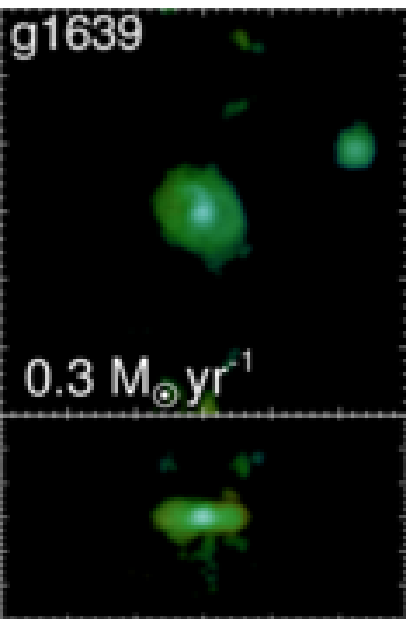}
\includegraphics[scale=\figsz]{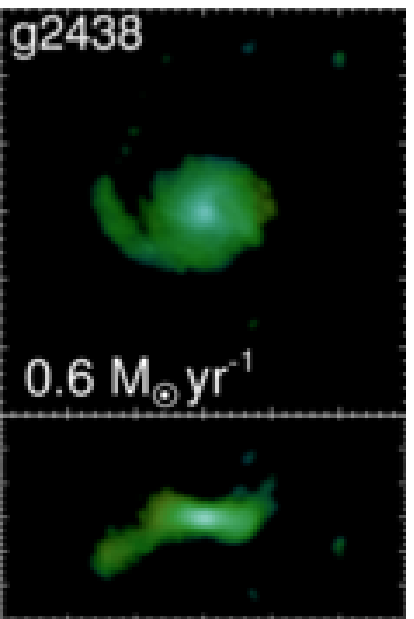}
\includegraphics[scale=\figsz]{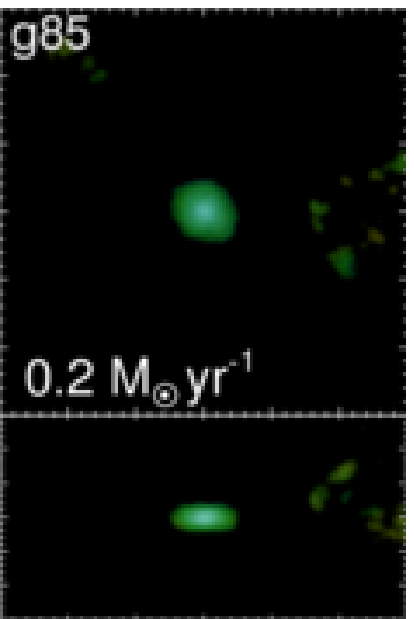}
\includegraphics[scale=\figsz]{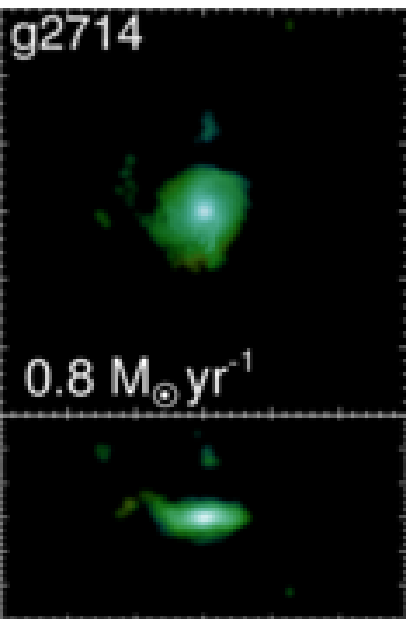}
\includegraphics[scale=\figsz]{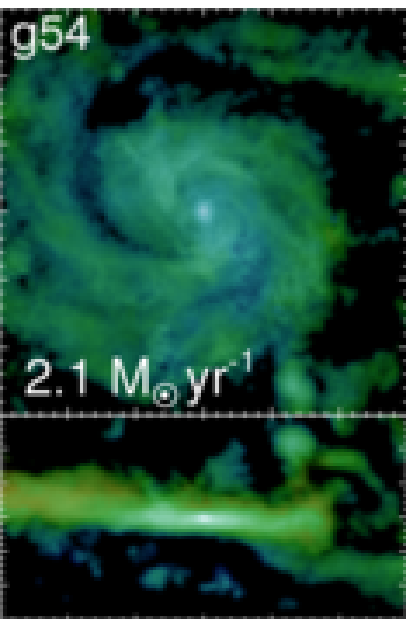}
\includegraphics[scale=\figsz]{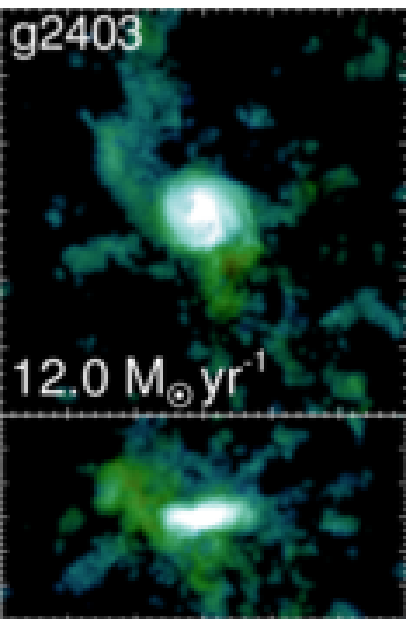}
\includegraphics[scale=\figsz]{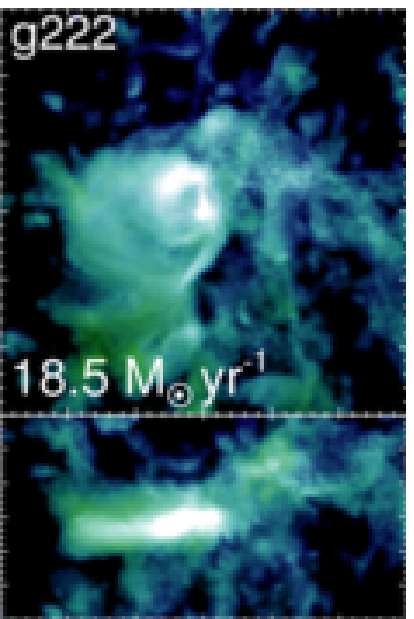}

 \smallskip

\includegraphics[scale=\figsz]{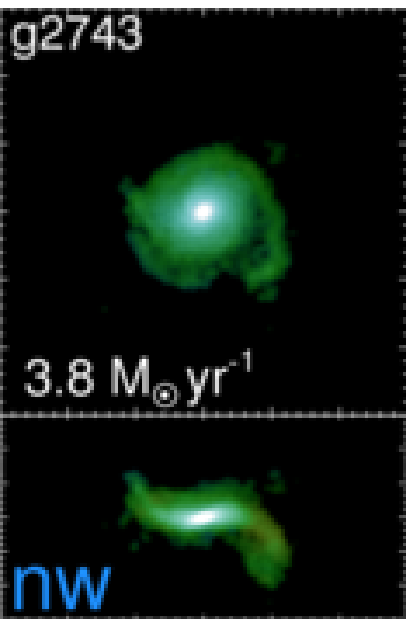}
\includegraphics[scale=\figsz]{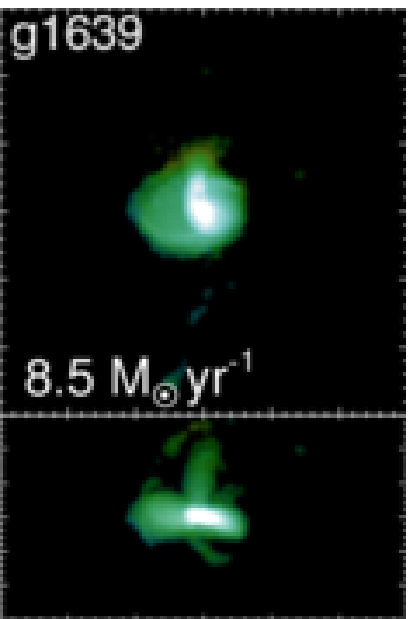}
\includegraphics[scale=\figsz]{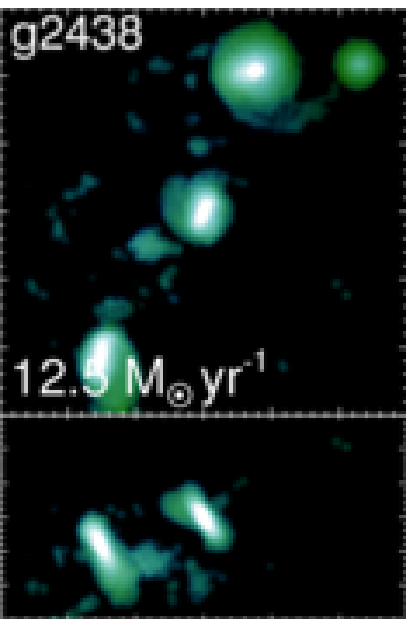}
\includegraphics[scale=\figsz]{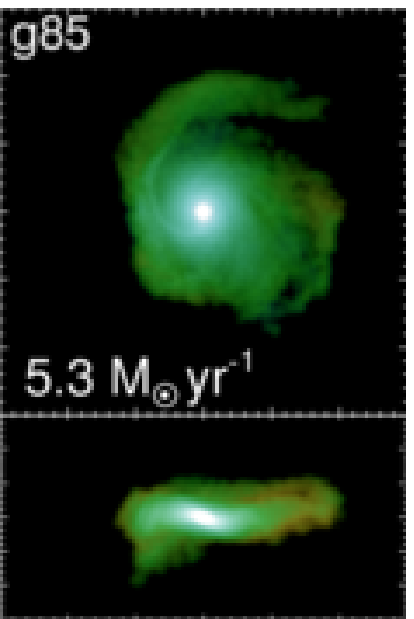}
\includegraphics[scale=\figsz]{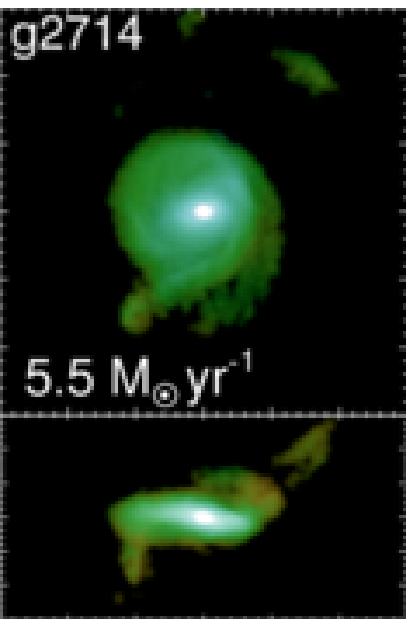}
\includegraphics[scale=\figsz]{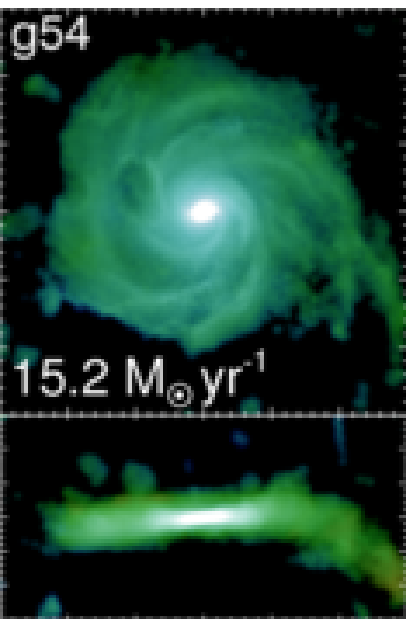}
\includegraphics[scale=\figsz]{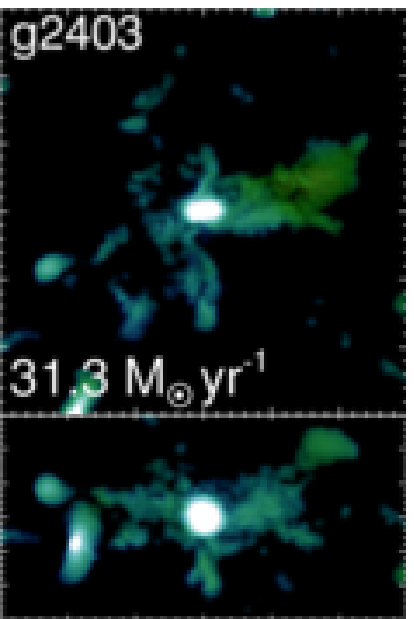}
\includegraphics[scale=\figsz]{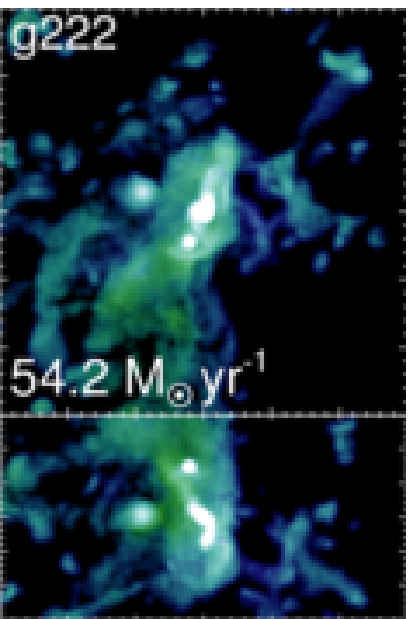}
 \end{center}
\caption{Star formation rate surface density maps color-coded according to the projected density-weighted temperature for our eight re-simulated galaxies at $z =2$.  For each galaxy, we show the projected gas distributions corresponding to simulations including momentum-driven winds ({\it top}), constant winds ({\it middle}), and no winds ({\it bottom}).  Face-on and edge-on views of each galaxy correspond to the total angular momentum of the gas component.  Galaxies are ordered from left to right for increasing stellar mass according to simulations with momentum-driven winds (with total halo masses ranging from $\sim 10^{11}$ to $2 \times 10^{12}$\,\Msun).  Total SFRs are listed for all galaxies and wind models.  The region shown is 30 kpc (physical) across.}
\label{fig:gasima}
\end{figure*}

The resulting zoom simulations have (high-resolution) gas particle mass $m_{\rm
gas} \approx 2.3 \times 10^5$\,\Msun, dark matter particle mass
$m_{\rm DM} \approx 1.2 \times 10^6$\,\Msun, and softening length
$\epsilon \approx 0.47\,h^{-1}$\,kpc comoving ($\sim 224$\,pc
physical at $z = 2$), equivalent to $2 \times 1024^3$ particles
homogeneously distributed in a $[24\,\hmpc]^3$ box.   Additionally,
we run similar zoom simulations with a factor of 2 lower spatial resolution and
a factor of eight lower mass resolution in order to test our results
for numerical convergence.   
A total of 215 snapshot files logarithmically spaced in the redshift range $z = 2$--11 were produced for each simulation, corresponding to time intervals ranging from $\sim 5$ to 25\,Myr.

Galaxies were identified by means
of the Spline Kernel Interpolative Denmax algorithm ({\sc
skid}\footnote[1]{http://www-hpcc.astro.washington.edu/tools/skid.html}) at
all available snapshots independently, and are thus defined as bound
groups of star-forming gas particles (i.e., gas particles with
densities above the threshold for star formation) and star particles
\citep[see][]{ker05}.  
We used a spherical overdensity algorithm to associate each {\sc skid}-identified
galaxy with a dark matter halo, where the virial radius was defined to enclose a mean
density given by \citet{kitayama96}.  Overlapping halos were
grouped together so that every final halo has a central
galaxy and a number of satellite galaxies by construction.  

The sample of galaxies presented here was defined by selecting the eight most massive central galaxies located within the re-simulated volumes, with the additional requirement of having no contamination by low-resolution dark matter particles within their virial radius; several central galaxies were located near the boundaries of the high-resolution region and were rejected because of contamination.      
Even though this sample is not mass nor volume complete, our galaxies are representative of ``normal" $z = 2$ systems, with similar gas fractions and SFRs relative to similar mass galaxies from the parent population.

The full evolution of each $z = 2$ galaxy was reconstructed
by identifying its most massive progenitor at
all available redshift snapshots.  Galaxies were also identified across
simulations with different wind models to allow for a detailed model
comparison.  Following \citet{ang13}, we calculated structural and kinematic properties 
of each galaxy relative to the position of its most bound gas particle, which
is a more meaningful and more stable definition of the nominal center of the galaxy 
compared to that computed by {\sc skid}, especially during close galaxy encounters and galaxy
mergers.

\begin{figure*}
\begin{center}
\includegraphics[scale=0.45]{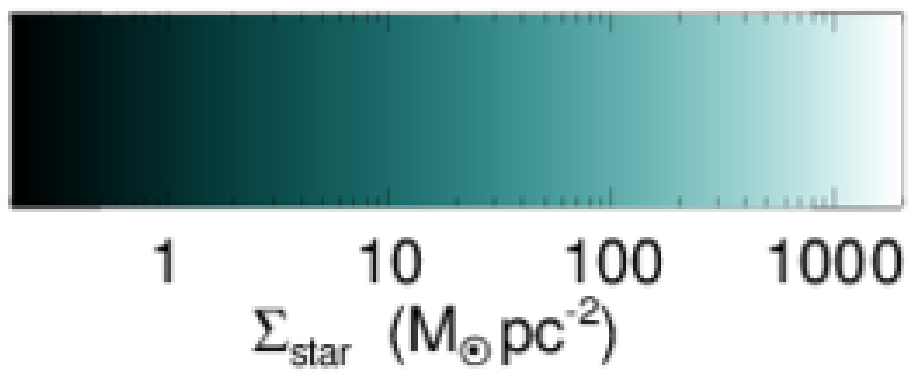}

\includegraphics[scale=\figsz]{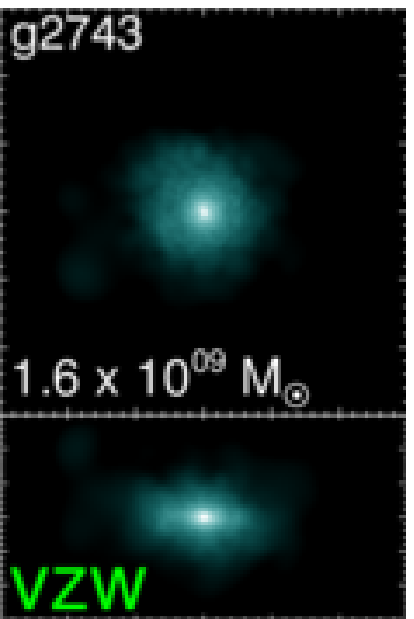}
\includegraphics[scale=\figsz]{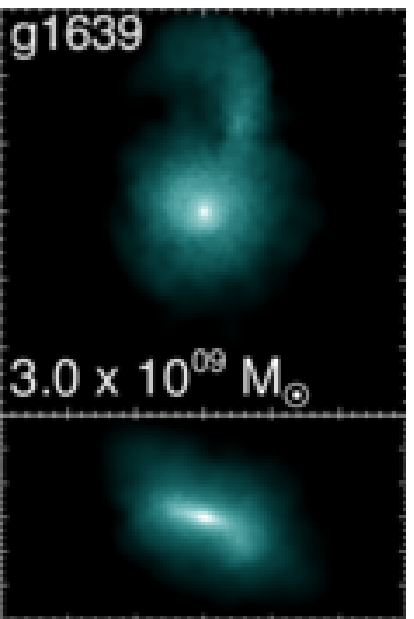}
\includegraphics[scale=\figsz]{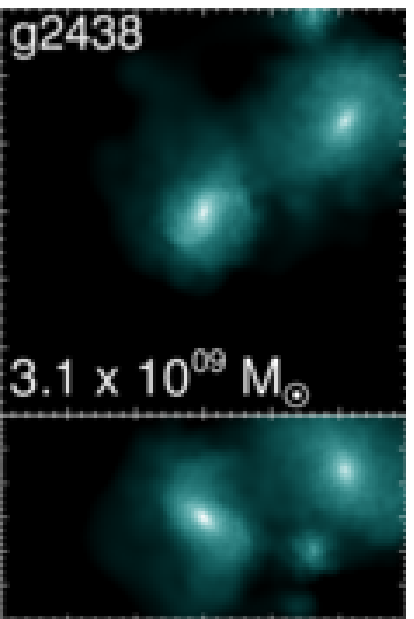}
\includegraphics[scale=\figsz]{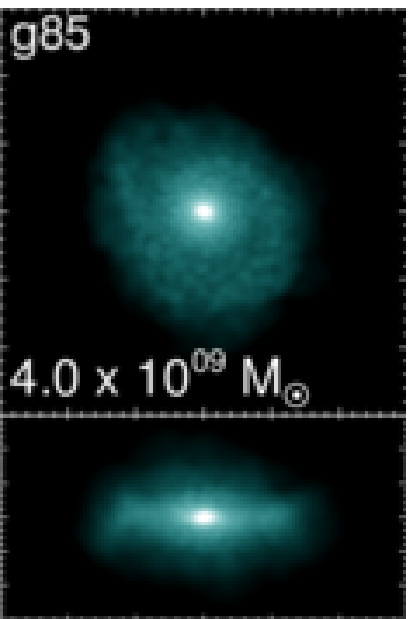}
\includegraphics[scale=\figsz]{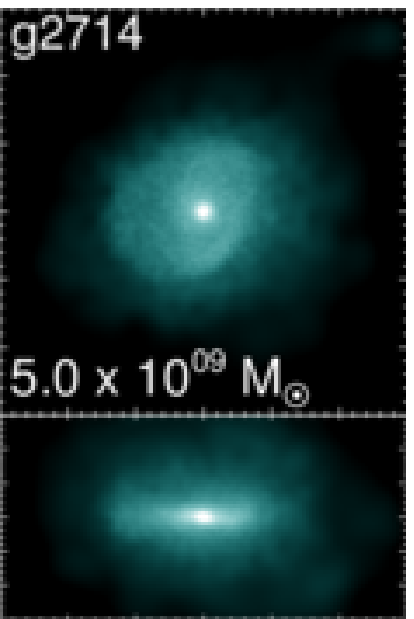}
\includegraphics[scale=\figsz]{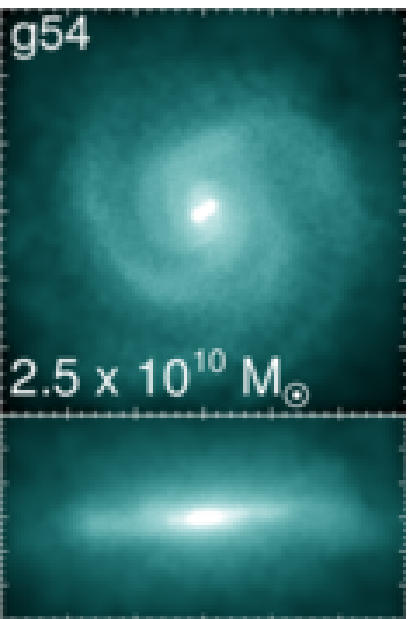}
\includegraphics[scale=\figsz]{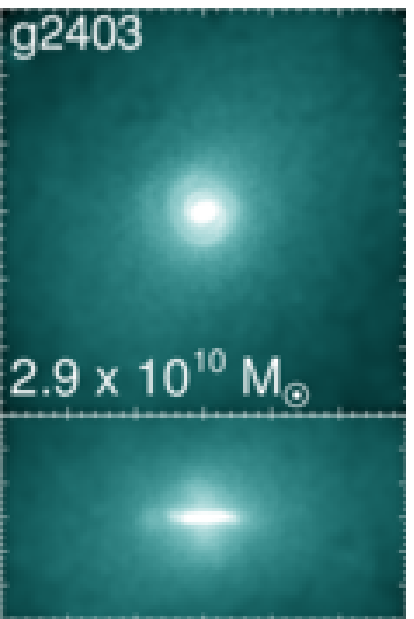}
\includegraphics[scale=\figsz]{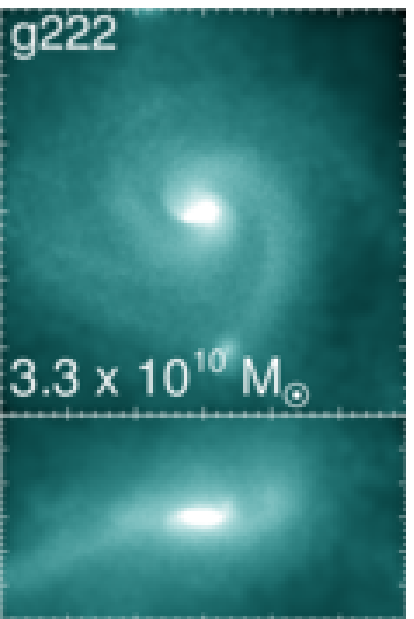}

 \smallskip

\includegraphics[scale=\figsz]{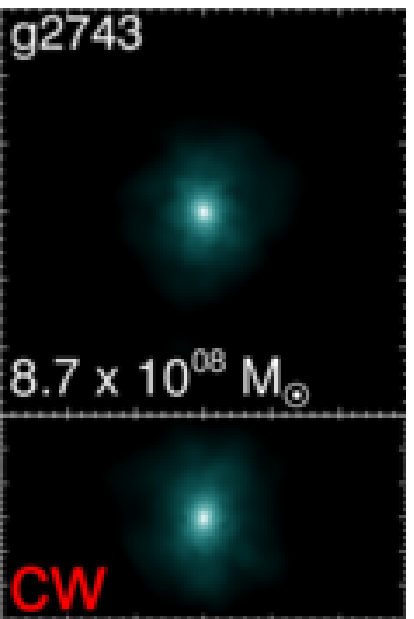}
\includegraphics[scale=\figsz]{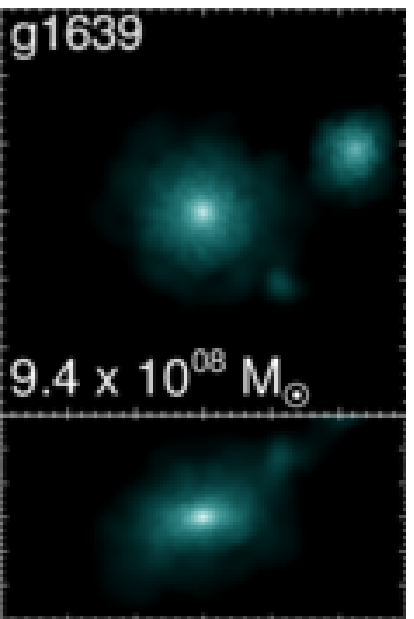}
\includegraphics[scale=\figsz]{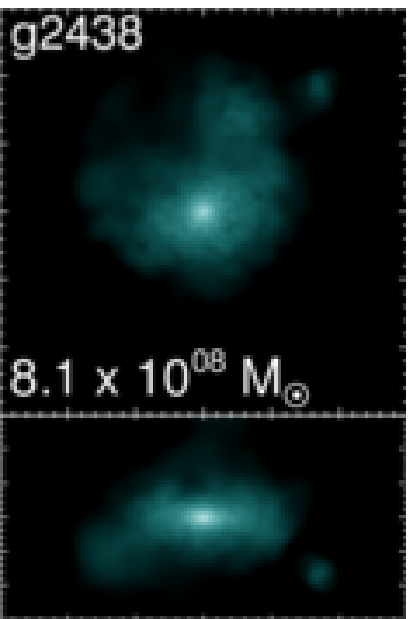}
\includegraphics[scale=\figsz]{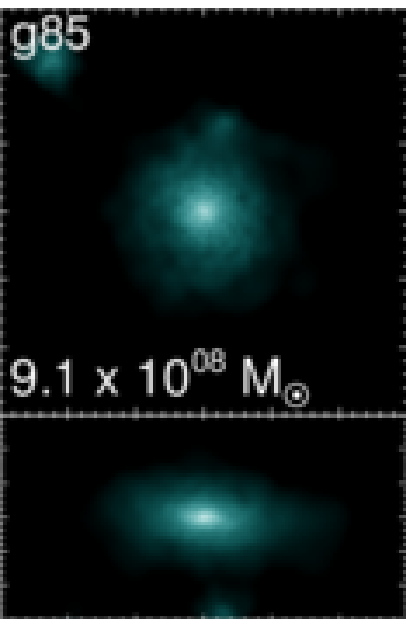}
\includegraphics[scale=\figsz]{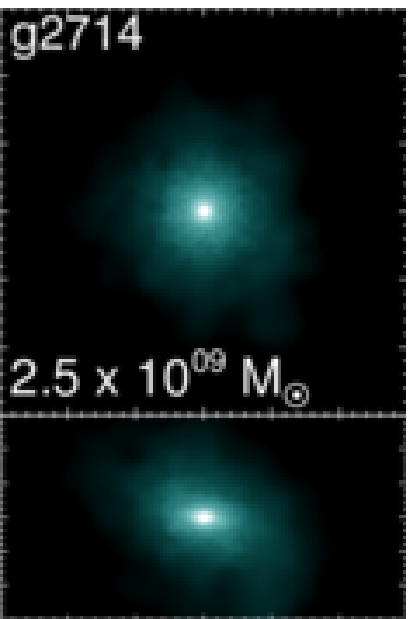}
\includegraphics[scale=\figsz]{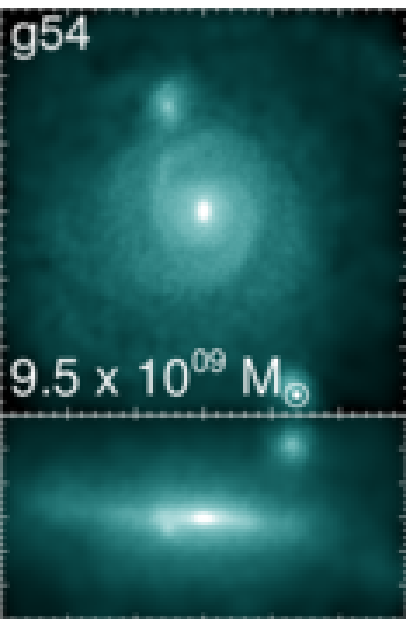}
\includegraphics[scale=\figsz]{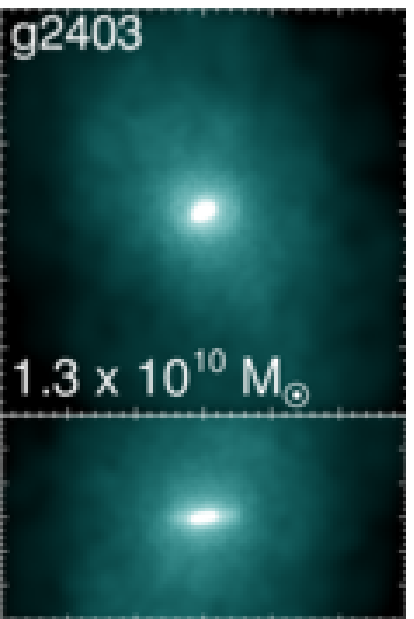}
\includegraphics[scale=\figsz]{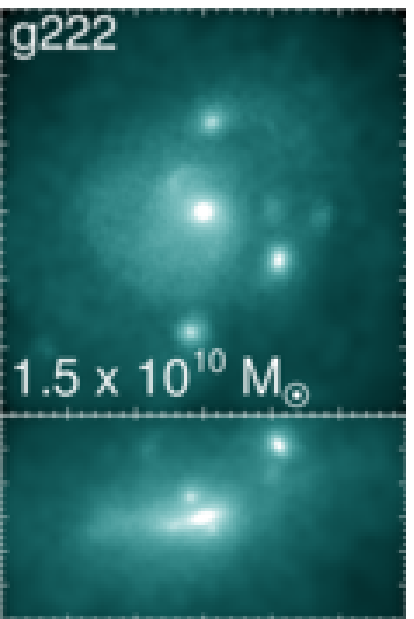}

 \smallskip

\includegraphics[scale=\figsz]{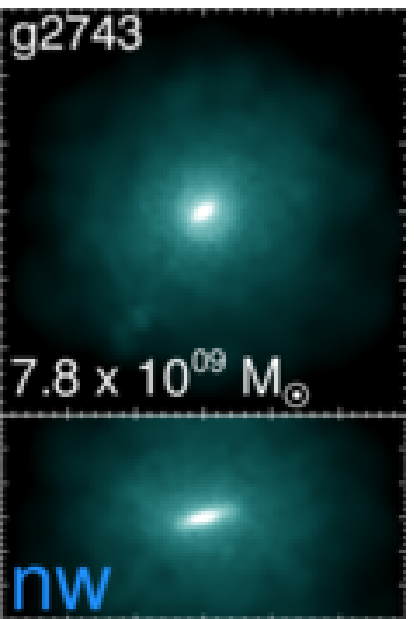}
\includegraphics[scale=\figsz]{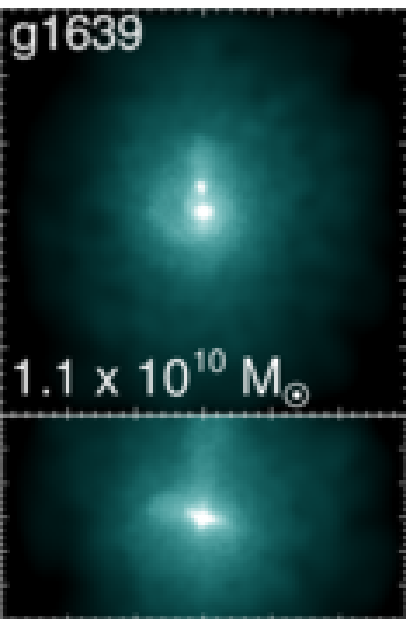}
\includegraphics[scale=\figsz]{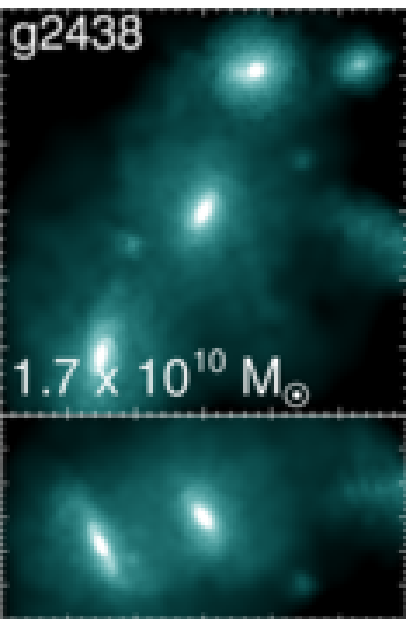}
\includegraphics[scale=\figsz]{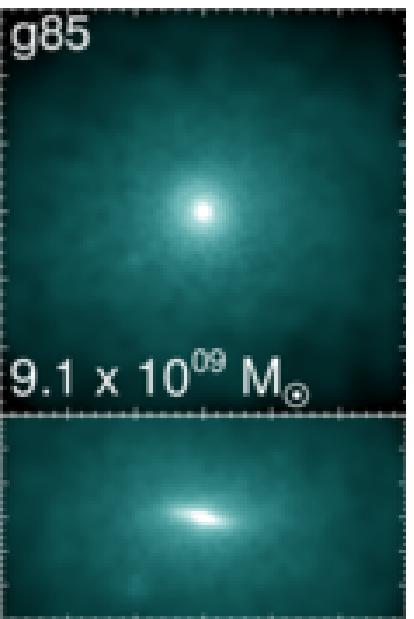}
\includegraphics[scale=\figsz]{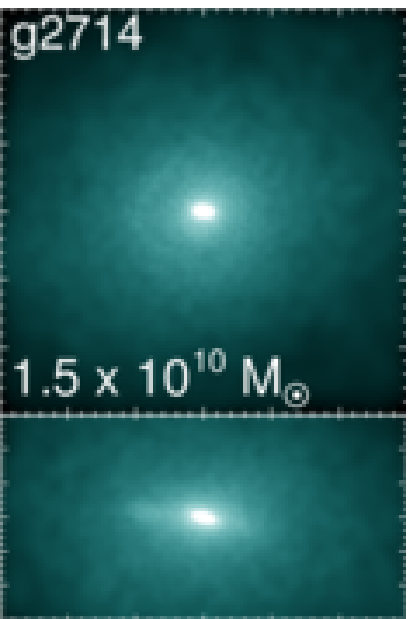}
\includegraphics[scale=\figsz]{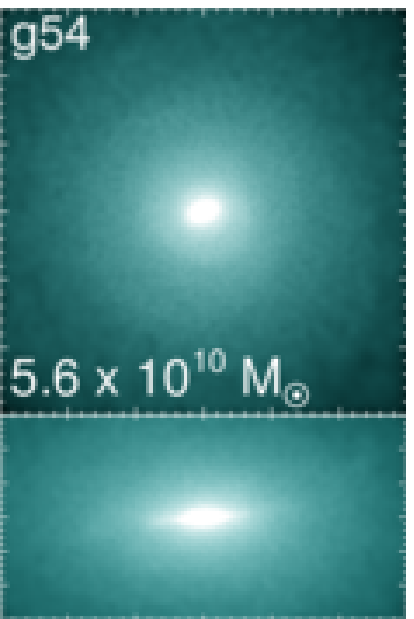}
\includegraphics[scale=\figsz]{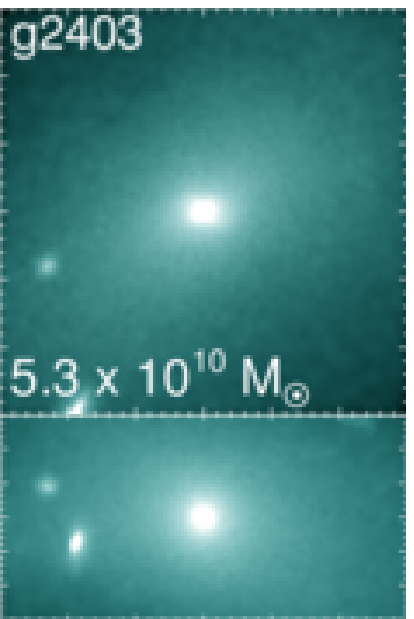}
\includegraphics[scale=\figsz]{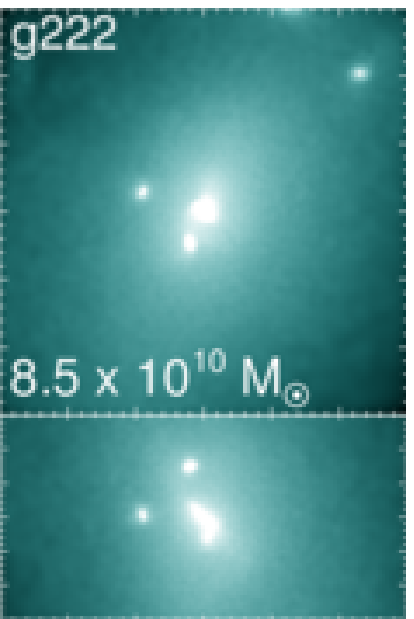}
 \end{center}
\caption{Same as Figure~\ref{fig:gasima} for the projected stellar surface density.  Total stellar masses are listed for all galaxies and wind models.}
\label{fig:starima}
\end{figure*}

\section{Galaxy Sample}\label{sec:samp}

We begin by describing the sample of eight galaxies used in this
work.  At $z = 2$, the masses of dark matter halos range
from $\sim 10^{11}$ to $2 \times 10^{12}$\,\Msun~and are typically
resolved with $\sim 2 \times 10^{5}$ to $4 \times 10^{6}$ particles.
As we describe in detail throughout this paper, galaxy properties
are highly dependent on the model adopted for galactic outflows.
For our fiducial simulations including momentum-driven winds (vzw
model), galaxies have stellar masses ranging from $\sim 1.6 \times
10^{9}$ to $3.3 \times 10^{10}$\,\Msun~and SFRs in the range $\sim
1$--40\,\Msun\,yr$^{-1}$ at $z = 2$.  Figure~\ref{fig:gasima} shows
face-on and edge-on views of the star-forming gas of galaxies at $z = 2$.  
Our galaxy sample spans a wide range
of sizes and morphologies, with the star-forming gas extending up
to scales of 2--15\,kpc from their centers.  Some galaxies are
located in high density environments and undergo frequent interactions
and mergers (e.g., galaxy g222) while others are
characterized by a smoother evolution, in lower density environments
(e.g., galaxy g54).  Despite this, most galaxies appear
to be rotationally supported disks at $z = 2$.  Figure~\ref{fig:starima}
shows the projected stellar distribution corresponding to the same galaxies.  
Stellar disks can also be visually identified
for most galaxies but in this case a higher fraction of stars
seems to correspond to a spheroidal component.

Galaxies are identified across the various wind simulations via their 
halos, making it possible to analyze the effects
of galactic winds in a galaxy-by-galaxy basis.  However, when
comparing the morphologies of galaxies across wind models it is
important to note that galaxy interactions and mergers are inherently
random processes.  Small deviations of orbital parameters in different
simulations can result in rather different morphologies at a given
time, and therefore it is not trivial to compare morphological
features such as spiral arms or tidal tails among the different
models (e.g., galaxies g1639, g2438, g85, and g54).  Despite this, a quick
visual inspection of Figures~\ref{fig:gasima} and~\ref{fig:starima}
reveals significant differences between wind prescriptions.  Constant
winds are very efficient in removing gas from galaxies and result
in lower mass galaxies with typically lower SFR surface densities
compared to the momentum-driven winds.  This is particularly evident
in the low mass range, where
the formation of large-scale disks of star-forming gas is highly
suppressed (e.g., galaxies g2743 and g85).  In contrast, simulations with no winds usually form
more massive and more concentrated galaxies compared to the vzw
model, with most of the star formation happening in the central
regions of galaxies.

\section{The Effects of Winds}\label{sec:eff}

\begin{figure*}
\begin{center}
\includegraphics[scale=0.54]{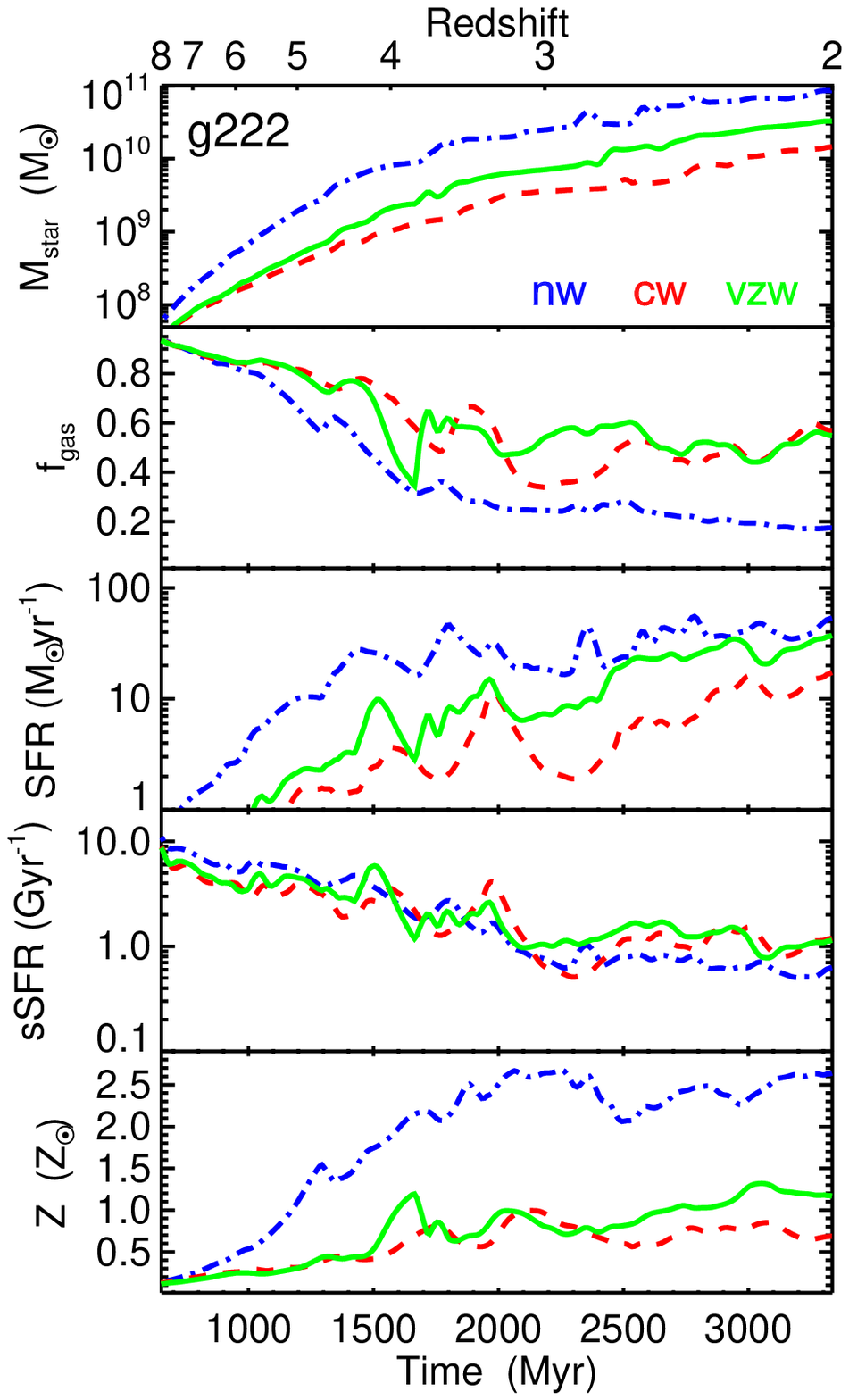}
\includegraphics[scale=0.54]{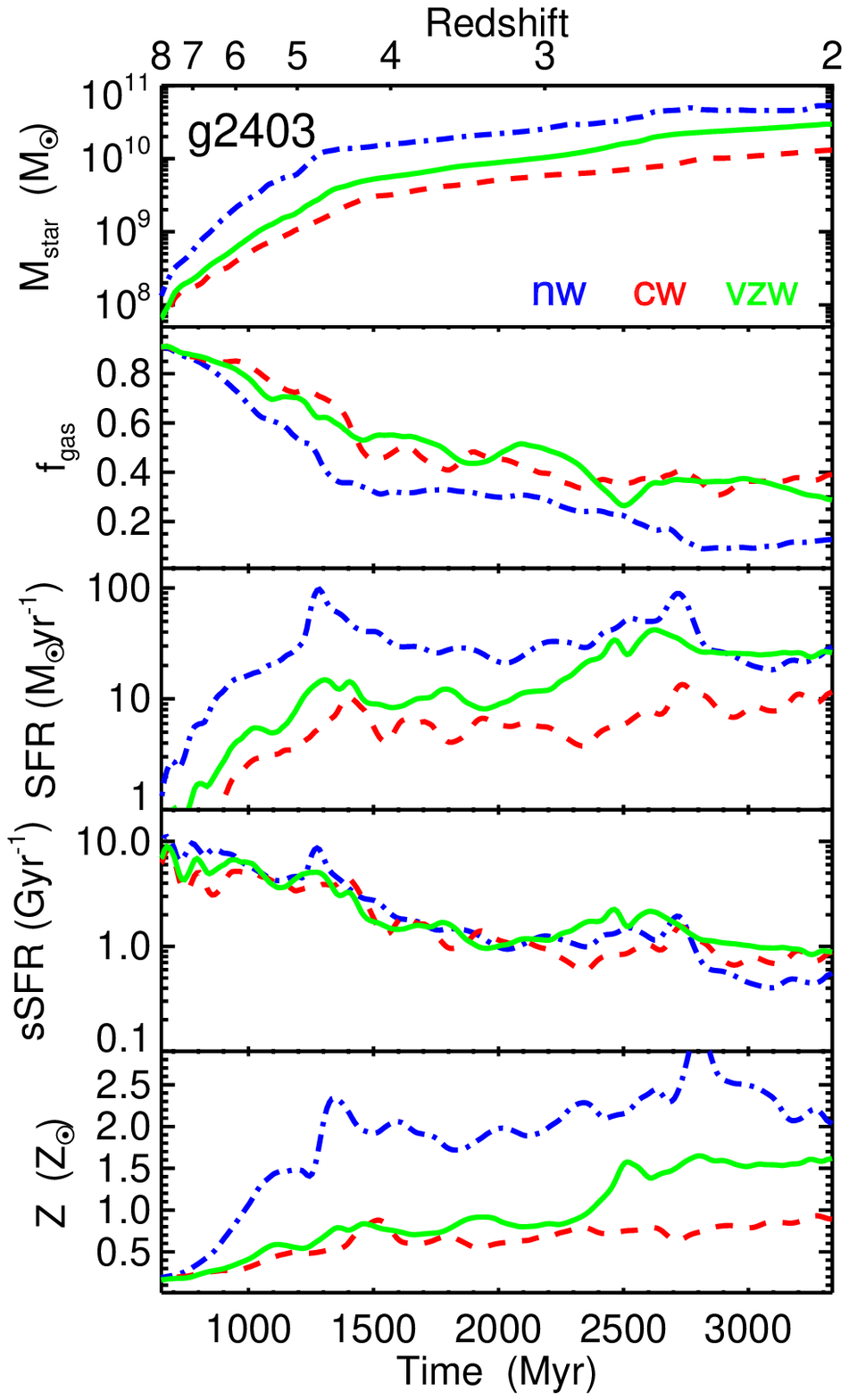}
\includegraphics[scale=0.54]{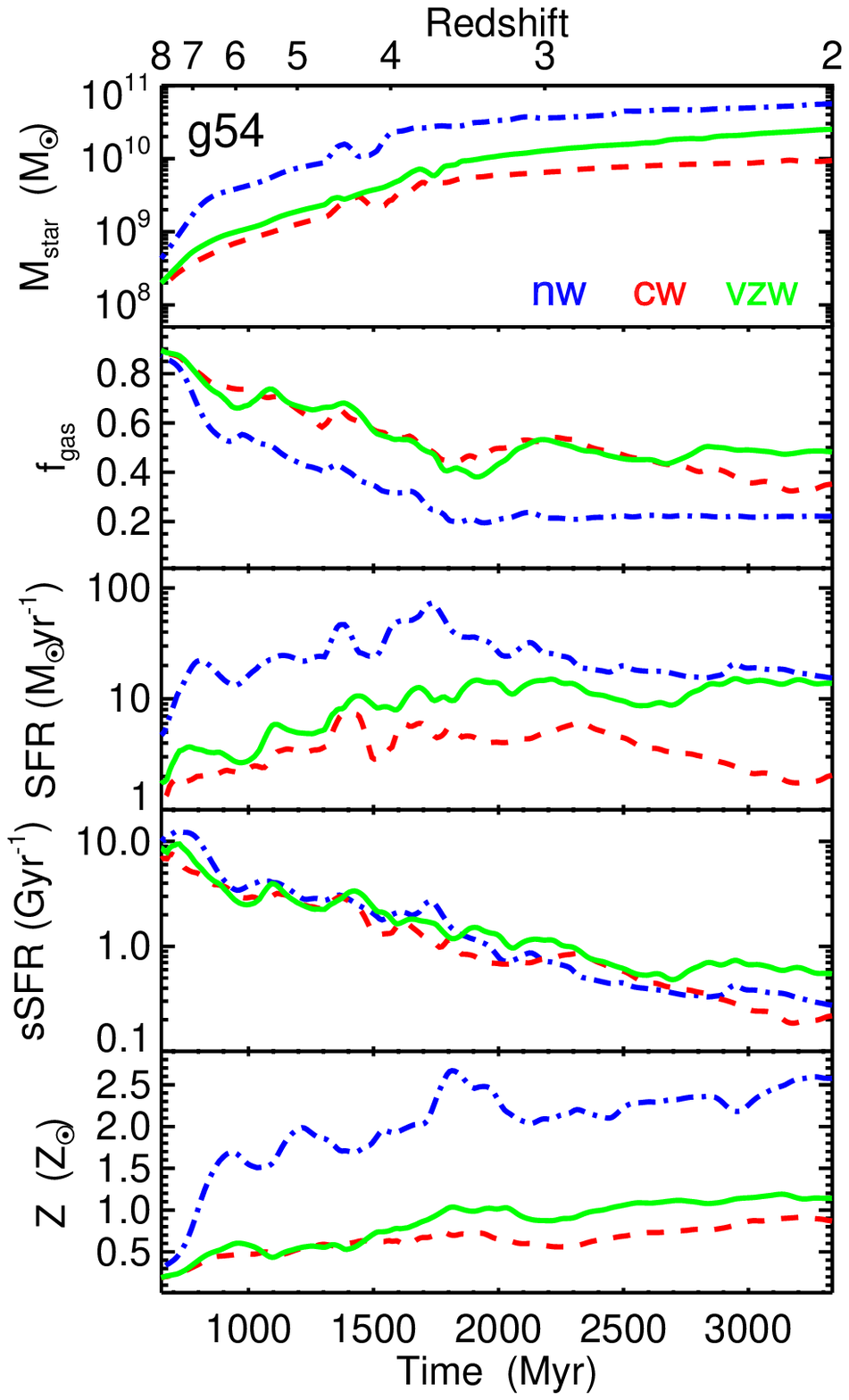}
\end{center}
\caption{Time evolution of galaxies g222, g2403, and g54 from $z = 8$ to $z = 2$ for simulations including models for momentum-driven winds (green solid lines), constant winds (red dashed lines), and no winds (blue dot-dashed lines).  {\it From top to bottom}: stellar mass, gas fraction, star formation rate, specific star formation rate, and average gas phase metallicity (SFR-weighted).  The time evolution of all physical quantities has been smoothed over time intervals of $\sim 50$\,Myr.}
\label{fig:evol}
\end{figure*}

\begin{figure*}
\begin{center}
\includegraphics[scale=0.54]{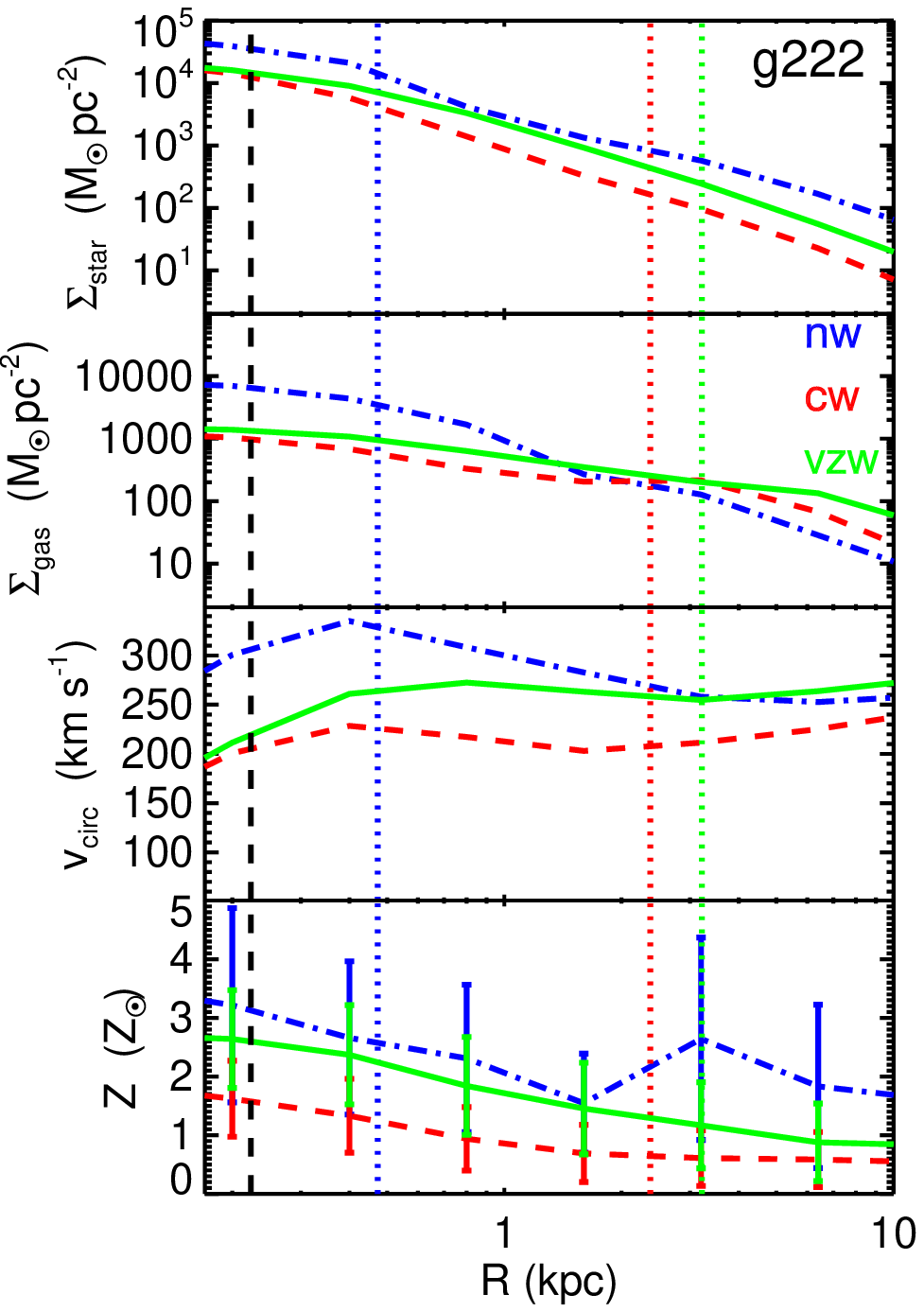}
\includegraphics[scale=0.54]{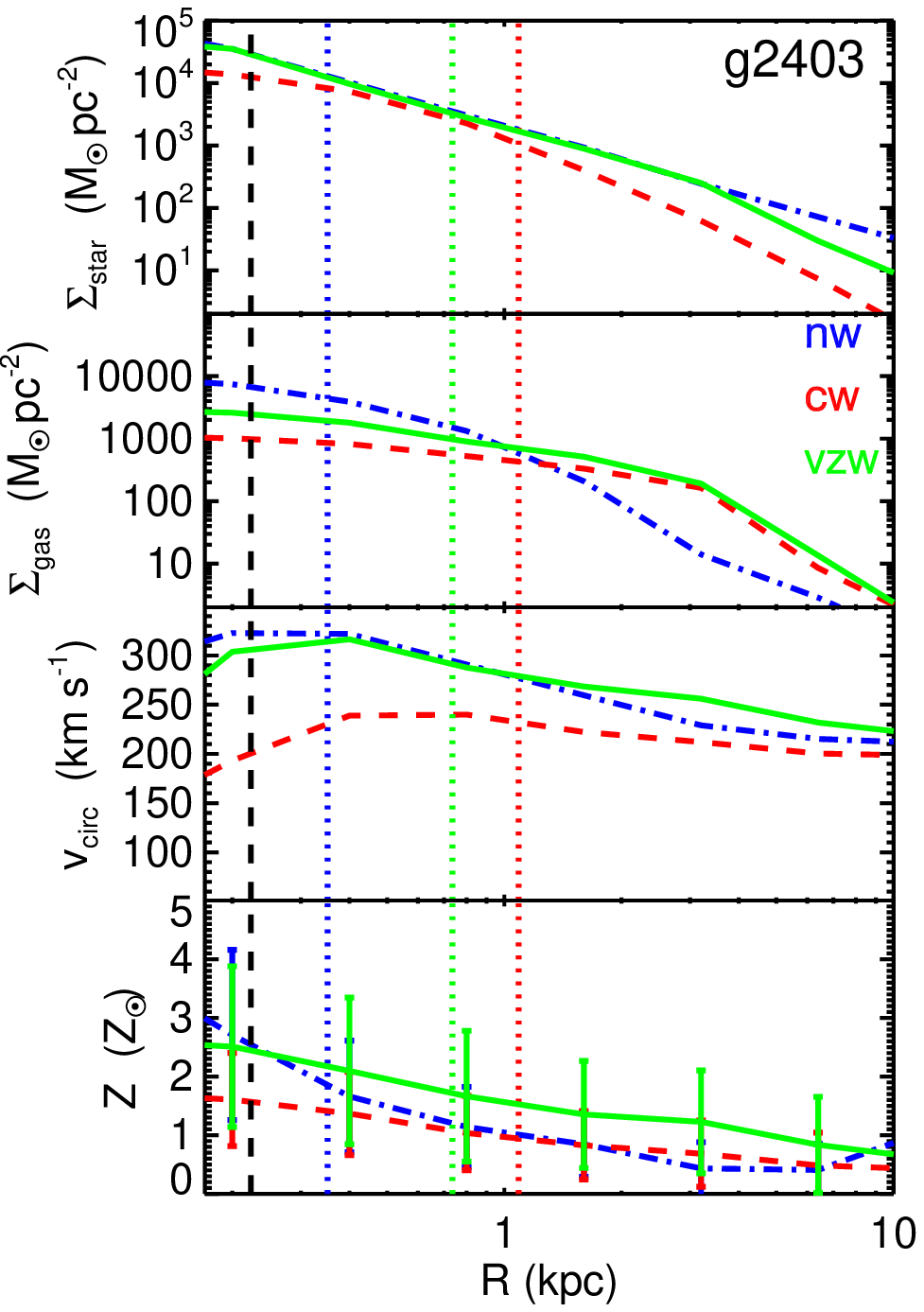}
\includegraphics[scale=0.54]{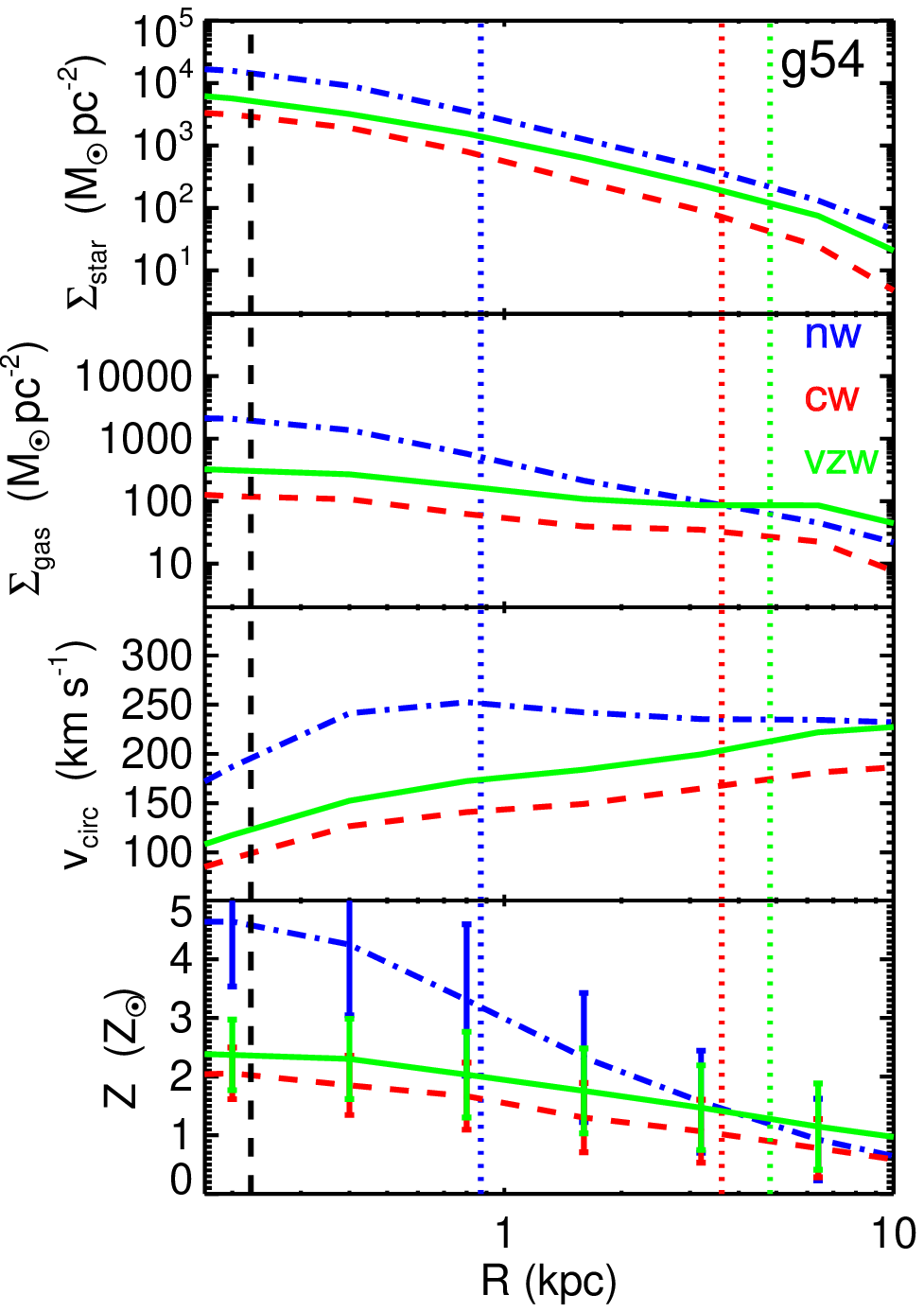}
\end{center}
\caption{Radial profiles of galaxies g222, g2403, and g54 at $z = 2$ for simulations including models for momentum-driven winds (green solid lines), constant winds (red dashed lines), and no winds (blue dot-dashed lines).  {\it From top to bottom}: azimuthally averaged face-on stellar and gas surface densities, circular velocity (calculated as $v_{\rm circ}^{2} = GM_{\rm enc}(r)/r$ for the enclosed mass $M_{\rm enc}(r)$ within radius $r$), and average gas phase metallicity (SFR-weighted).  Error bars in the bottom panel indicate the spread of the particle distribution within each radial bin for each galaxy.  Black vertical dashed lines show the softening length of the simulation at $z = 2$ ($\epsilon \approx 224$\,pc physical) while the green, red, and blue vertical doted lines correspond to the radius enclosing half of the total SFR for the different wind prescriptions.}
\label{fig:prf}
\end{figure*}

\subsection{Time Evolution}

Each galaxy is identified not only across simulations with different
wind models but also back in time at all available snapshots, as the
central galaxy in the resimulated halo.
Figure~\ref{fig:evol} shows examples of the time evolution of
galaxies in terms of their stellar mass, gas fraction, SFR, and
metallicity.  Since we are interested in the evolution of galaxies
over cosmological time scales, the time evolution of all physical
quantities has been smoothed over time intervals of $\sim 50$\,Myr.
Major mergers can still be identified as abrupt changes in the
stellar mass of galaxies (e.g., galaxy g222 between redshifts $z =
2$--3 and g54 at $z \approx 4$).  These events are generally followed
by a temporary increase in the SFR of galaxies but their overall
evolution is dominated by smooth accretion, as has been reported
for simulations of high redshift galaxies elsewhere
\citep[e.g.,][]{dek09,dav10}.

Simulations with no winds clearly result in higher stellar masses
for all galaxies at all times.  
Very early on, the SFR of galaxies increases very rapidly for
simulations with no winds.  When galactic outflows are included,
galaxies regulate themselves resulting in significantly lower SFRs
relative to the no wind model at $z > 3$.  
For momentum-driven winds, recycling of the outflowing gas back into galaxies occurs on a time scale less than the {\it Hubble} time, providing an additional gas supply which is not available in simulations with no winds \citep{opp10}.  This additional gas supply results in comparatively higher SFRs toward the end of the simulation at $z = 2$. 
Constant winds are, however, more efficient in removing cold gas from galaxies owing to the higher outflow velocities ($v_{\rm w} = 680$\,km\,s$^{-1}$) for comparatively similar mass loading factors ($\eta = 2$)---for the three most massive systems in our galaxy sample (g222, g2403, and g54), we find typical wind velocities $v_{\rm w} \approx 500$--600\,km\,s$^{-1}$ and mass loadings $\eta \approx 1.4$--1.9 in the redshift range $z = 8 \rightarrow 2$ for the momentum conserving model.  This results in systematically lower SFRs for the constant wind model at all times.  

Interestingly, specific SFRs follow a common trend regardless of the wind model adopted:
galaxies have the highest specific SFRs very early on (as high as
10\,Gyr$^{-1}$ at $z = 8$) and then decrease monotonically with
decreasing redshift.  To first order, outflows reduce SFRs by a factor $1/(1+\eta)$ relative to halo accretion rates \citep{dav12}, which leads to specific SFRs that remain the same across wind models (SFRs and stellar masses are reduced by the same amount).  Wind recycling accretion, however, yields increasing specific SFRs at later times, especially in the momentum-driven wind model.

Gas fractions are systematically lower for the no wind simulations compared to either
the constant wind or the momentum-driven wind models.  While galaxies g222, g2403, and g54 have gas fractions below 20\,\% for simulations with no winds at $z = 2$, galactic outflows are able to maintain gas fractions a factor of two higher at the end of the simulation, in better agreement with observations \citep{tacconi10}.
Gas fractions may be conveniently expressed in terms of the depletion time $t_{\rm dep} \equiv M_{\rm gas}/{\rm SFR}$ and the specific SFR as $f_{\rm gas} \equiv [(1+(t_{\rm dep}\,{\rm sSFR})^{-1})]^{-1}$ \citep{dav12}.  Since specific SFRs are roughly insensitive to wind model early on and the depletion time is shorter in higher stellar mass galaxies \citep[$t_{\rm dep} \propto M_{*}^{-0.3}$;][]{dav11b}, no wind simulations yield lower gas fractions.   

Finally, metallicities are significantly higher for all galaxies in the no
wind model, owing to a greater conversion of baryons into stars
and a lack of ejection of metals into the IGM \citep{hirschmann13}.
These results are qualitatively consistent with those obtained using
lower-resolution non-zoom cosmological runs~\citep{dav11a,dav11b}.

\subsection{Radial Profiles}

The radial profiles of galaxies reveal interesting differences among
wind models, as shown in Figure~\ref{fig:prf} for galaxies g222,
g2403, and g54 at $z = 2$.   Face-on projected stellar and gas surface densities,
circular velocity, and metallicity as a function of radial distance
from the centers of galaxies have been calculated as average values
within logarithmically spaced radial bins.  The most massive galaxies
(g222 and g2403) reach resolved stellar and gas surface densities in the
range $\Sigma_{\rm star} = 10^{4}$--$10^{5}$\,\Msun\,pc$^{2}$ and $\Sigma_{\rm gas}
= 10^{3}$--$10^{4}$\,\Msun\,pc$^{2}$ respectively, with differences
up to an order of magnitude among wind models.  At their centers,
stellar surface densities of most galaxies reach values well
above $10^{3}$\,\Msun\,pc$^{2}$ despite gravitational forces being
softened at scales $\sim \epsilon \approx 224$\,pc (Figure~\ref{fig:starima}).

\begin{figure*}
\begin{center}
\includegraphics[scale=0.34]{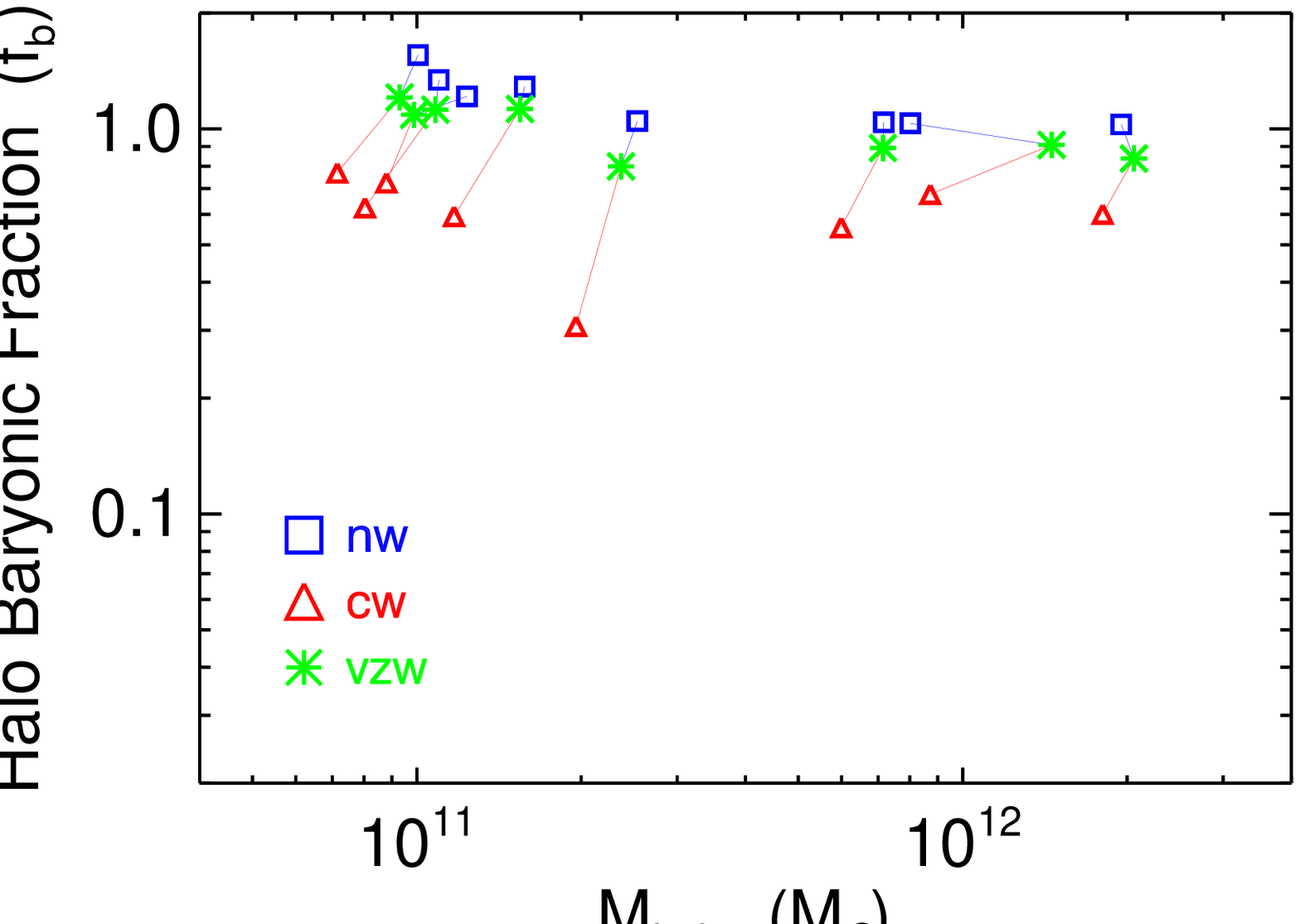}
\includegraphics[scale=0.34]{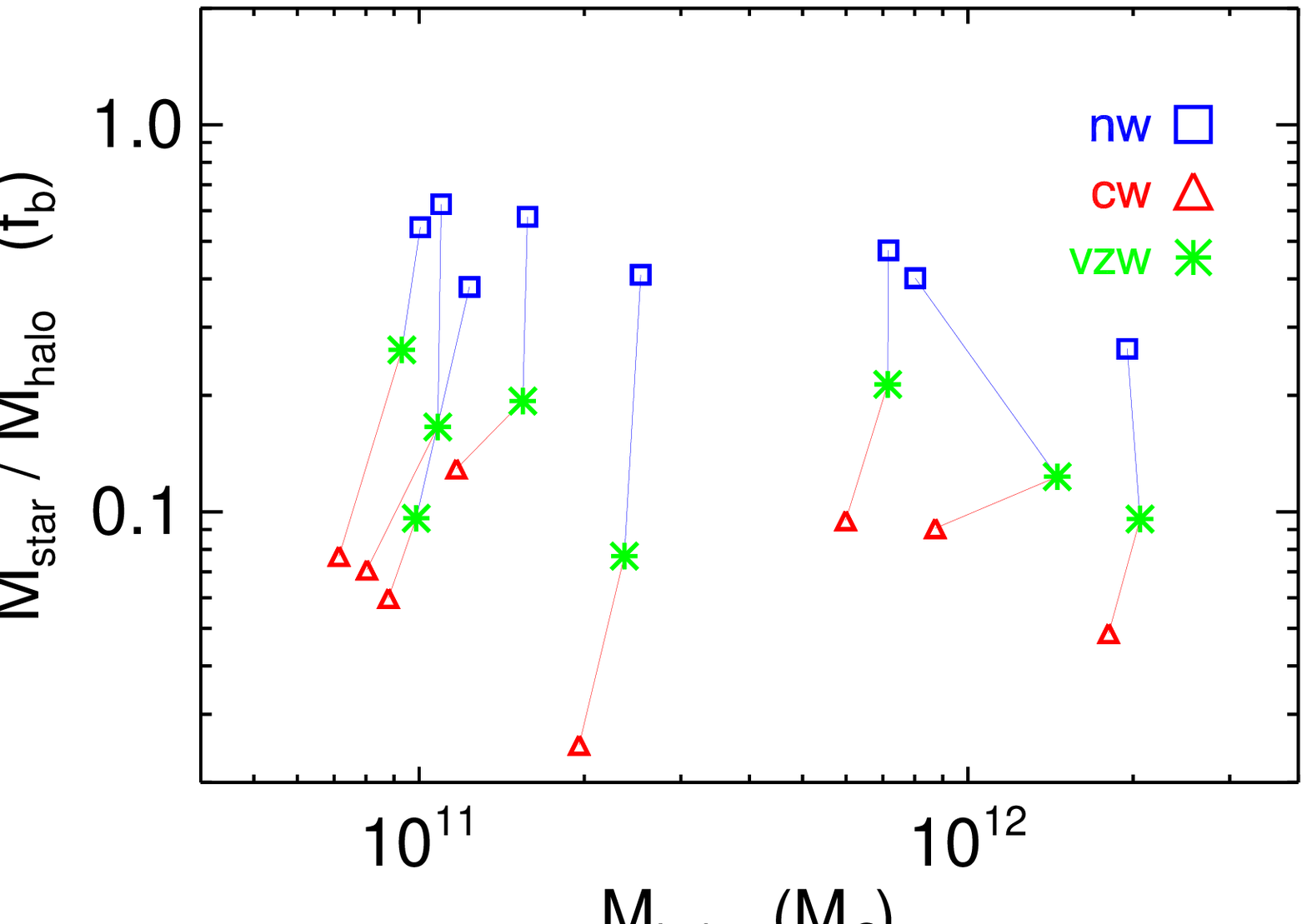}
\includegraphics[scale=0.34]{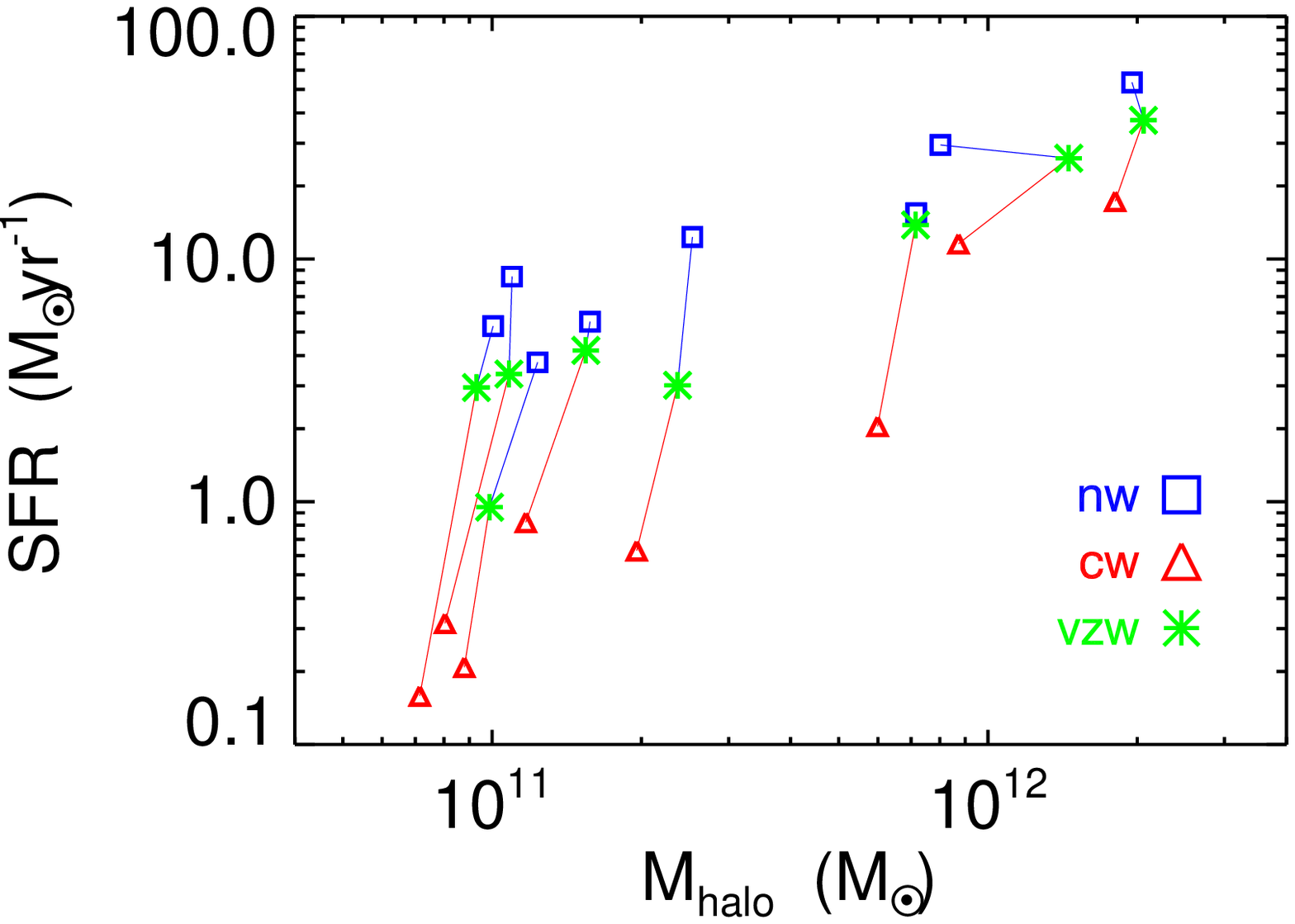}
\end{center}
\caption{Effects of galactic winds on global properties of $z = 2$ galaxies and their parent halos as a function of halo mass.  {\it Left}: total halo baryonic fraction in units of the cosmological baryonic fraction ($f_{b} = 0.165$).  {\it Middle}: ratio of central stellar mass to halo mass in units of $f_{b}$.  {\it Right}: central galaxy star formation rate.  Galaxies from simulations including momentum-driven winds (vzw), constant winds (cw), and no winds (nw) are shown as green star symbols, red triangles, and blue squares respectively.  Red and blue solid lines connect galaxies from the vzw simulations with their galaxy counterparts from the cw and nw simulations respectively.}
\label{fig:vshalo}
\end{figure*}

\begin{figure*}
\begin{center}
\includegraphics[scale=0.5]{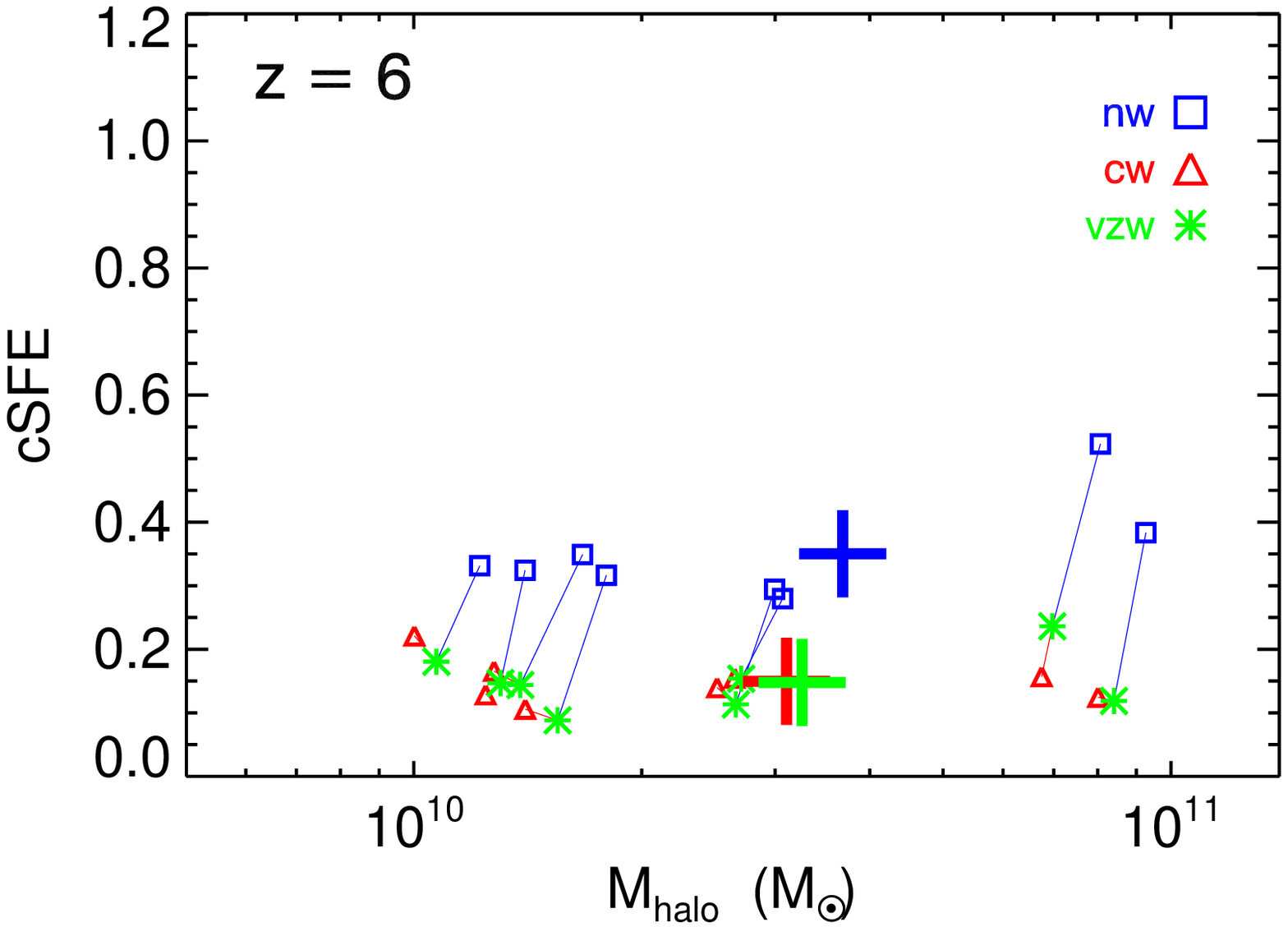}
\includegraphics[scale=0.5]{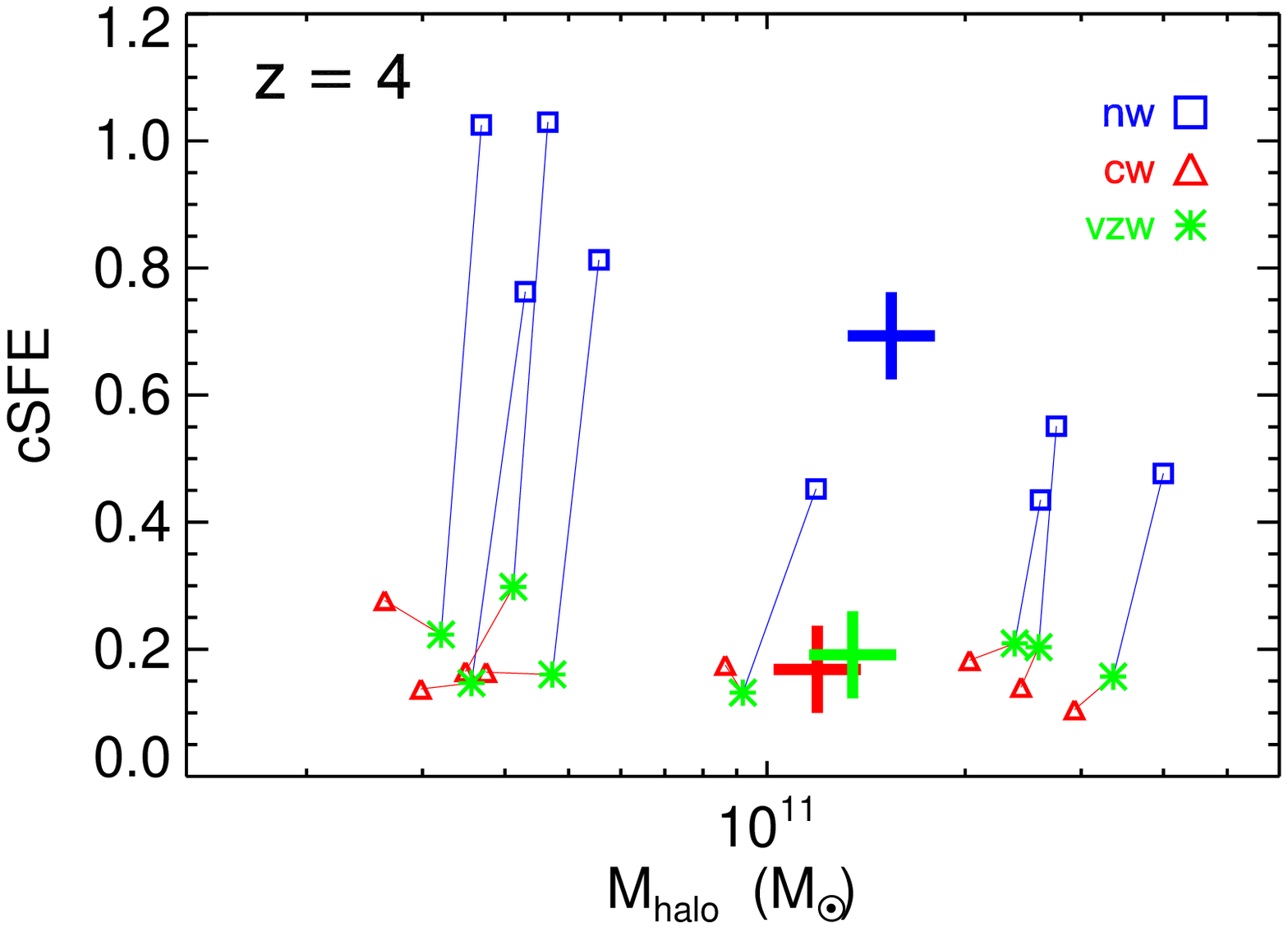}
\includegraphics[scale=0.5]{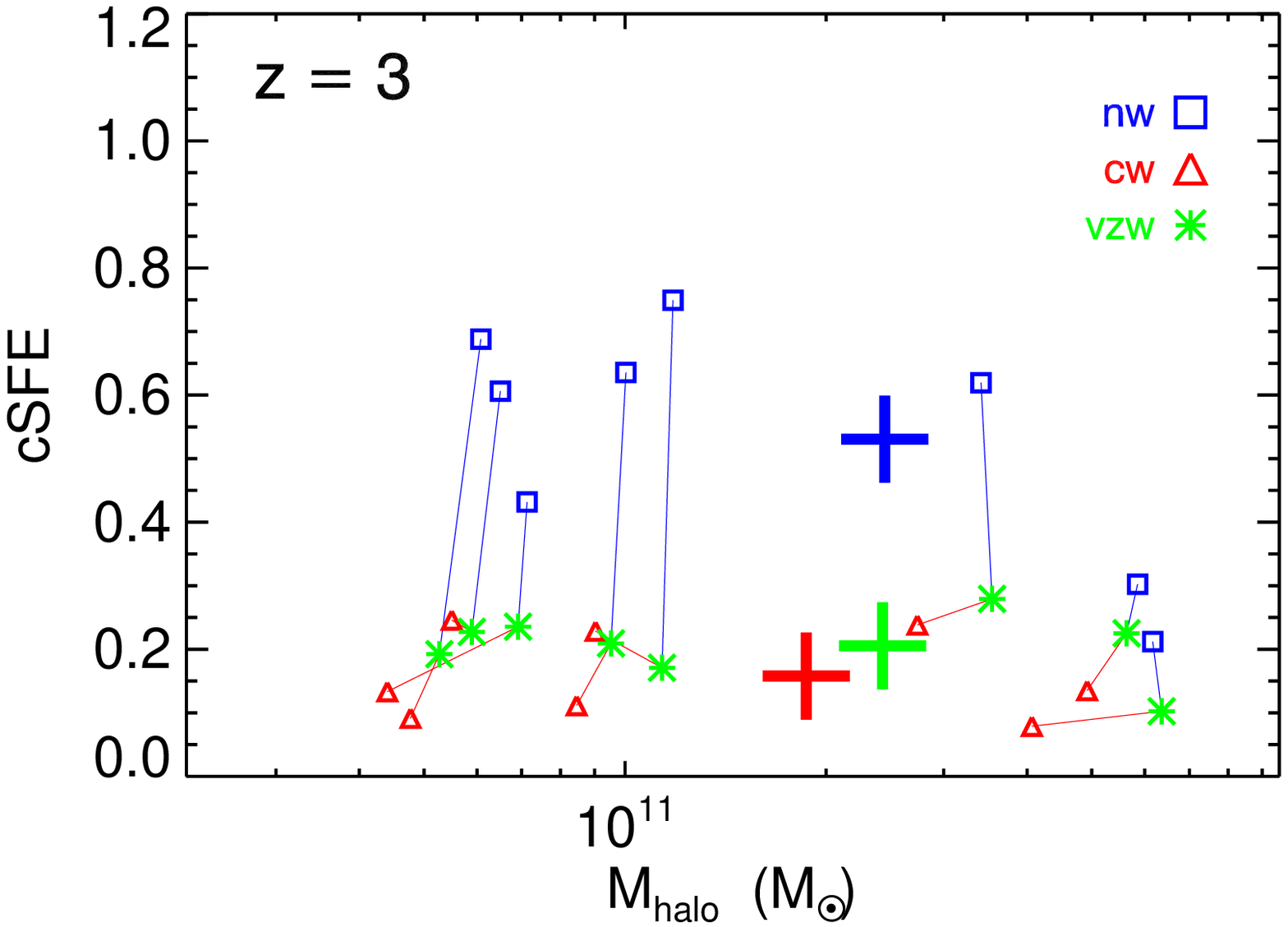}
\includegraphics[scale=0.5]{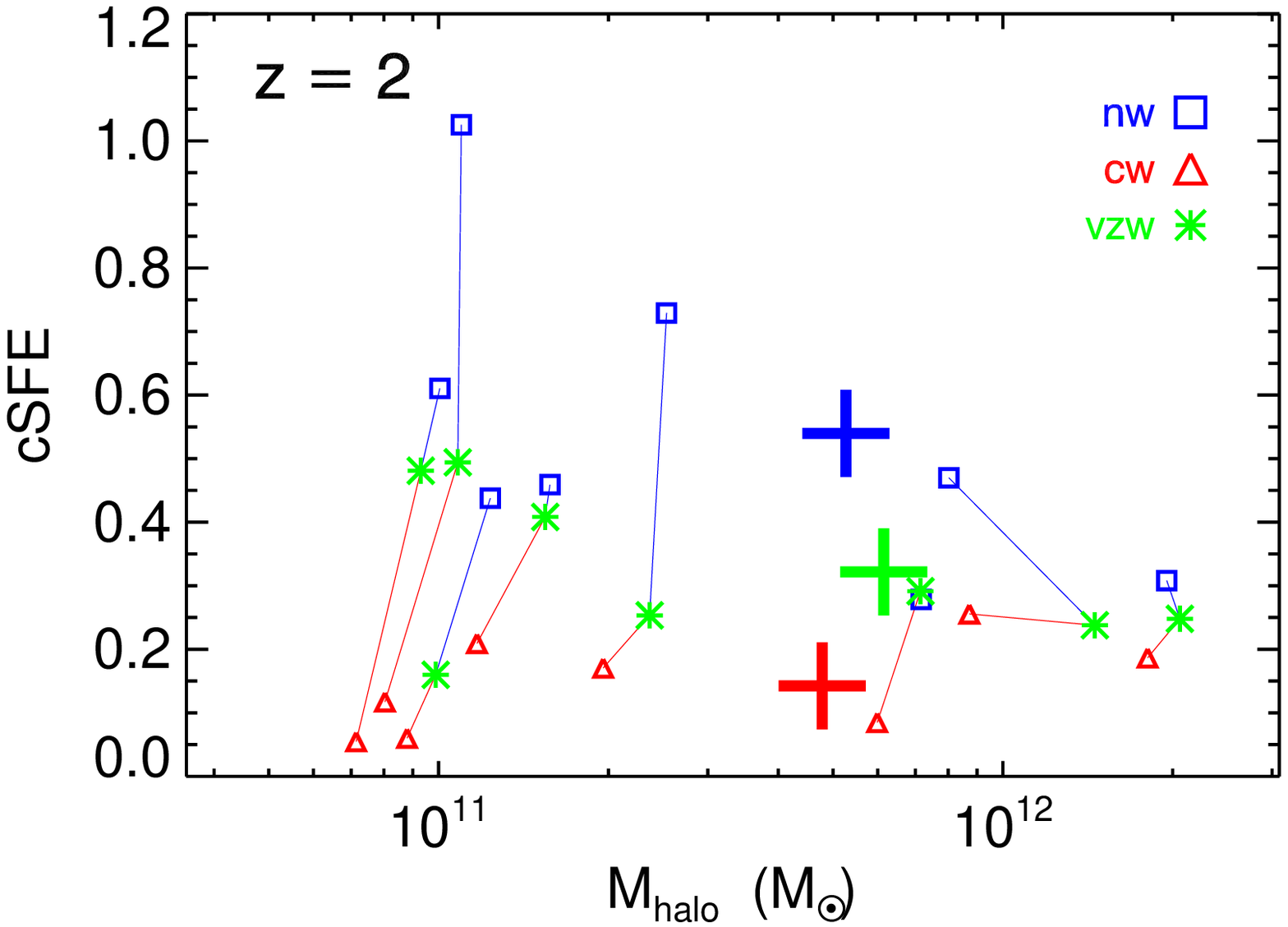}
\end{center}
\caption{Evolution of the cosmic star formation efficiency, i.e., the ratio of the central galaxy SFR to the total halo baryonic accretion rate \citep{dek09} from $z = 6$ to $z = 2$ as a function of halo mass for all simulated galaxies and for the different wind prescriptions.  Lines and colors are as in Figure~\ref{fig:vshalo}.  Green, red, and blue crosses show the average cSFE and halo mass for the momentum-driven winds, constant winds, and no wind models respectively.}
\label{fig:cSFE}
\end{figure*}

No wind simulations usually result in higher stellar surface densities
at all radii compared to the vzw and cw wind models, and often steeper
density gradients at the centers of galaxies.  The effects of
galactic winds are, however, more evident in the distribution of
the star-forming gas: nw galaxies contain most of their gas within
the inner kiloparsec, reaching gas surface densities which can be
factors of a few above the cw and vzw models.  
Galactic winds yield more extended disks by removing gas from their
centers and having it re-accreted at larger scales.  
Wind recycling, especially in the vzw model, results in 
significant amounts of gas extending up to a few kiloparsecs away from the
galaxy center (see galaxies g222 and g54 in Figure~\ref{fig:prf}).  As we
show in the next sections, this result is common to most of our
simulated galaxies.  Galaxy g2403 is one of the few galaxies that
present a very compact gas distribution even in the presence of
momentum-driven winds, similar to galaxies from simulations with no galactic
outflows.

Most of our simulated galaxies are characterized by flat rotation
curves extending up to a few kiloparsecs.  The no wind model usually
produces more massive and more compact galaxies and this results
in higher circular velocities that peak at smaller radii compared
to the vzw and cw models.  Simulations with winds produce more
smoothly rising rotation curves, with lower circular velocities at
small radii that tend to approach values similar to the nw model
at larger scales, where the contribution from the dark matter
component becomes dominant (see galaxies g222 and g54 in
Figure~\ref{fig:prf}; this is explored in more detail in Section~\ref{sec:rot}).  
Galaxy g2403 is again an atypical case, with 
very similar rotation curves for the nw and vzw wind models.

Finally, negative radial gas-phase metallicity gradients 
(higher metallicities at the centers of galaxies) are common in most
simulated galaxies, in general agreement with observations \citep{yuan11,swi12b,jones13}.
Nonetheless, inverted metallicity gradients have been observed in $z \sim 2$--3 galaxies and attributed to either galaxy interactions or the infall of metal-poor gas into the centers of galaxies \citep{cres10,queyrel12,troncoso13}.
In our simulations, abrupt changes in azimuthally averaged metallicities or even positive
metallicity gradients may eventually occur for galaxies undergoing 
mergers (e.g., galaxy g222 at $R \approx 3$--4\,kpc). 
Overall, we find that simulations with no winds usually result in higher
metallicities at all radii and also show steeper metallicity
gradients.  Galactic outflows serve to redistribute metal-enriched gas from
the central regions of galaxies over larger scales, resulting in
less steep metallicity gradients.

\subsection{Connecting Individual Halos Across Wind Models}

In order to understand the global effects of galactic winds it 
is useful to look at their effects on different galaxy properties
at fixed halo mass, since the latter is expected to be roughly
insensitive to baryonic processes.  In Figure~\ref{fig:vshalo}, we
plot the halo baryonic fraction, the central stellar mass to halo
mass fraction, and the SFR as a function of halo mass for all
simulated galaxies at $z = 2$.  Here, each vzw galaxy is connected
to the corresponding nw and cw galaxies to help identify both the
global effects of outflows on the galaxy population as well as the
effects on individual galaxies.  Total halo masses are very similar
for the nw and vzw models at $z = 2$, except occasionally
due to the slightly different timing of merger events and our adopted
definition of dark matter halos for each central galaxy (see
Section~\ref{sec:sim}).  Halo masses are, however, systematically
lower for simulations with constant winds, for which gas outflows 
are able to escape more easily from the halo potential well, 
all of which generally have escape velocities below 
the constant wind velocity of $v_w = 680$\,km\,s$^{-1}$.
Note that momentum-driven wind velocities are below that of the
cw model in all but the most massive galaxies, especially at high redshift
when galactic velocity dispersions are lower \citep{opp08}.

Figure~\ref{fig:vshalo} shows that halo baryonic fractions are only
slightly higher for the no wind model (and higher on average than
the cosmological baryonic fraction $f_{b} = 0.165$) compared to the
vzw model: most of the outflowing gas does not escape the halo potential
well for the momentum-driven wind model.  This
is despite the fact that the wind speed typically is comparable to
or exceeding the escape velocity (and scales with it as well); but
hydrodynamic (i.e., ram pressure) slowing is actually dominant in
many cases~\citep{opp08}.  In contrast, the high efficiency of
constant winds in expelling gas from halos results in systematically
lower halo baryonic fractions, a significant suppression of star
formation in the central galaxy, and the consequent reduction of
stellar masses for all galaxies in our sample.

Figure~\ref{fig:vshalo} (middle panel) confirms our
earlier expectations: for a given halo mass, central galaxy stellar
masses are systematically higher in the absence of galactic outflows.
Interestingly, SFRs for the nw and vzw models happen to be 
similar at $z = 2$ for most simulated galaxies.  However, as we
have seen in Figure~\ref{fig:evol} for galaxies g222, g2403, and g54,
momentum-driven winds result in rather different star formation
histories, with significant suppression of star formation early on
and enhanced activity due to wind recycling at later times \citep{opp10}.
The wind recycling channel does not act on the cw simulations 
for the range of halo masses probed here, since these halos generally
have escape velocities lower than the assumed wind speed.

Figure~\ref{fig:cSFE} illustrates the effects of galactic winds in
the star formation histories of our simulated galaxies by showing
the evolution of the cosmic star formation efficiency (cSFE) in the
redshift range $z = 2$--6.  The cSFE is the ratio
of the SFR of the central galaxy to the total halo baryonic accretion
rate given in \citet{dek09}:  
\begin{equation}\label{eq:mdotvir}
\dot{M}_{\rm acc} \simeq 6.6\, f_{\rm b}\, \left ( \frac{M_{\rm halo}}{10^{12}M_{\odot}} \right )^{1.15} (1+z)^{2.25}\, M_{\odot}\,{\rm yr}^{-1}.
\end{equation}
At $z = 6$, the average cSFE of
galaxies for simulations including outflows (cw and vzw models) is
$\sim 0.15$, about a factor of two lower than simulations with no
winds.  From $z = 6$ to $z = 4$, the average cSFE increases rapidly
up to $\sim 0.69$ for nw simulations, and then decreases down to
$\sim 0.54$ at $z = 2$.  Simulations with constant winds seem to
follow a similar tendency, with the average cSFE peaking at $z \sim
4$--5 and decreasing at lower redshifts, but with significantly
lower values (roughly $\times 3$, as expected from the $(1+\eta)$
suppression of star formation).
Momentum-driven winds result, however, in an overall
increase of cSFEs from $z = 6$ to $z = 2$, owing to the recycling
of gas that was launched into winds at earlier times.  The vzw model
causes an effective delay in star formation by ejecting significant amounts of gas
from small, early galaxies and having it reaccreted by $z\sim 2$.
This coincidentally produces average SFRs similar to simulations 
with no winds at $z\approx 2$.
Note that trends for individual galaxies mimic the trends for the overall cosmic star formation history in these wind models, with momentum-driven winds generally producing a later peak in cosmic SFR than no-wind or constant-wind cases~\citep{opp06}.

\subsection{Rotation Curves}\label{sec:rot}

\begin{figure}
\begin{center}
\includegraphics[scale=0.6]{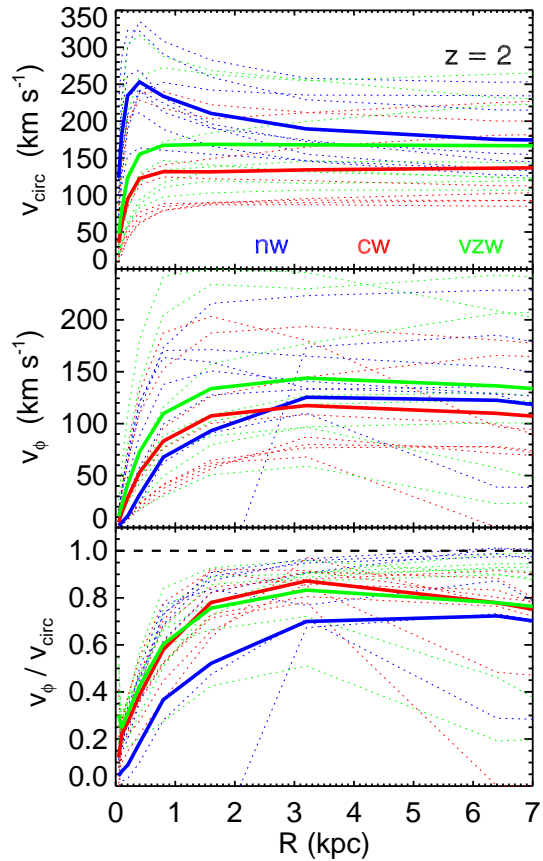}
\end{center}
\caption{{\it Top}: circular velocity ($v_{\rm circ}$) as a function of radial distance from the center of simulated galaxies at $z = 2$.  We compute $v_{\rm circ}^{2} = GM_{\rm enc}(r)/r$ for the total enclosed mass $M_{\rm enc}(r)$ within distance $r$.  {\it Middle}: azimuthal velocity ($v_{\rm \phi}$) of the gas particles with respect to the total angular momentum of the galaxy averaged within logarithmically spaced radial bins (SFR-weighted).  {\it Bottom}: ratio of the average azimuthal velocity of the gas to the circular velocity as a function of radial distance from the center of galaxies.  Green, red, and blue dotted lines show rotation curves of individual galaxies (or the ratio $v_{\rm \phi} / v_{\rm circ}$) for the momentum-driven winds, constant winds, and no wind models respectively.  Thick solid lines show the rotation curve (or $v_{\rm \phi} / v_{\rm circ}$) averaged over all galaxies corresponding to each wind model.}
\label{fig:vrot}
\end{figure}

\begin{figure}
\begin{center}
\includegraphics[scale=0.6]{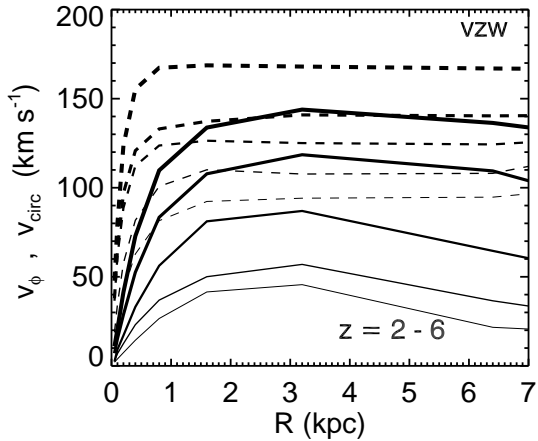}
\end{center}
\caption{Build up of the galaxy rotation curves as a function of redshift from $z = 6$ (thin grey line) to $z = 2$ (thick black line) for simulations including momentum-driven winds.  Dashed lines show the circular velocity ($v_{\rm circ}$) averaged over all galaxies (as in Figure~\ref{fig:vrot}, top panel) while solid lines show the actual gas azimuthal velocity ($v_{\rm \phi}$) averaged within each radial bin (and again averaged over all vzw galaxies).}
\label{fig:buildup}
\end{figure}

Galaxy rotation curves contain substantial information about the
structure and kinematics of galaxies.  Figure~\ref{fig:vrot} (top panel) shows
the rotation curves of all simulated galaxies at $z = 2$
(dotted lines), color-coded by wind model.  Here, we calculate
the rotation velocity from the enclosed mass at a given radius,
which we call $v_{\rm circ}$.  Our
galaxies are characterized by flat rotation curves extending up to
several kpc scales, with asymptotic circular velocities ranging
from $\sim 100$ to 260\,km\,s$^{-1}$ at $z = 2$ for the vzw model.  
In Figure~\ref{fig:prf}, we showed for galaxies g222, g2403, and g54 
that galactic outflows can have a 
significant impact on the overall shape of their rotation curves.

The thick solid lines in Figure~\ref{fig:vrot} (top panel) show 
stacked rotation curves for each wind
model, where circular velocities have been averaged over all galaxies
as a function of radius.  Simulations with winds produce galaxies
with more gradually rising rotation curves compared to the more
centrally peaked rotation curves of galaxies with no winds.  The
average circular velocities for the nw and vzw models tend to
approach similar values at larger radii, where the contribution
from the dark matter component becomes dominant.  Simulations with
constant winds are more efficient in removing gas from halos,
resulting in systematically lower asymptotic circular velocities,
but their shape is similar to the momentum-driven winds.

The middle panel of Figure~\ref{fig:vrot} shows a more
observationally motivated way of calculating the rotation curve.  Rather
than taking the enclosed mass and assuming full rotational support,
here we directly calculate the SFR-weighted azimuthal velocities of gas particles
with respect to the total angular momentum of the galaxy, and average
them within logarithmically spaced radial bins, which we call $v_{\rm \phi}$.
This is analogous to an H$\alpha$ rotation curve.
The rotation curves derived in this way show a significantly more gradual
increase in circular velocity with radius in the inner parts, 
because of the increased dispersion support in the central regions.
This is particularly clear for simulations with no winds, 
which produce more compact galaxies and centrally peaked rotation curves 
based on the mass profiles.

The bottom panel of Figure~\ref{fig:vrot} shows the ratio of
$v_{\rm \phi}/v_{\rm circ}$.  We see that the gas azimuthal velocity
is, on average, lower than the inferred circular velocities.  
The gas reaches rotation velocities comparable to
the circular velocity at $\sim 3$\,kpc away from the centers of galaxies.
In the inner regions, the contribution of random motions to the
dynamical equilibrium of galaxies becomes comparable to the support from
ordered rotation.  The discrepancy between $v_{\rm \phi}$ and $v_{\rm circ}$ 
at scales $< 3$\,kpc occurs for all simulated galaxies and wind models, and it is well
resolved in our simulations.  Using $v_{\rm \phi}$ to
infer the mass profile of galaxies could, therefore, 
lead to significant misestimations \citep{vale07}.

Figure~\ref{fig:buildup} shows the build-up of rotation curves with
time by comparing stacked rotation curves from $z = 6$ to $z = 2$ for
our simulations with momentum-driven winds.  Solid lines show $v_{\rm \phi}$
(for the gas), and dashed lines show $v_{\rm circ}$.  We find that
the average asymptotic circular velocities increase from $\sim 90$
to 170\,km\,s$^{-1}$ with decreasing redshift,
as expected because our galaxies are becoming more massive.
For $v_{\rm circ}$, the peak of the (average) rotation curve is always at $\sim 1$--2\,kpc,
perhaps moving inward to lower redshifts. In contrast, the location of the peak
of $v_{\rm \phi}$ occurs at 3--4\,kpc, and does not vary much with redshift. 
Comparing $v_{\rm \phi}$ to $v_{\rm circ}$, we see that 
gas azimuthal velocities are, on average, 
lower than the inferred circular velocities at all redshifts.
The ratio $v_{\rm \phi}/v_{\rm circ}$ evaluated at $R \approx 3$\,kpc increases
from $\sim 0.48$ at $z = 6$ to $\sim 0.86$ at $z = 2$, indicating 
that galaxies become, on average, progressively more
rotationally supported with time (and therefore with increasing mass) in the redshift
range $z = 6 \rightarrow 2$.

\section{Comparison to Observations}\label{sec:obs}

The high resolution afforded by {\it Hubble} probing both young
stars and older stellar populations, together with adaptive optics-enhanced
spectral studies from the ground, have opened up new windows into
high-redshift galaxy assembly.  Here we examine both qualitatively
and quantitatively how our simulated galaxies compare with the
latest observations of $z\sim 2$ galaxies.

\subsection{Qualitative Galaxy Morphologies}

\begin{figure*}
\begin{center}

\includegraphics[scale=0.5]{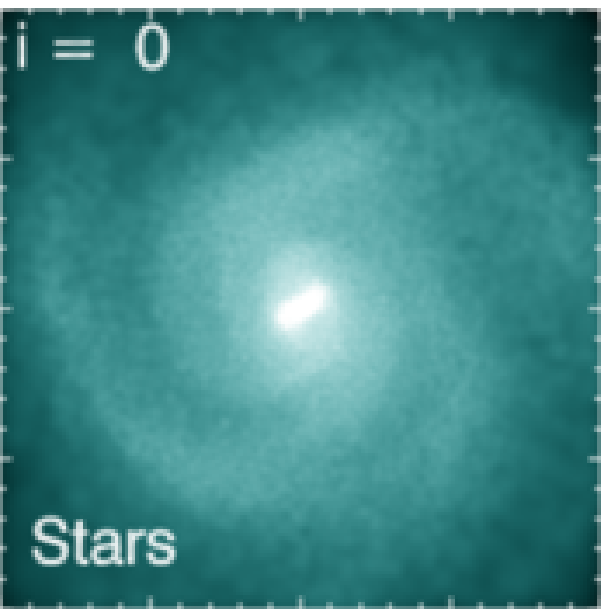}
\includegraphics[scale=0.5]{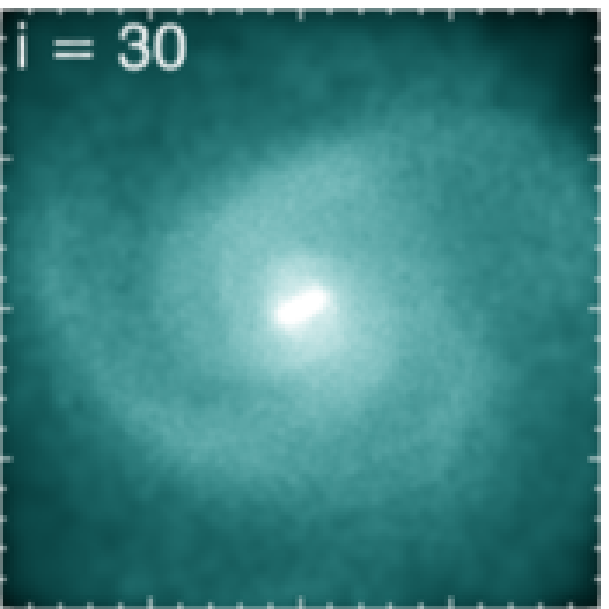}
\includegraphics[scale=0.5]{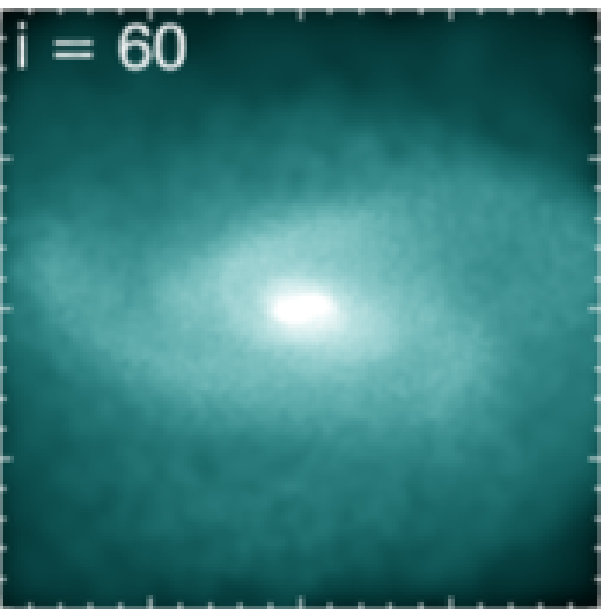}
\includegraphics[scale=0.5]{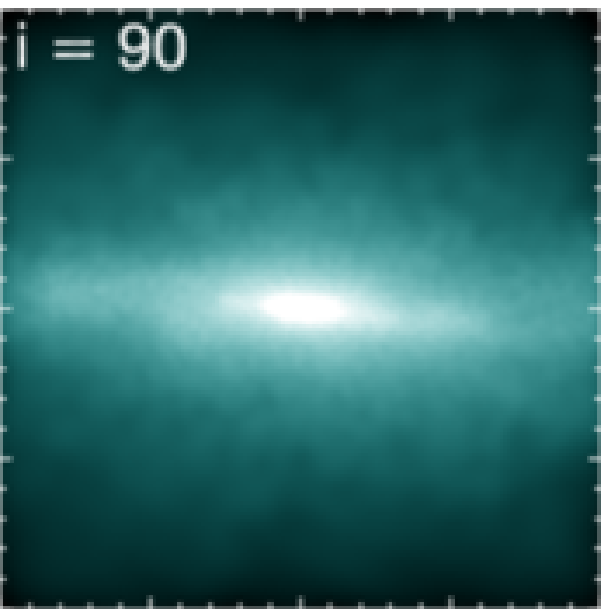}
\includegraphics[scale=0.3]{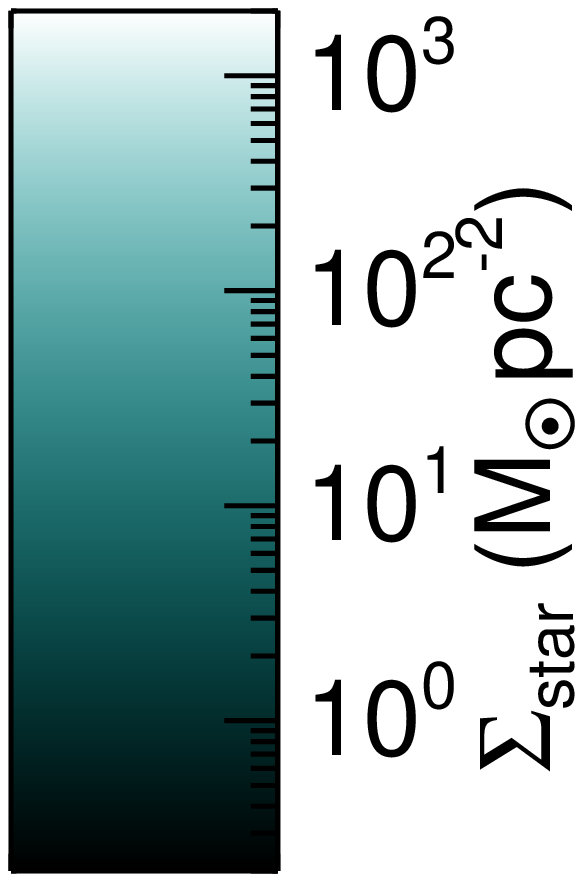}

\includegraphics[scale=0.5]{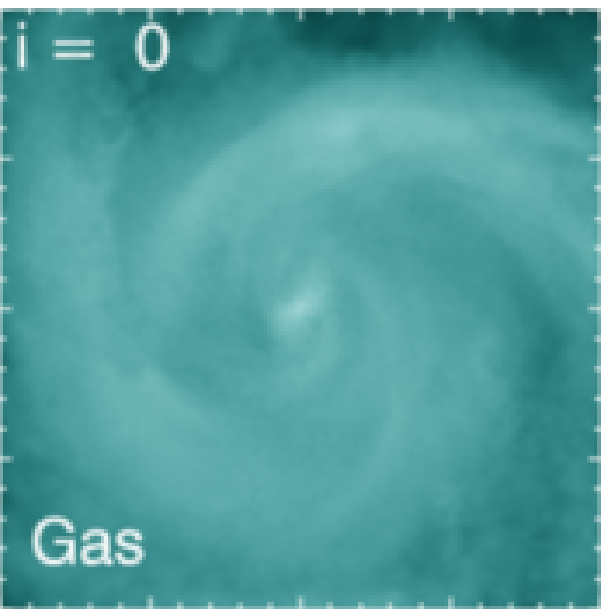}
\includegraphics[scale=0.5]{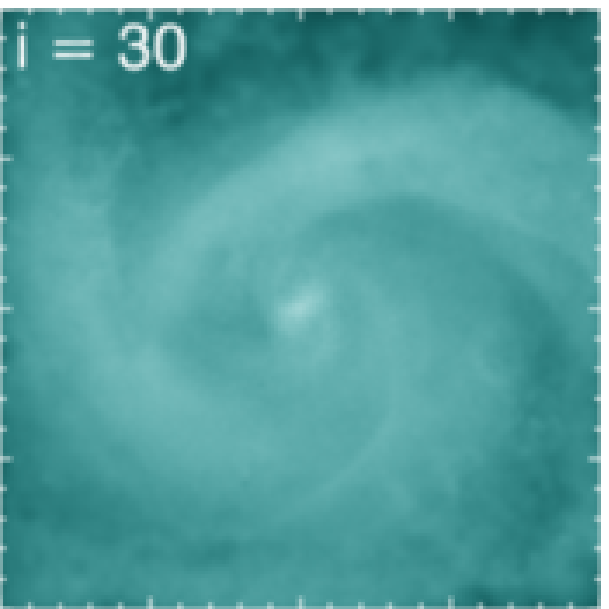}
\includegraphics[scale=0.5]{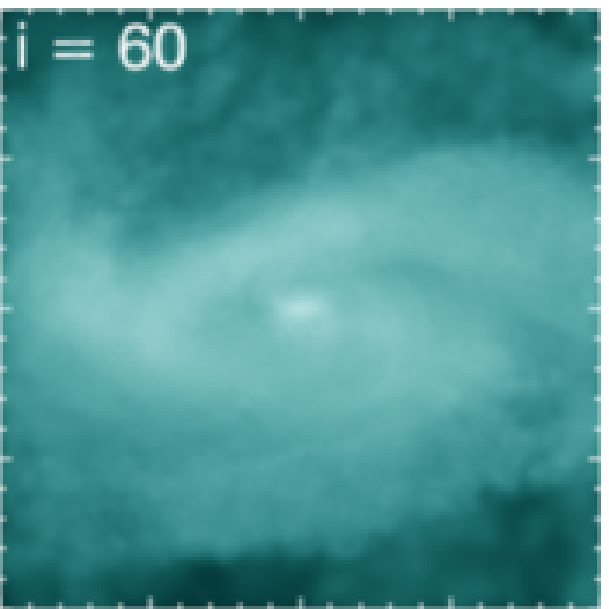}
\includegraphics[scale=0.5]{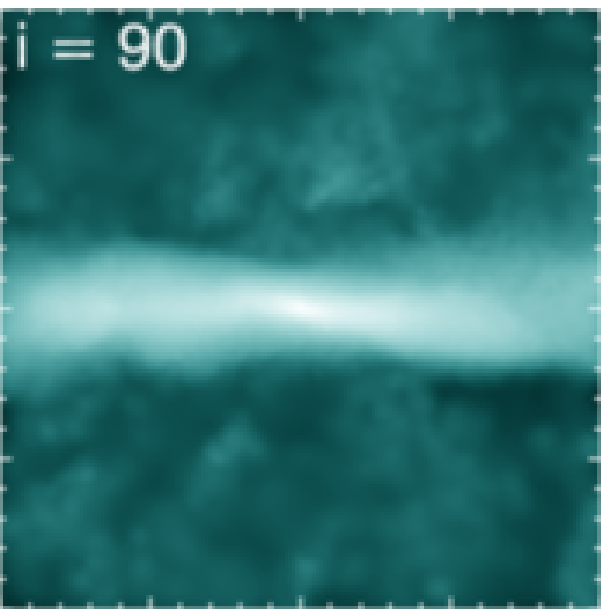}
\includegraphics[scale=0.3]{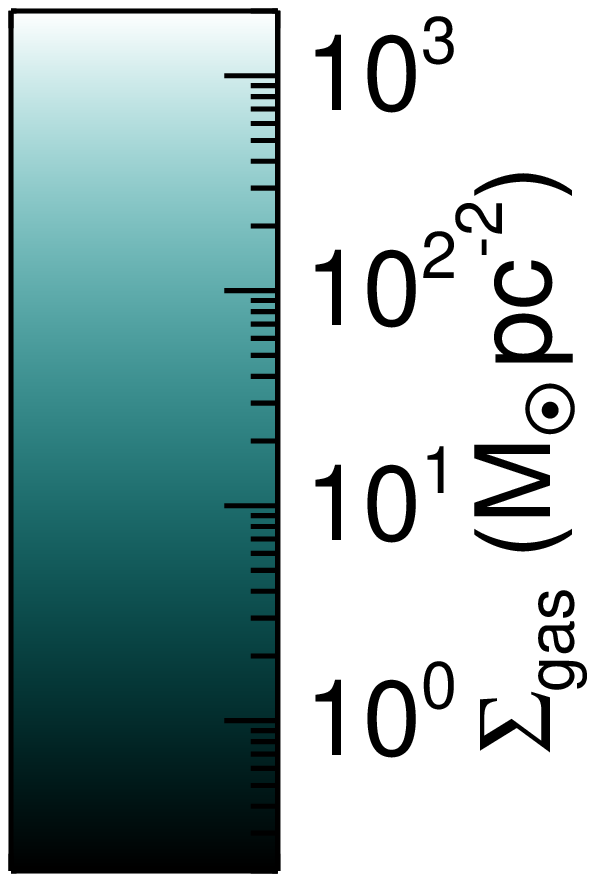}

\includegraphics[scale=0.5]{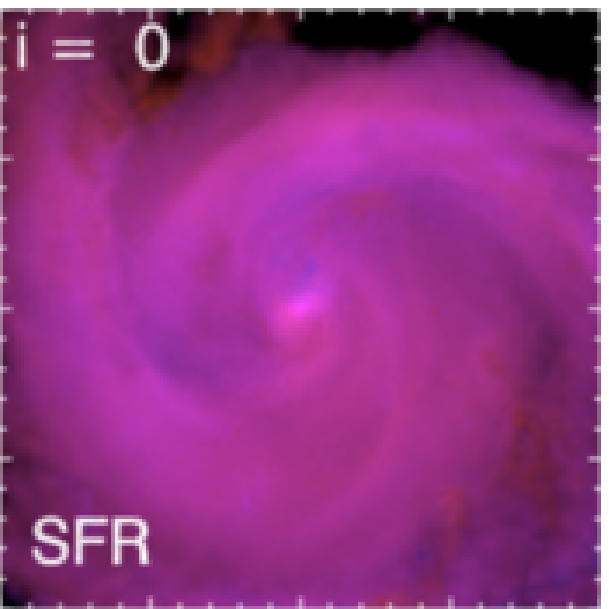}
\includegraphics[scale=0.5]{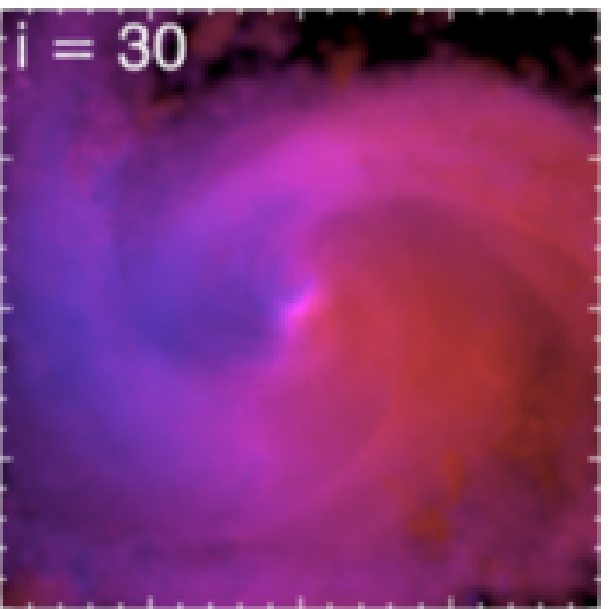}
\includegraphics[scale=0.5]{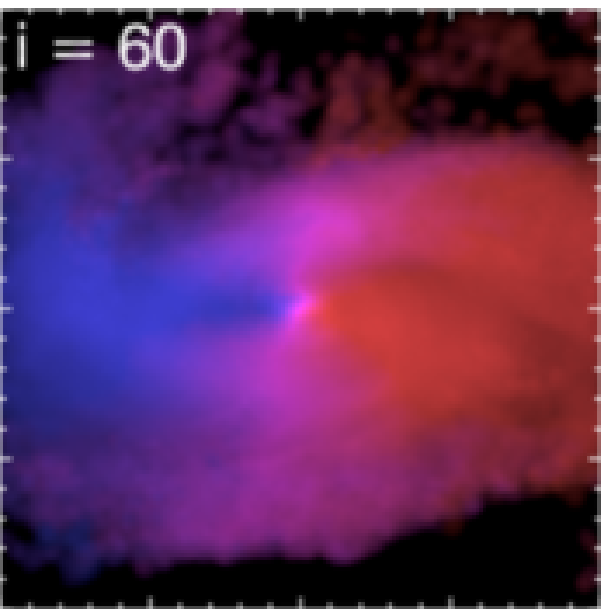}
\includegraphics[scale=0.5]{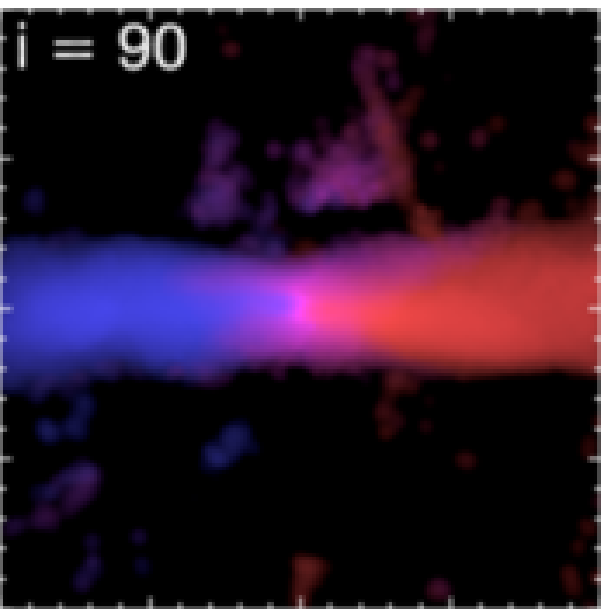}
\includegraphics[scale=0.4]{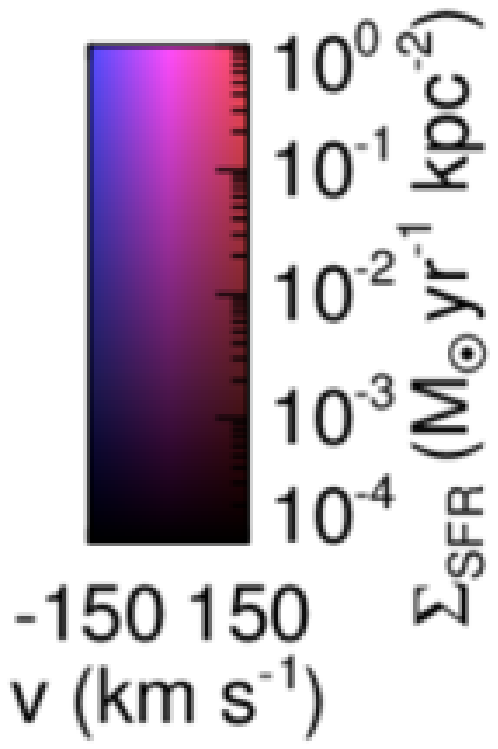}

 \bigskip
 \bigskip

\includegraphics[scale=0.3]{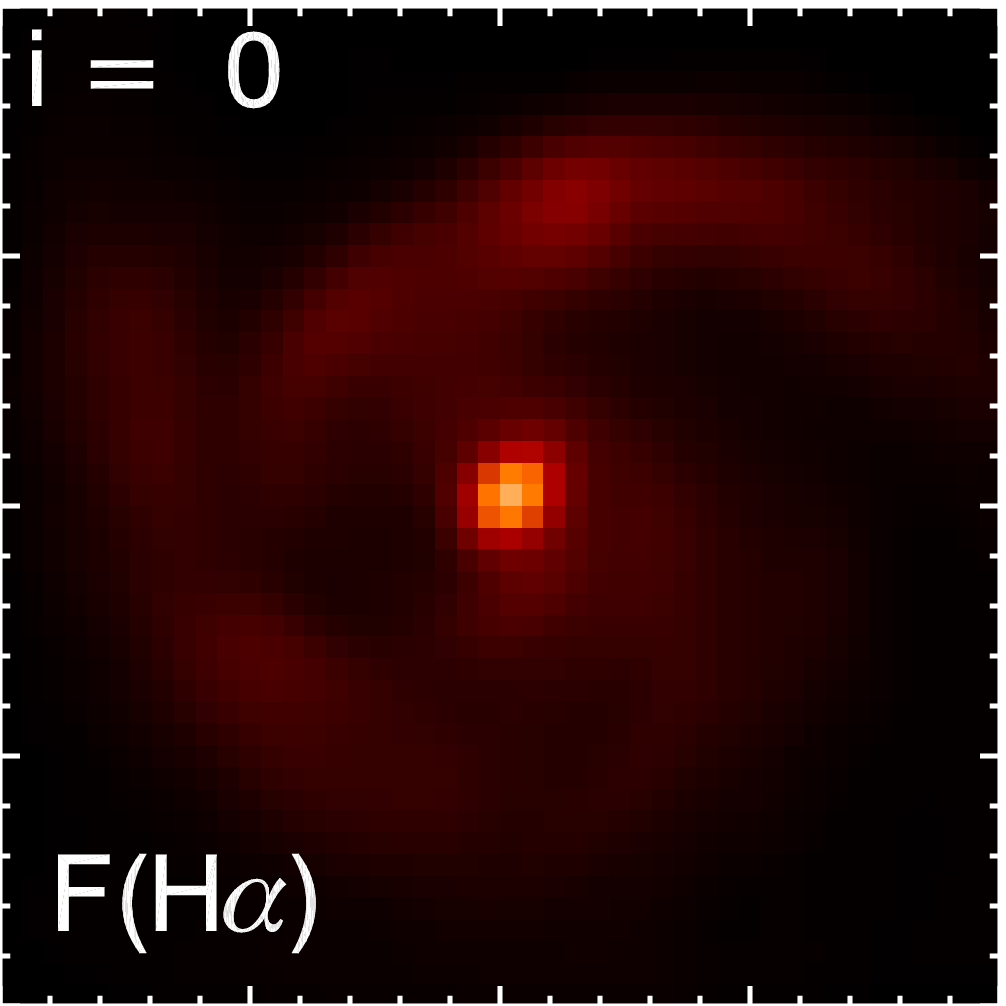}
\includegraphics[scale=0.3]{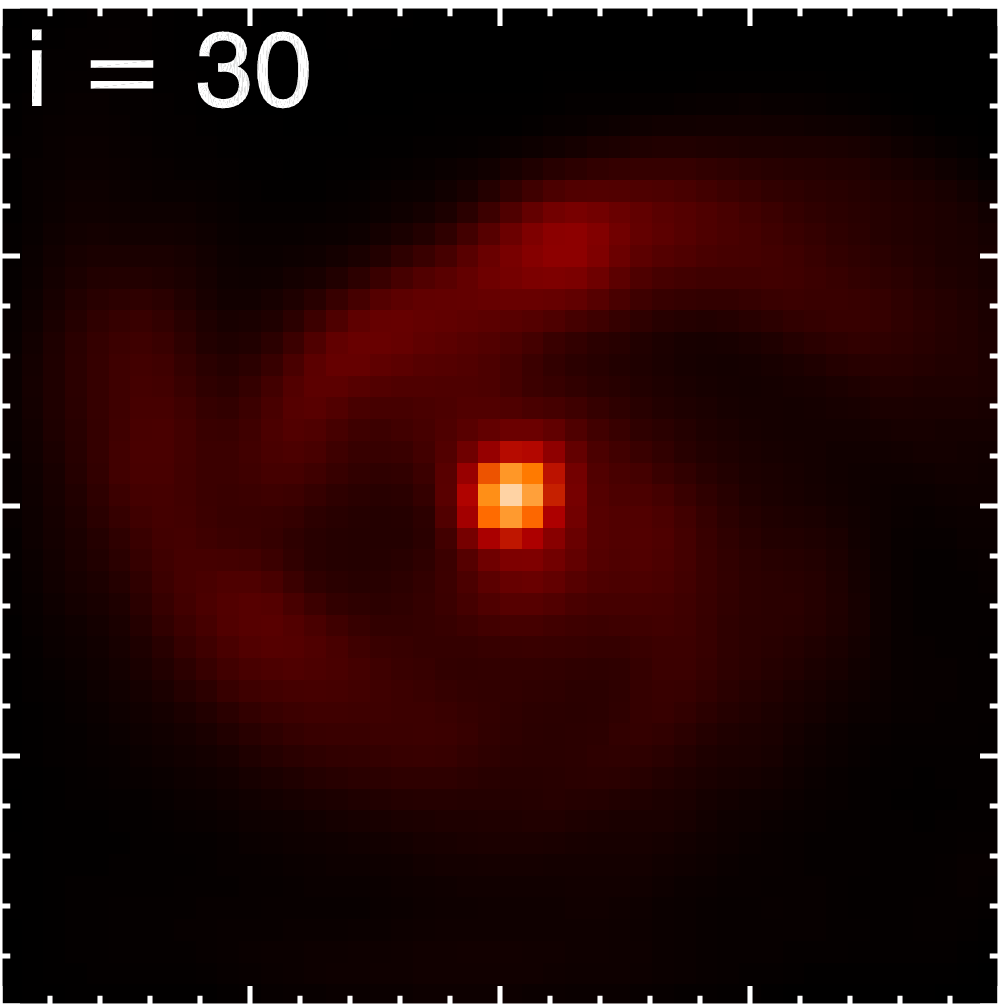}
\includegraphics[scale=0.3]{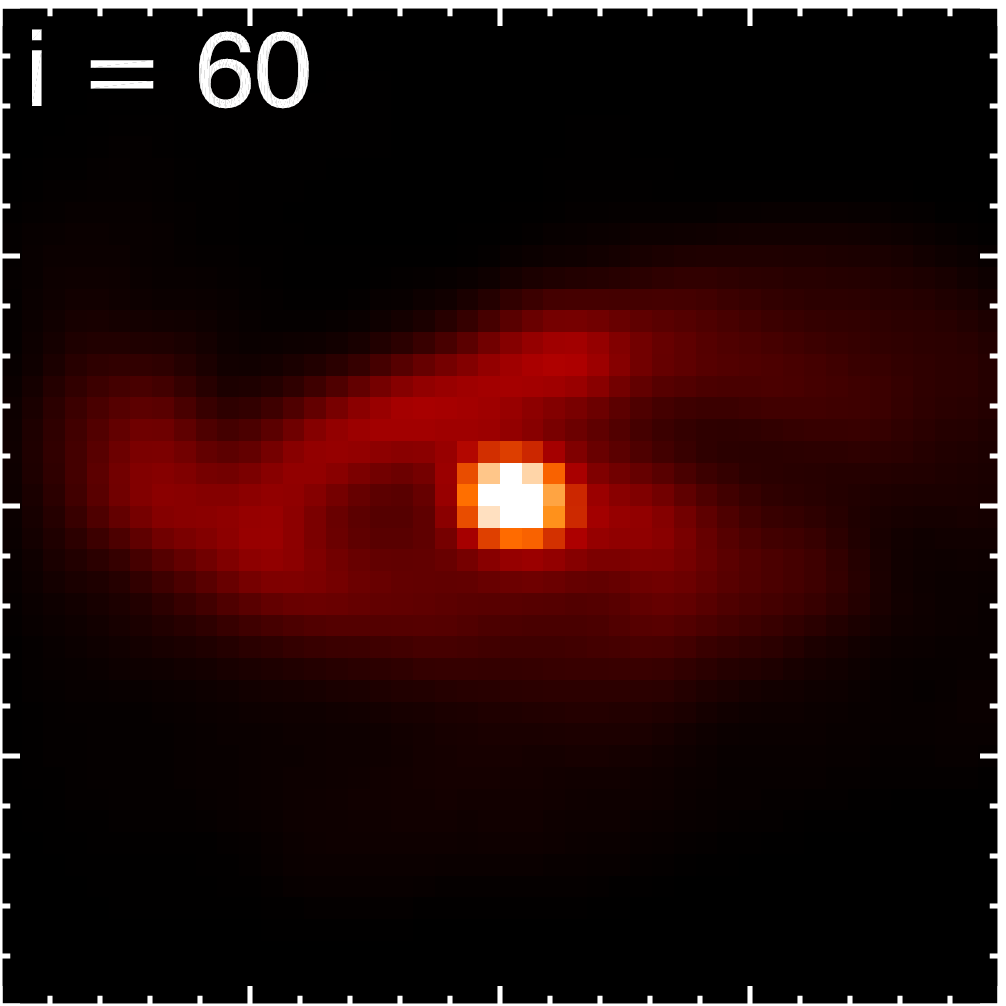}
\includegraphics[scale=0.3]{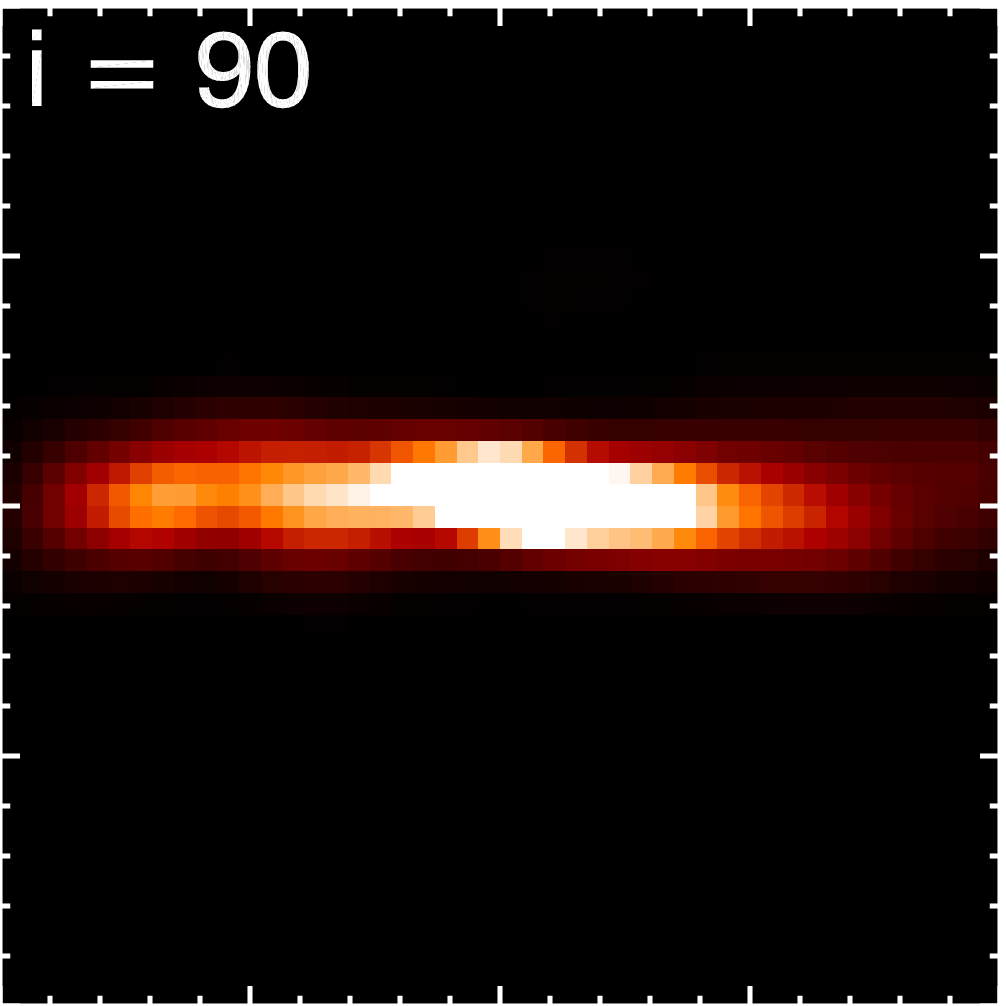}
\includegraphics[scale=0.4]{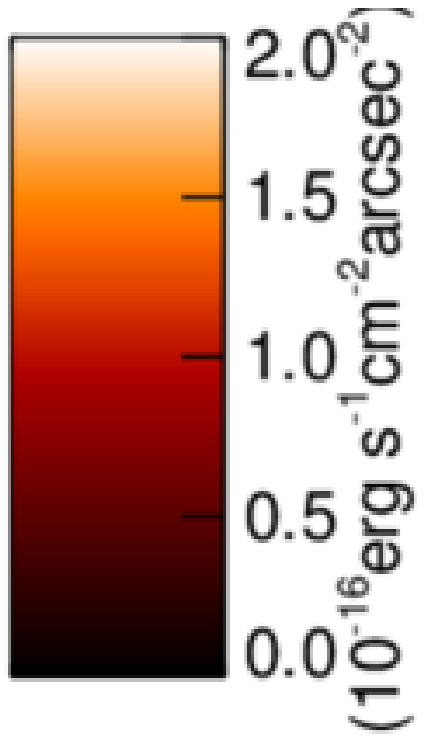}

\includegraphics[scale=0.3]{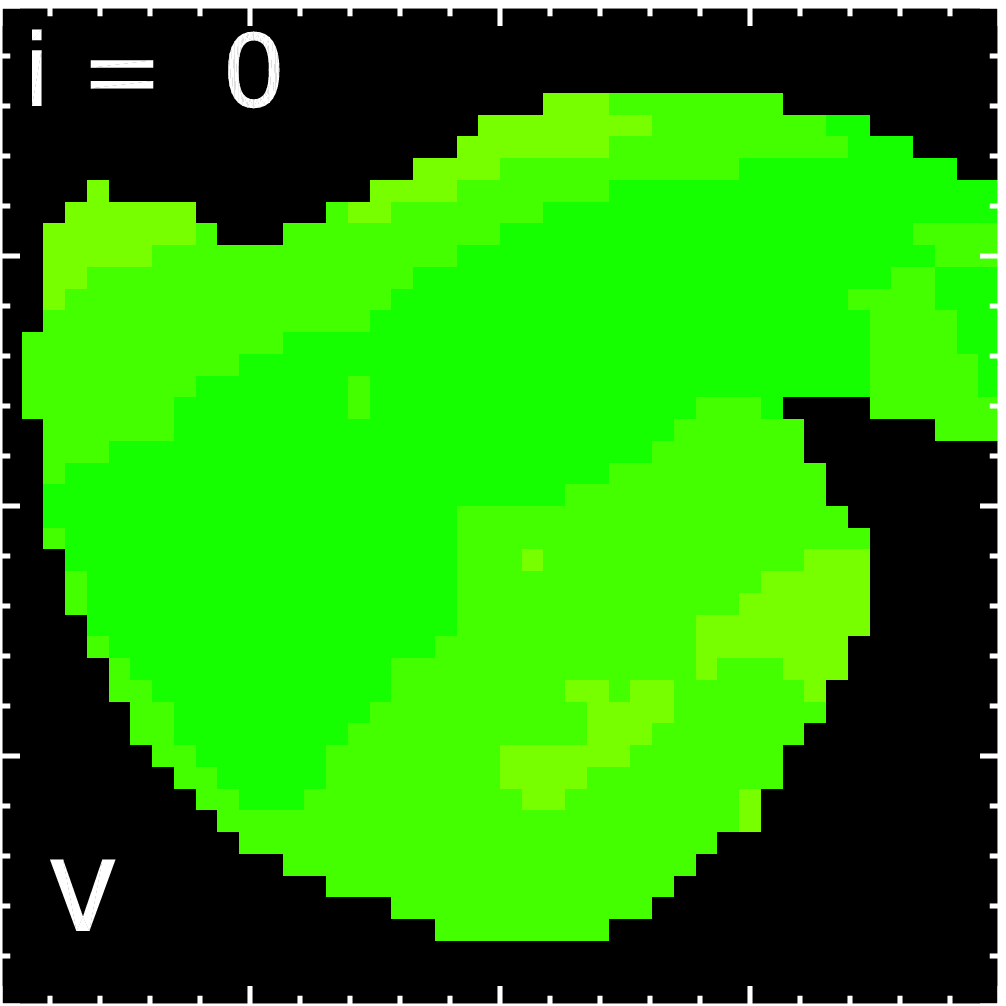}
\includegraphics[scale=0.3]{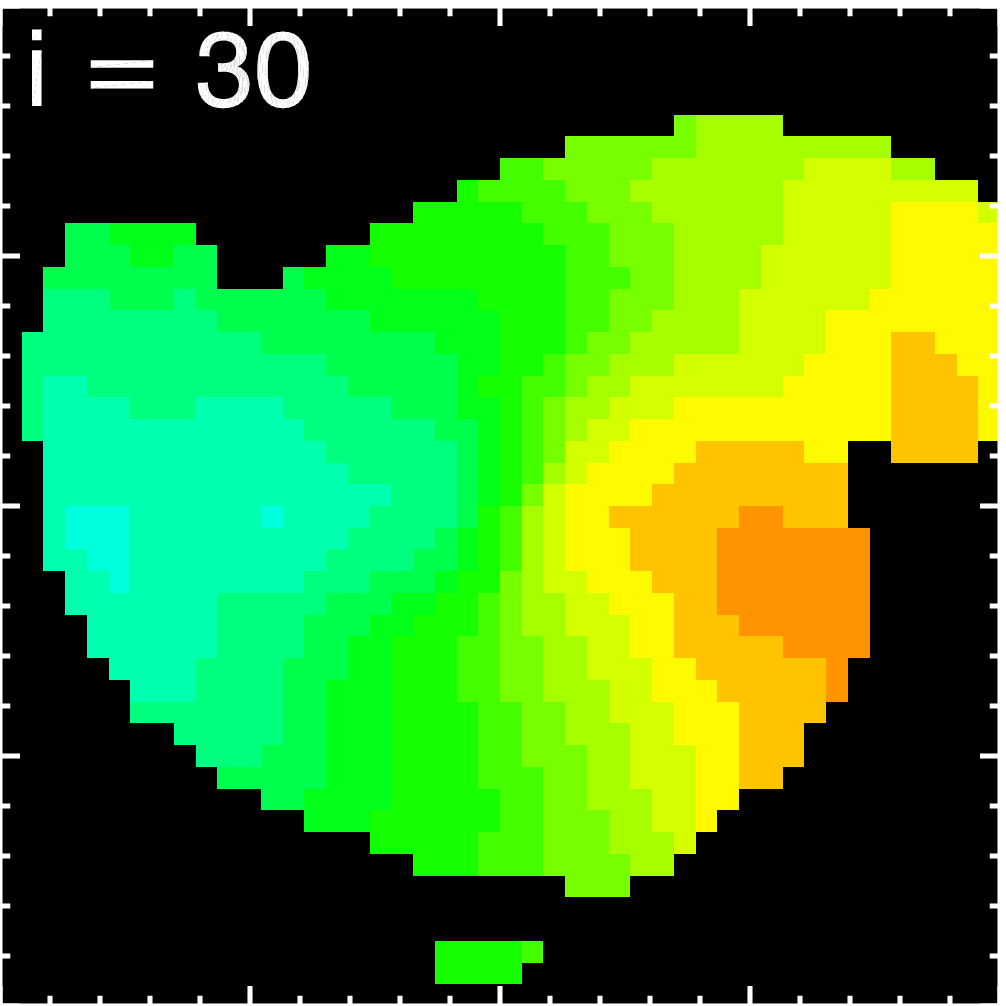}
\includegraphics[scale=0.3]{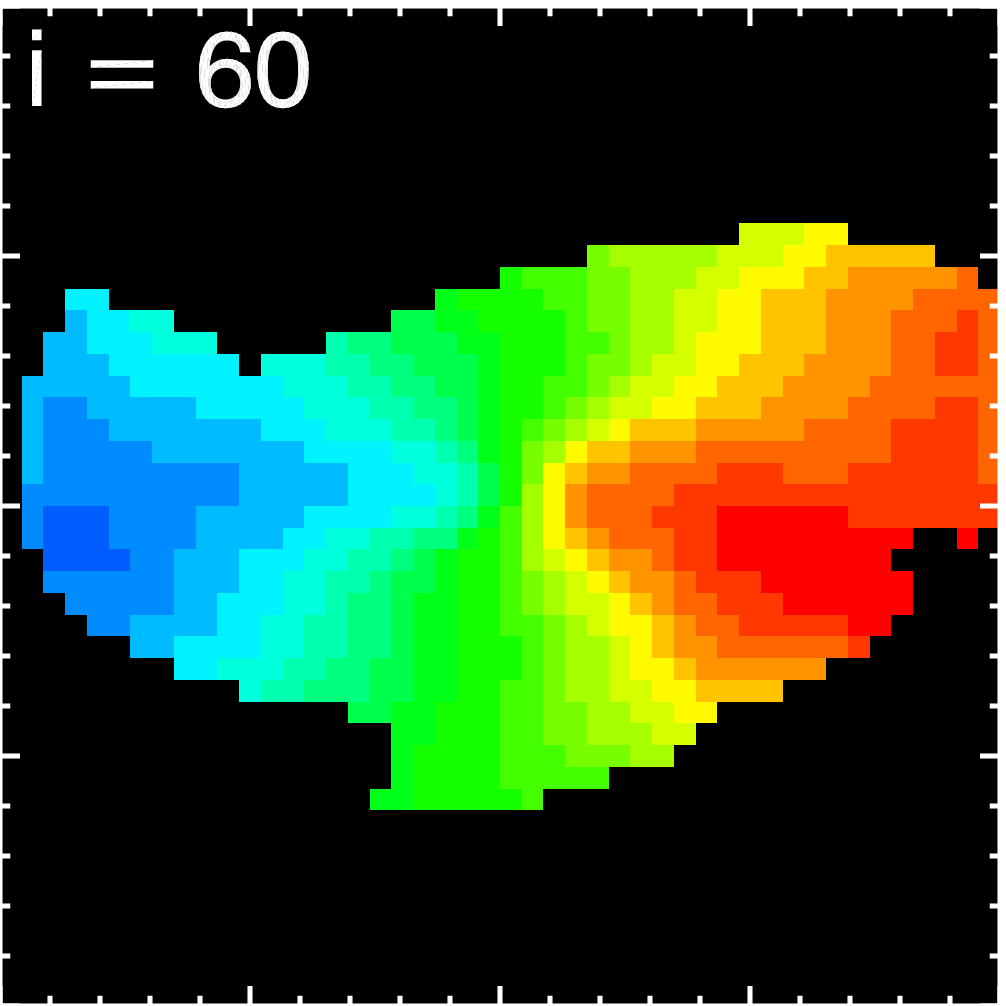}
\includegraphics[scale=0.3]{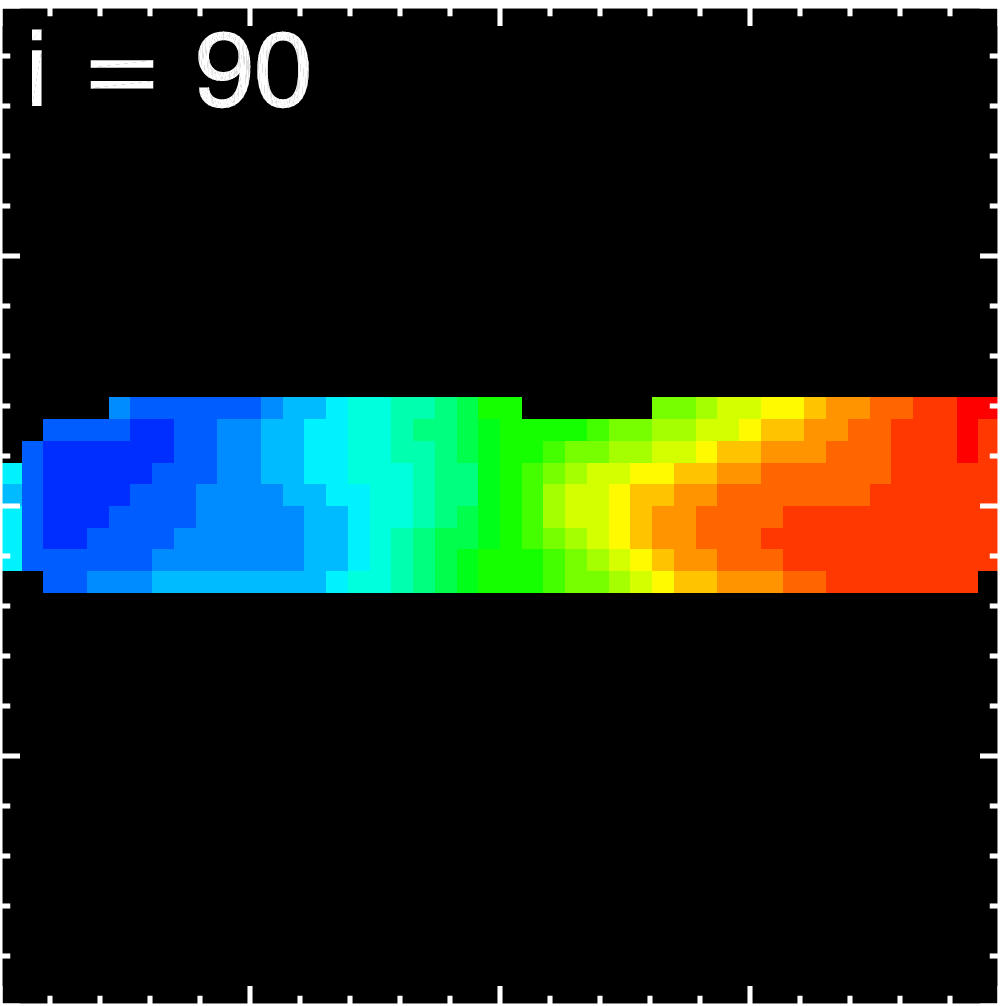}
\includegraphics[scale=0.4]{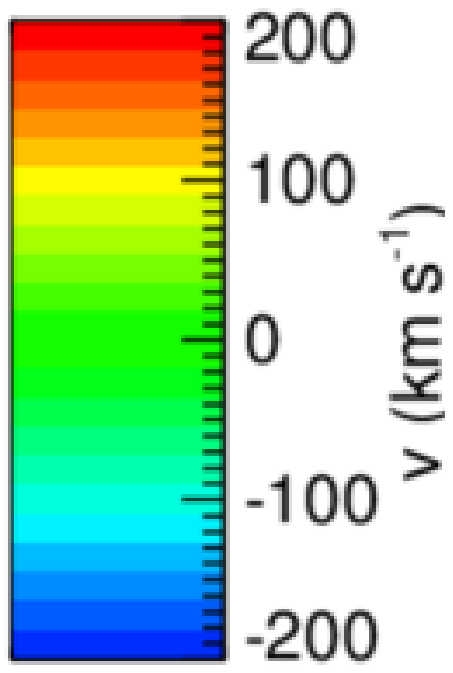}

\includegraphics[scale=0.3]{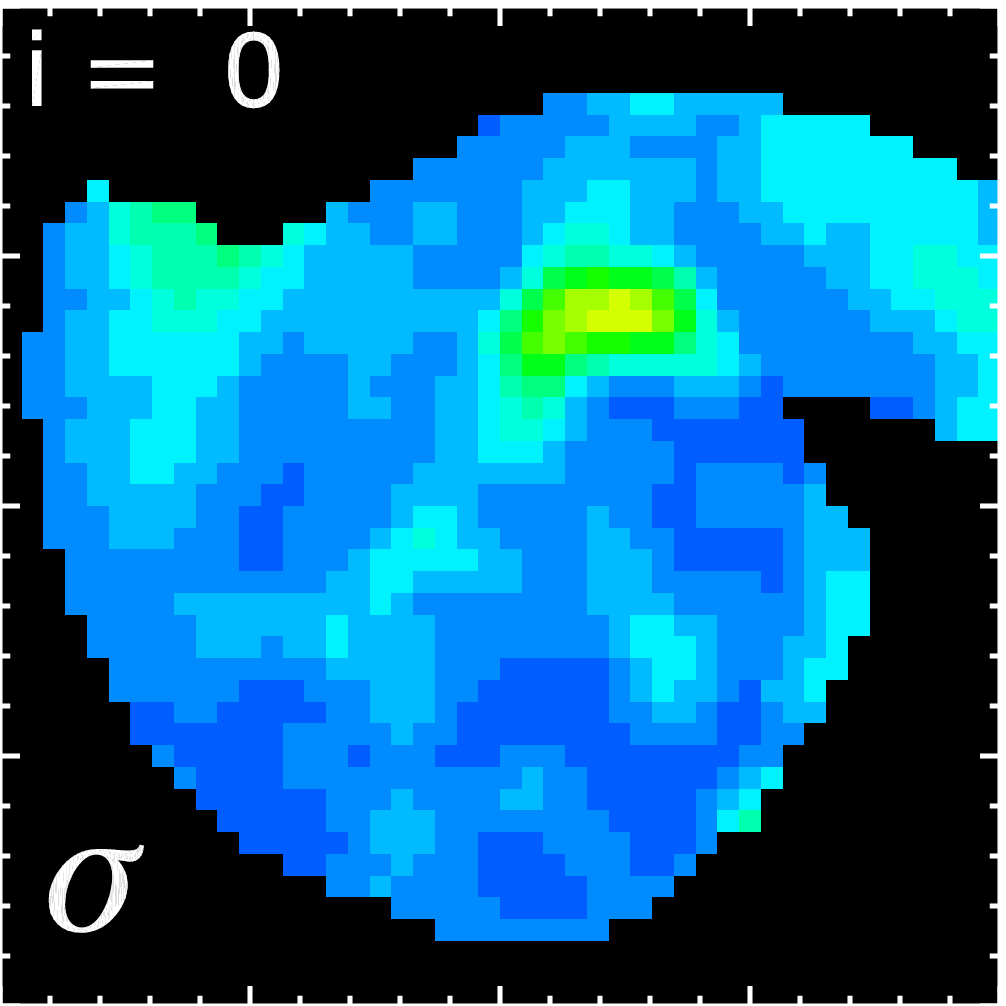}
\includegraphics[scale=0.3]{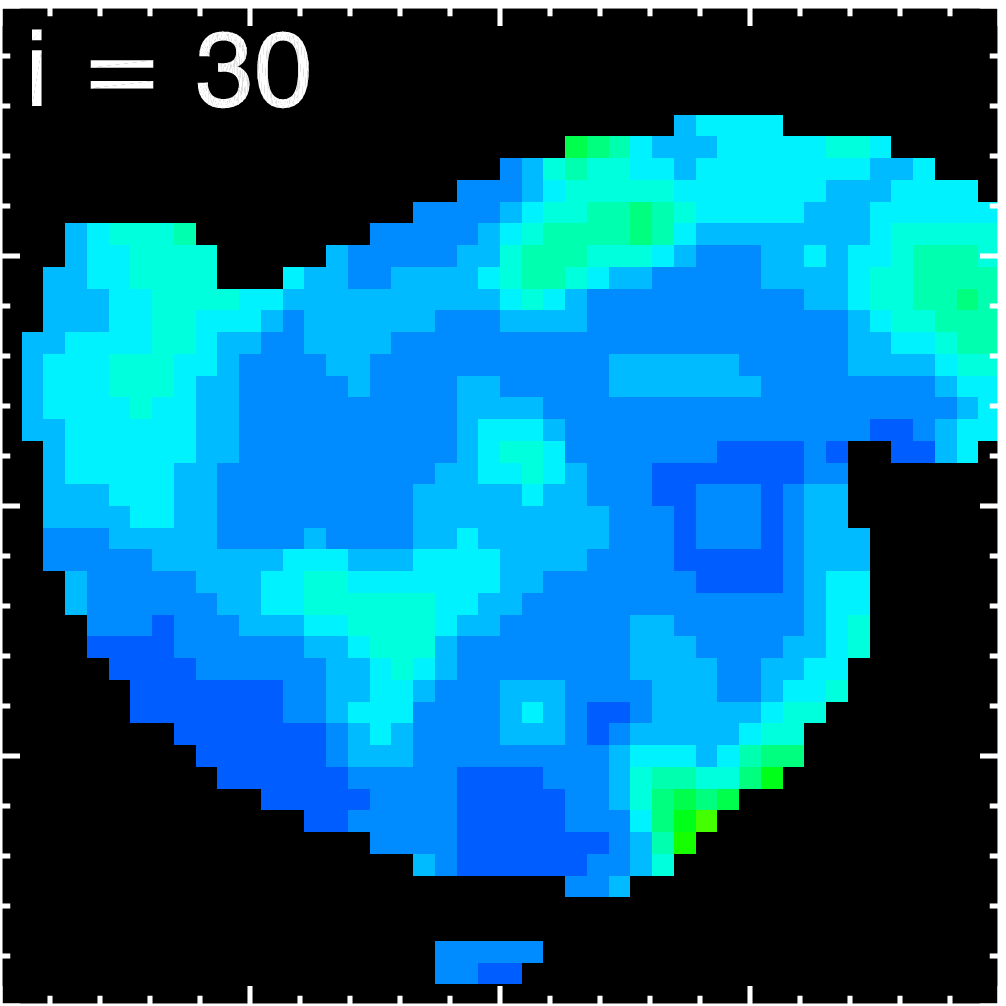}
\includegraphics[scale=0.3]{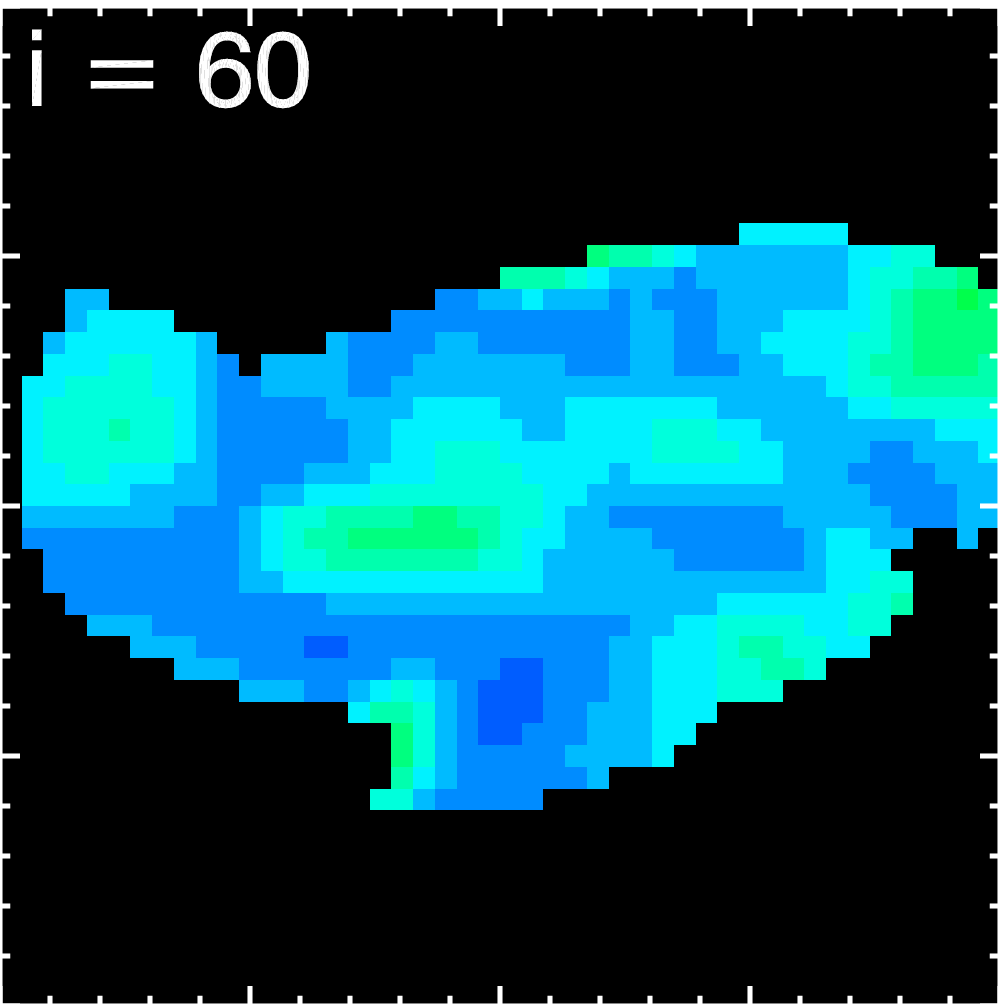}
\includegraphics[scale=0.3]{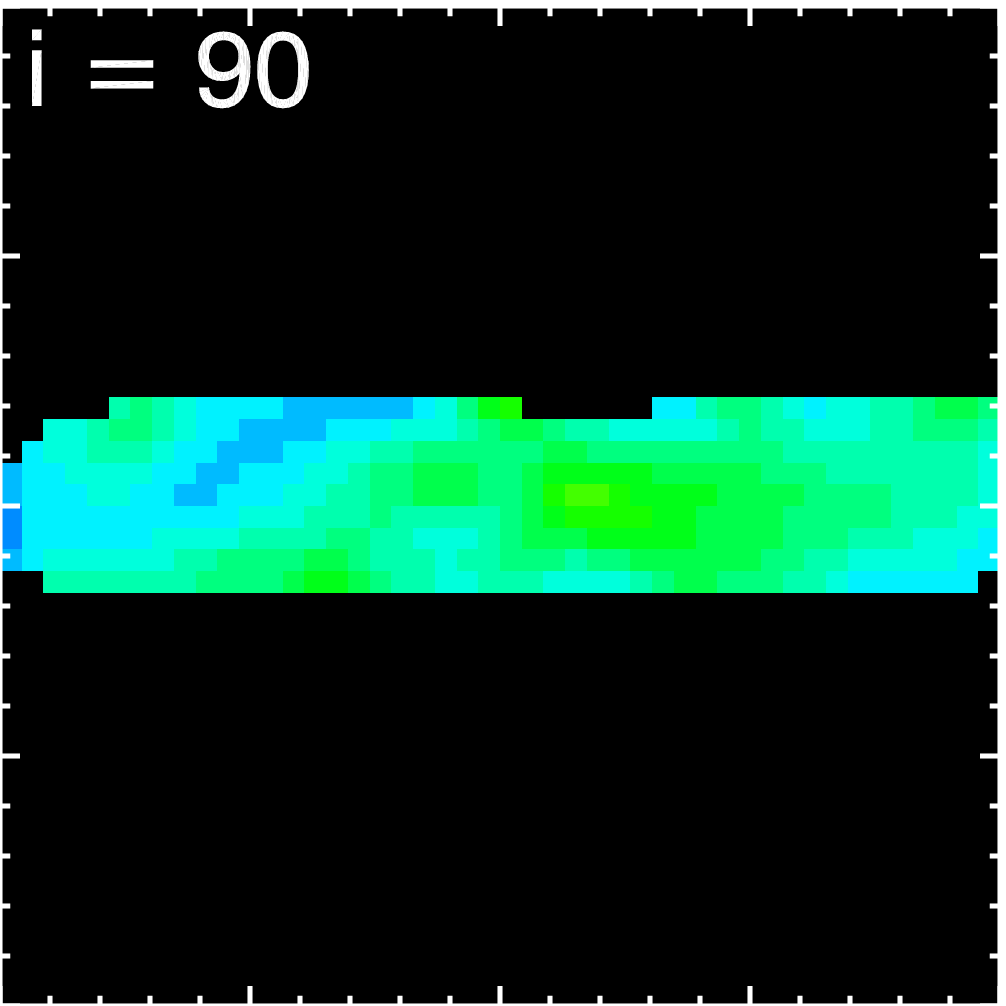}
\includegraphics[scale=0.4]{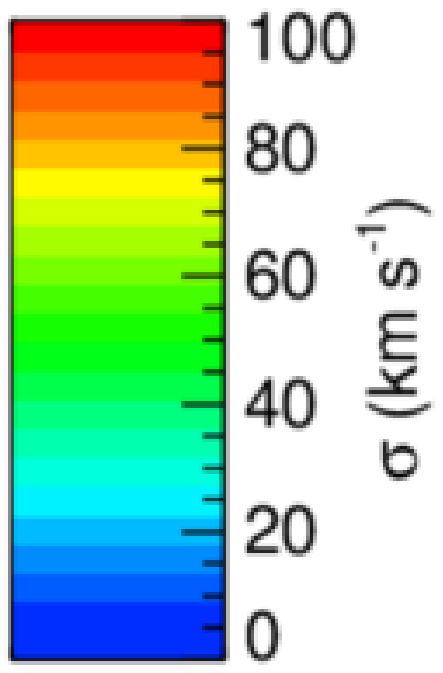}
\end{center}
\caption{Morphological and kinematic properties of galaxy g54 at $z = 2$ for our fiducial simulation including momentum-driven winds.   
Upper panels show (from top to bottom) the projected stellar surface density, gas surface density, and SFR surface density color coded by (SFR-weighted) line-of-sight velocity.
Lower panels show (from top to bottom) mock H$\alpha$ line intensity, line-of-sight velocity, and one-dimensional projected velocity dispersion maps that mimic integral field unit observations with SINFONI.  Two-dimensional projected images and mock maps are shown at different inclination angles, from direct face-on ({\it left}) to edge-on ({\it right}).  The region shown is 20\,kpc (physical) across.}
\label{fig:mock54}
\end{figure*}

\begin{figure*}
\begin{center}

\includegraphics[scale=0.5]{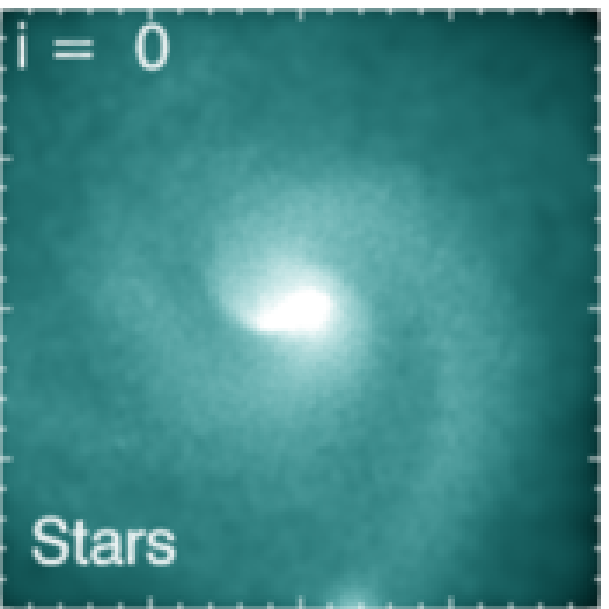}
\includegraphics[scale=0.5]{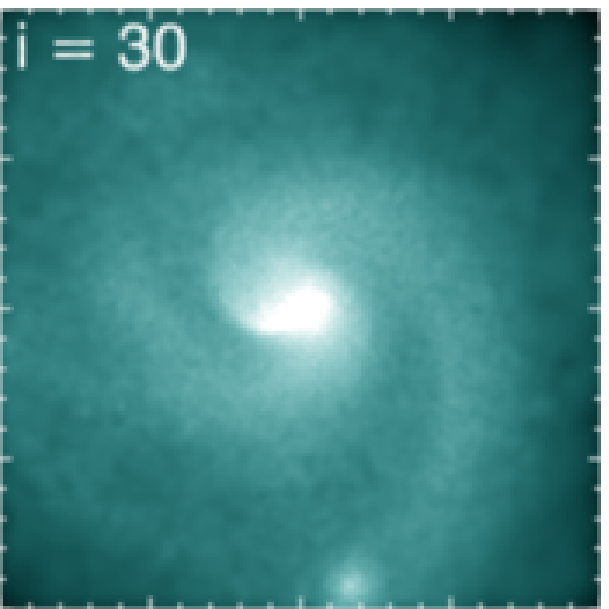}
\includegraphics[scale=0.5]{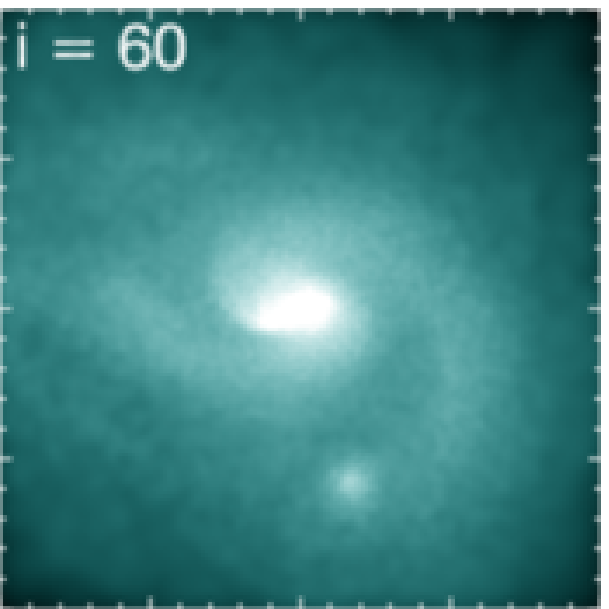}
\includegraphics[scale=0.5]{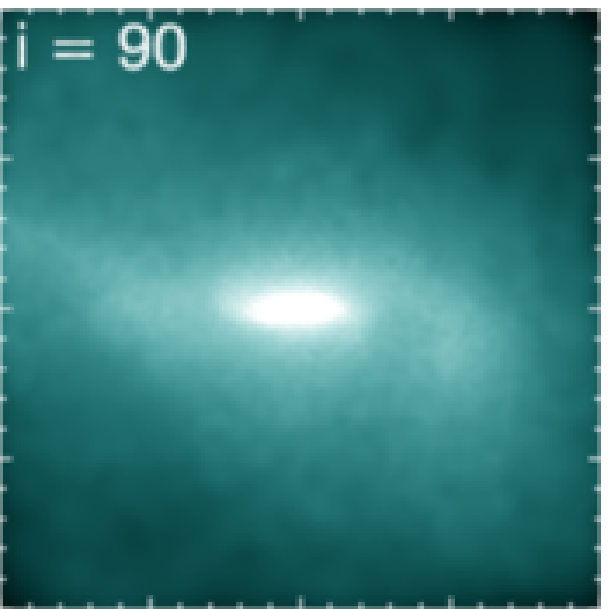}
\includegraphics[scale=0.3]{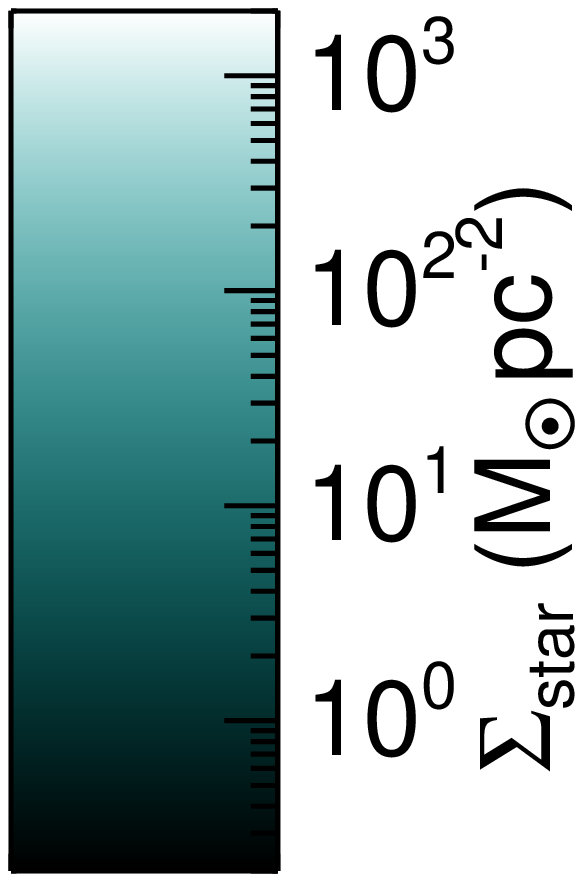}

\includegraphics[scale=0.5]{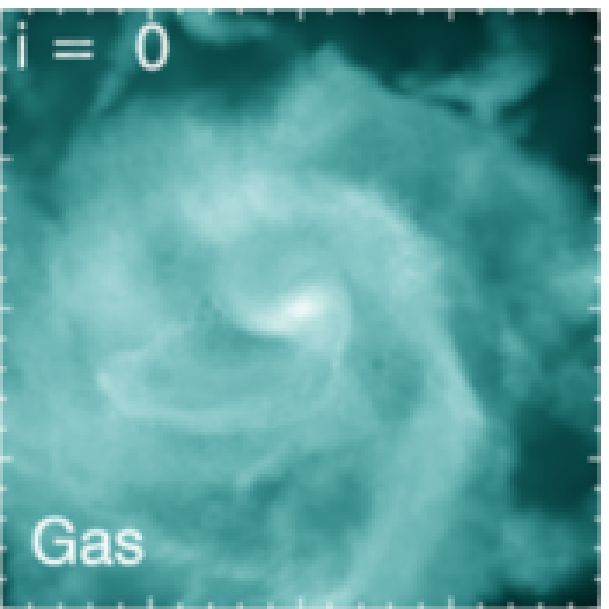}
\includegraphics[scale=0.5]{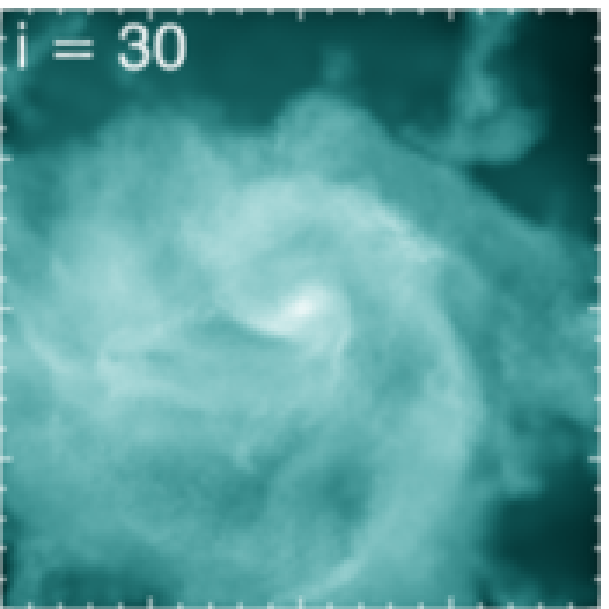}
\includegraphics[scale=0.5]{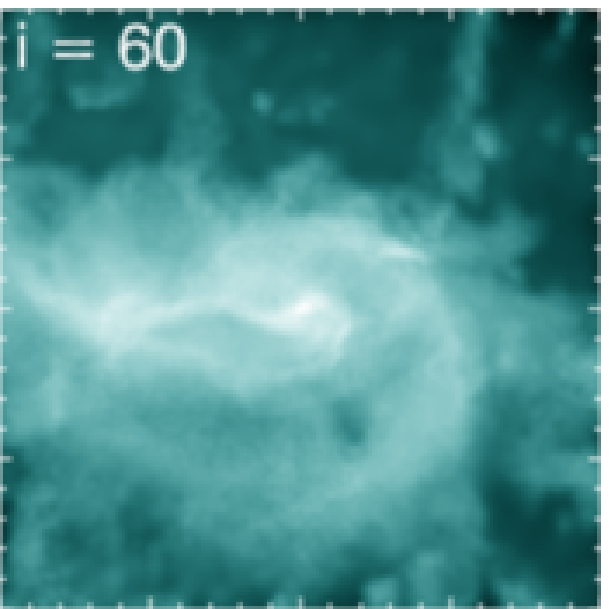}
\includegraphics[scale=0.5]{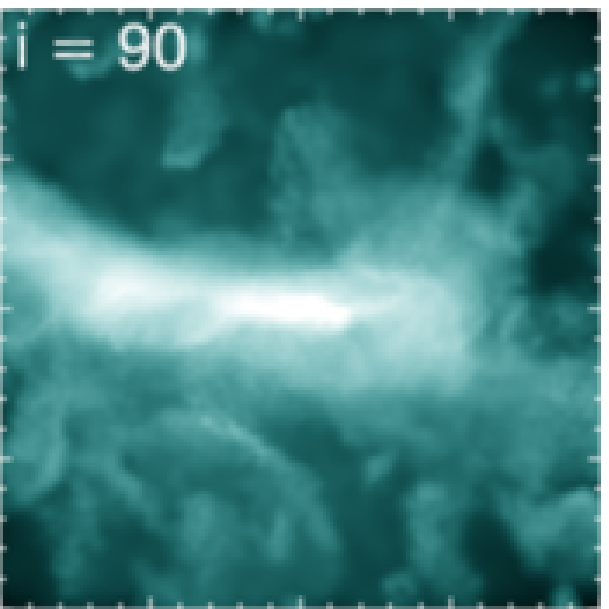}
\includegraphics[scale=0.3]{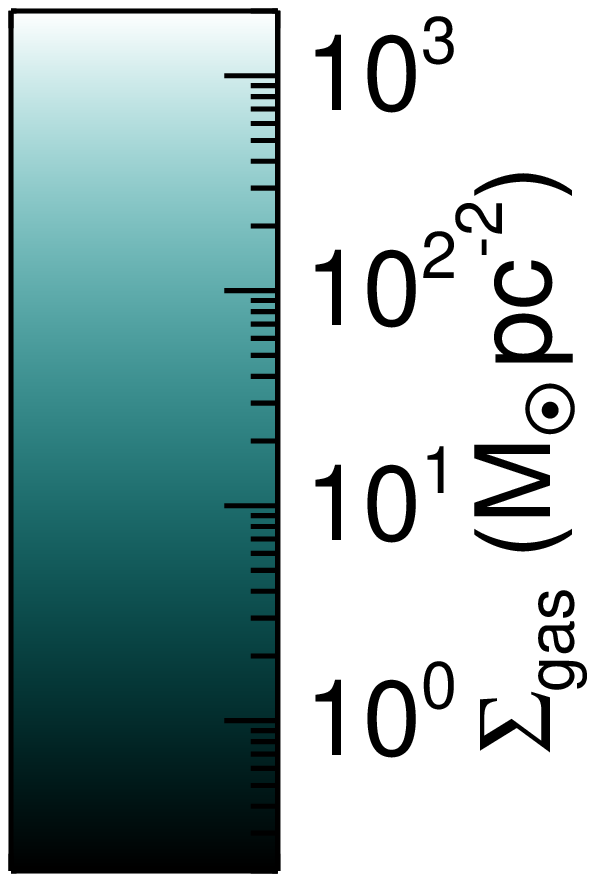}

\includegraphics[scale=0.5]{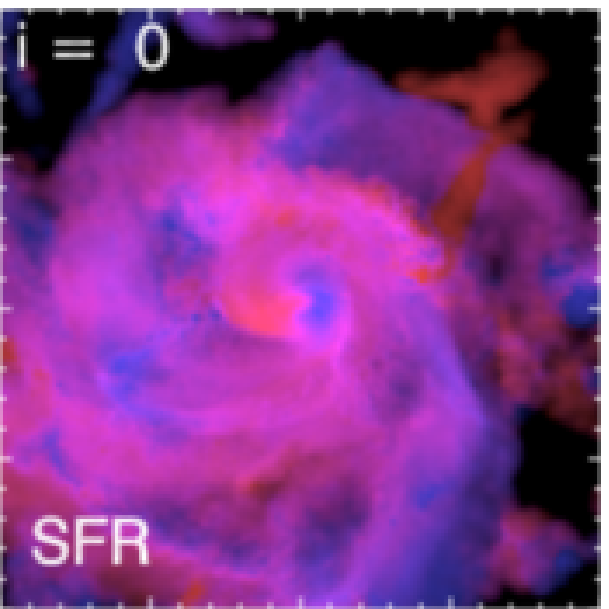}
\includegraphics[scale=0.5]{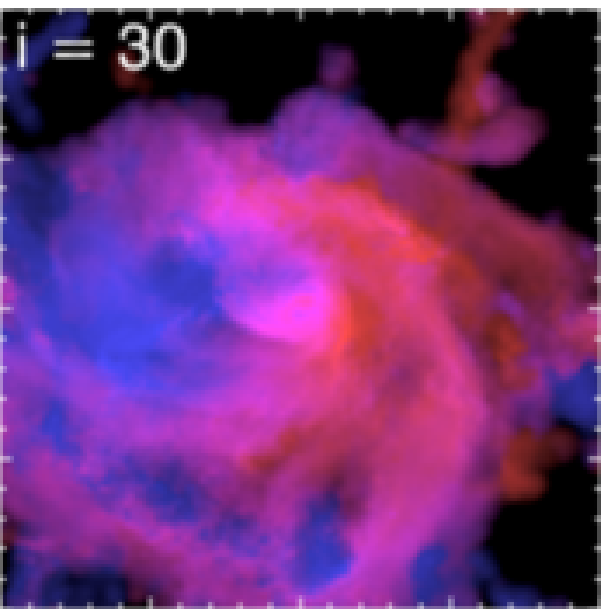}
\includegraphics[scale=0.5]{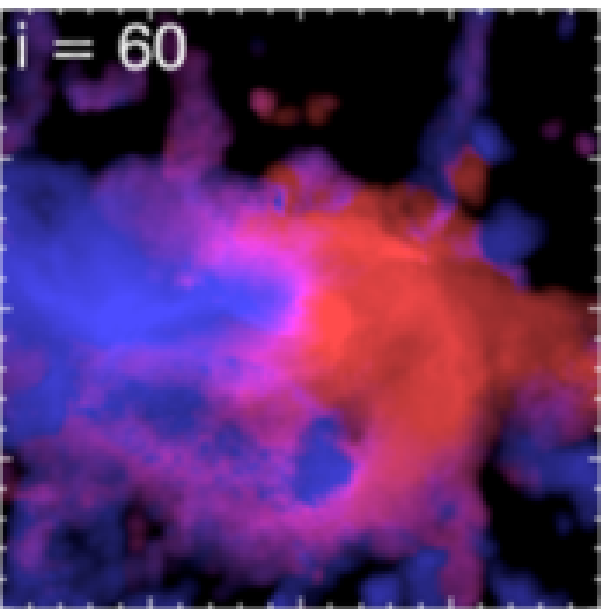}
\includegraphics[scale=0.5]{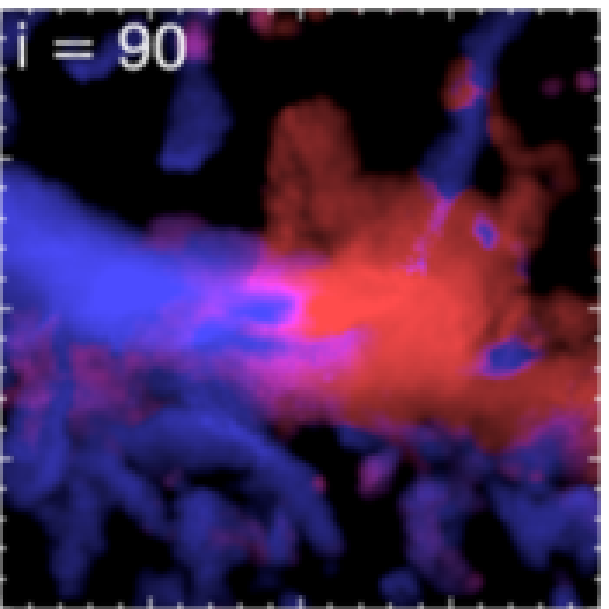}
\includegraphics[scale=0.4]{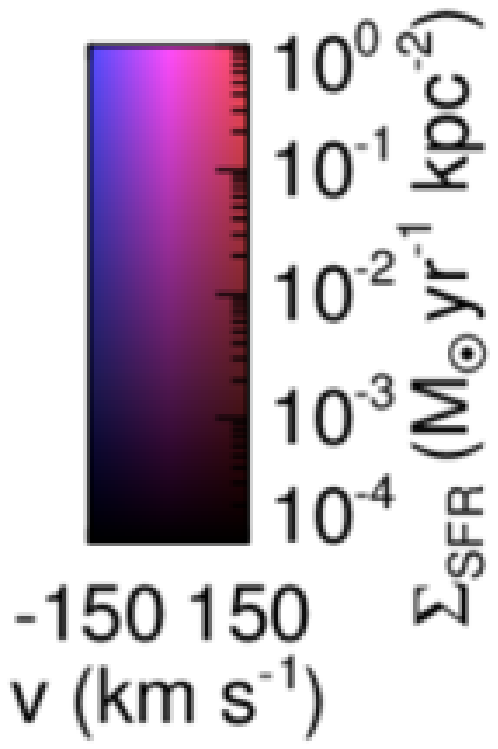}

 \bigskip
 \bigskip

\includegraphics[scale=0.3]{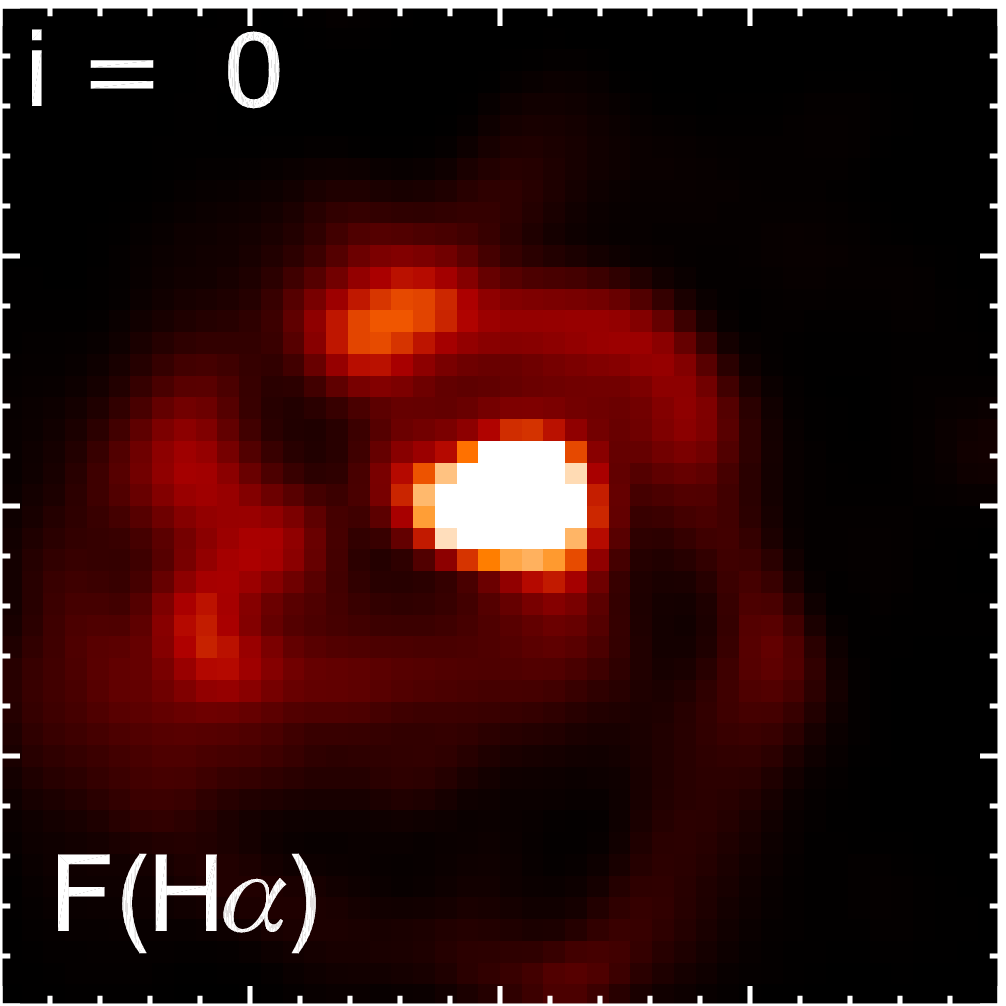}
\includegraphics[scale=0.3]{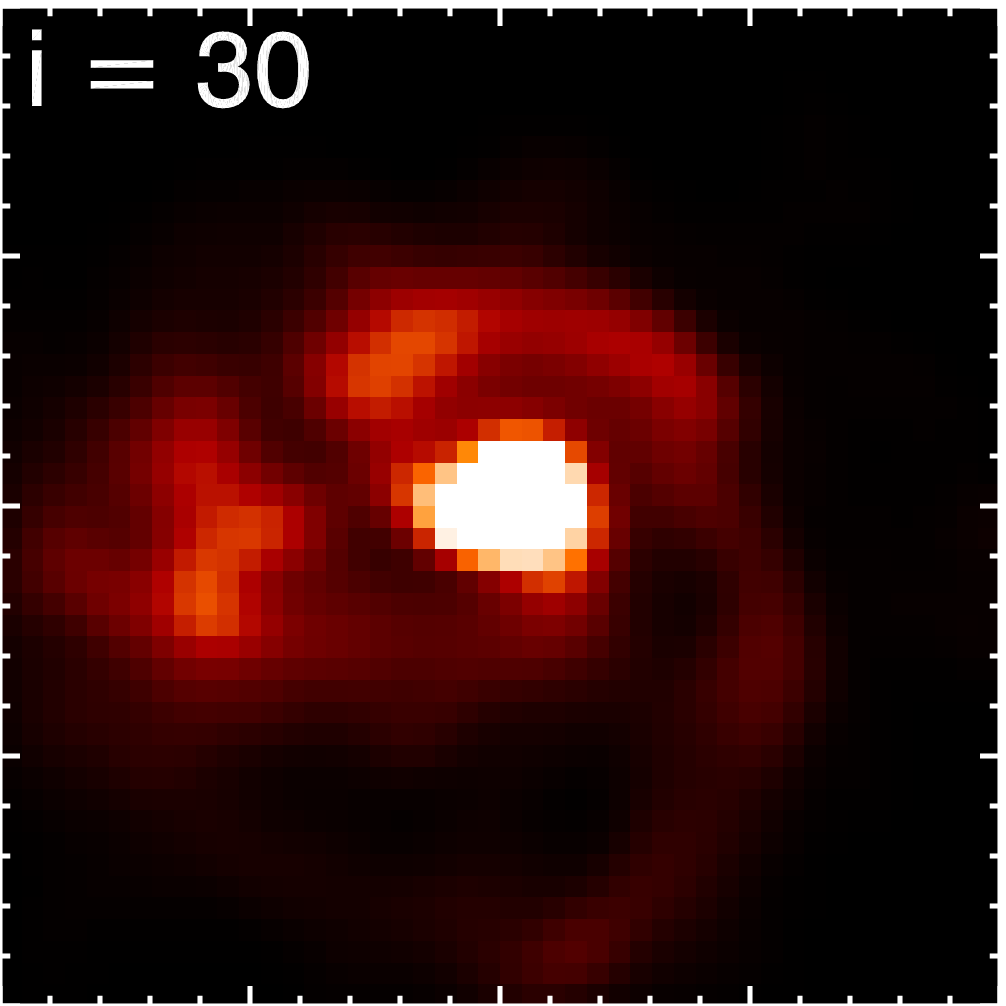}
\includegraphics[scale=0.3]{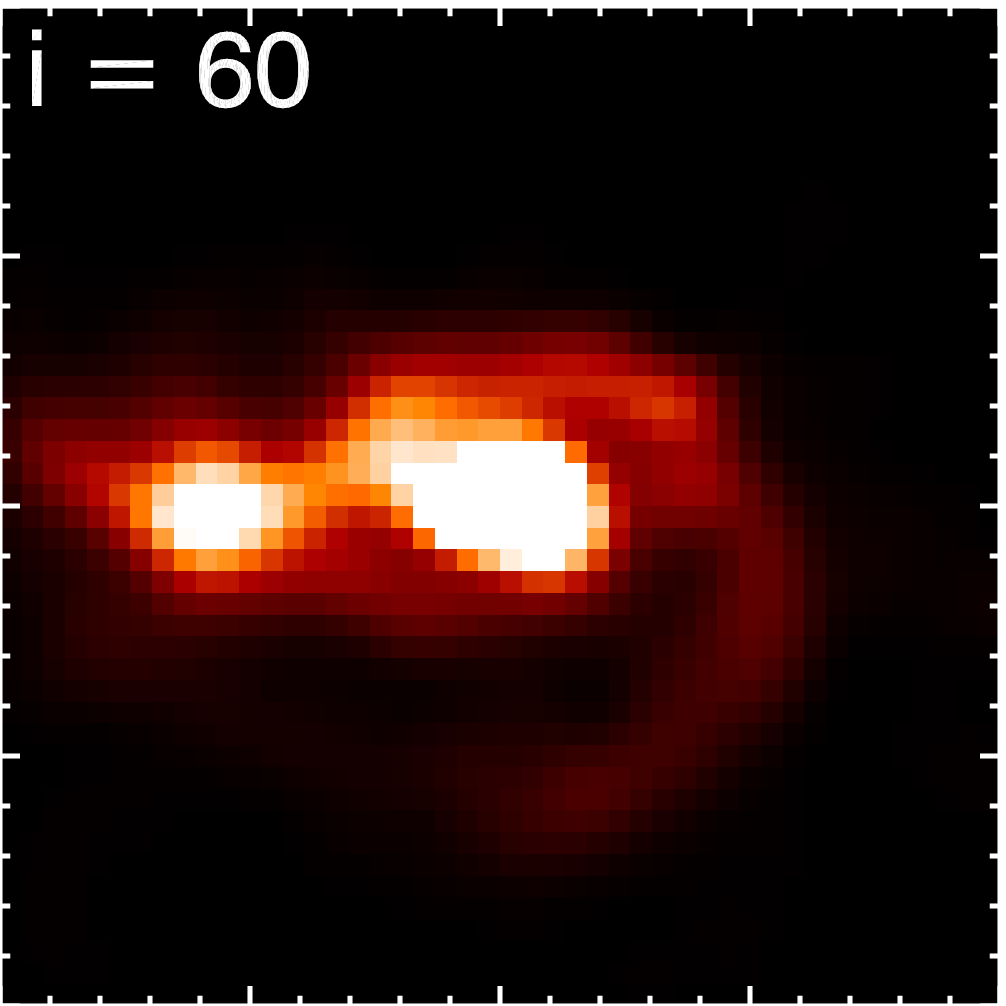}
\includegraphics[scale=0.3]{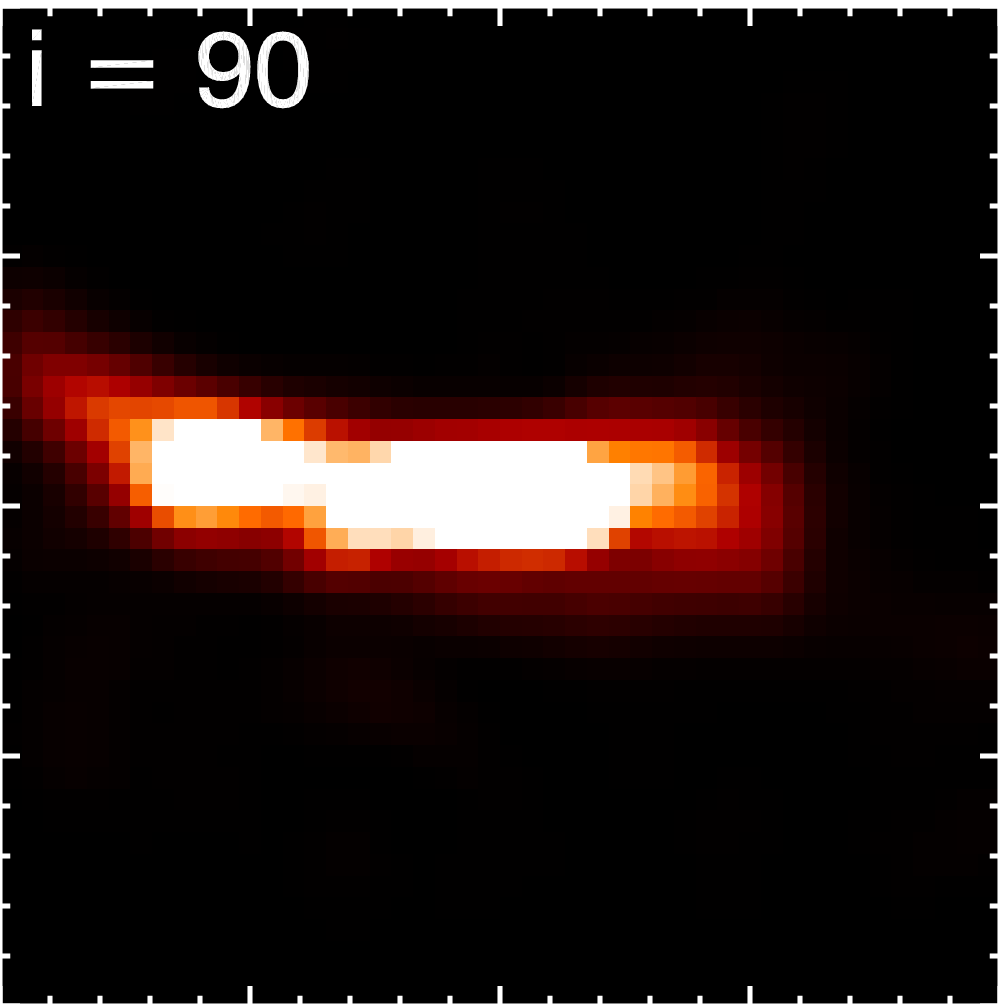}
\includegraphics[scale=0.4]{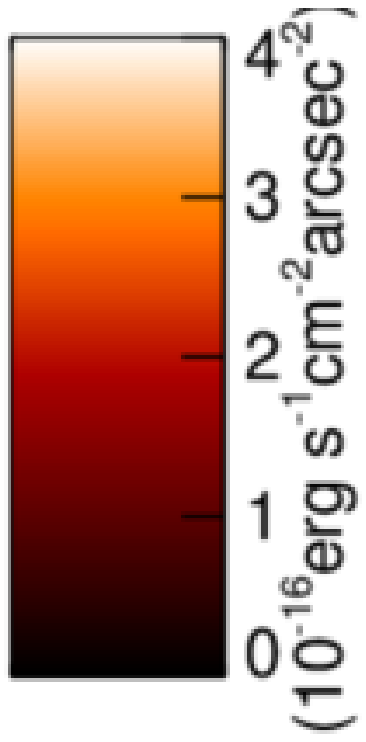}

\includegraphics[scale=0.3]{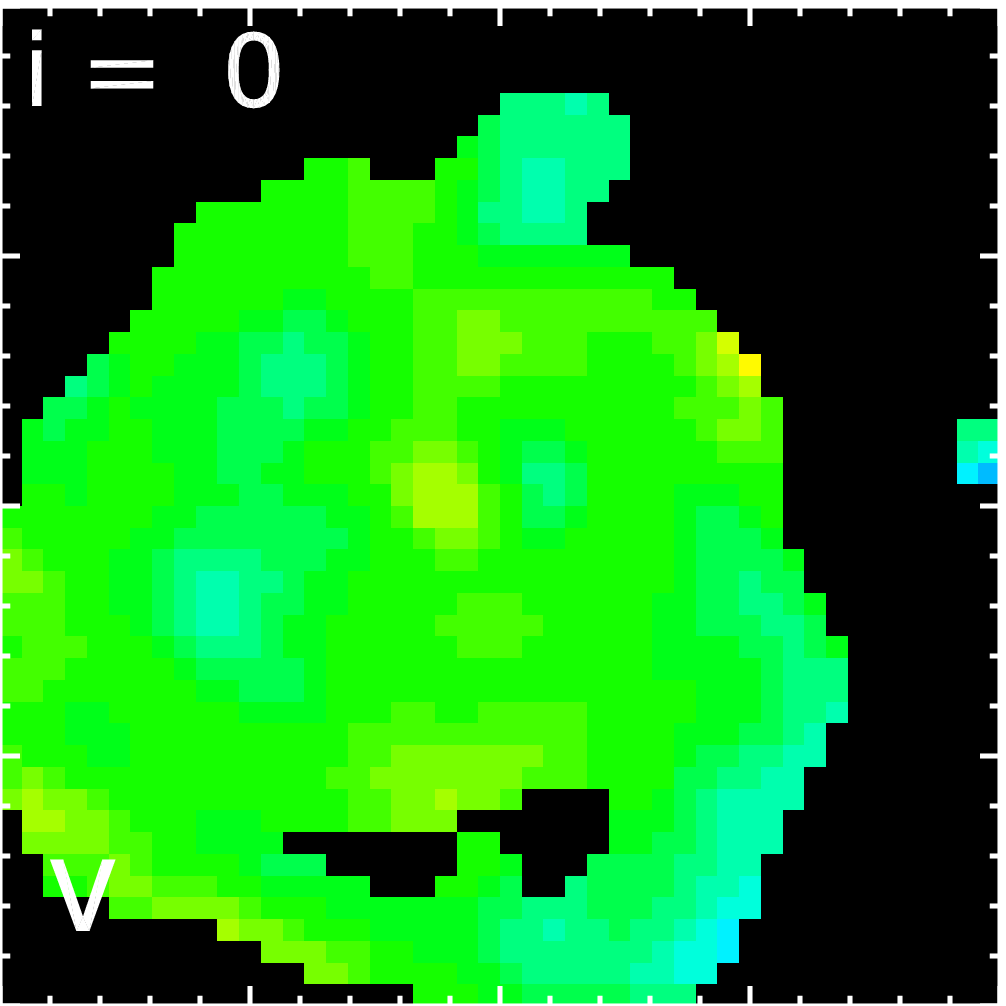}
\includegraphics[scale=0.3]{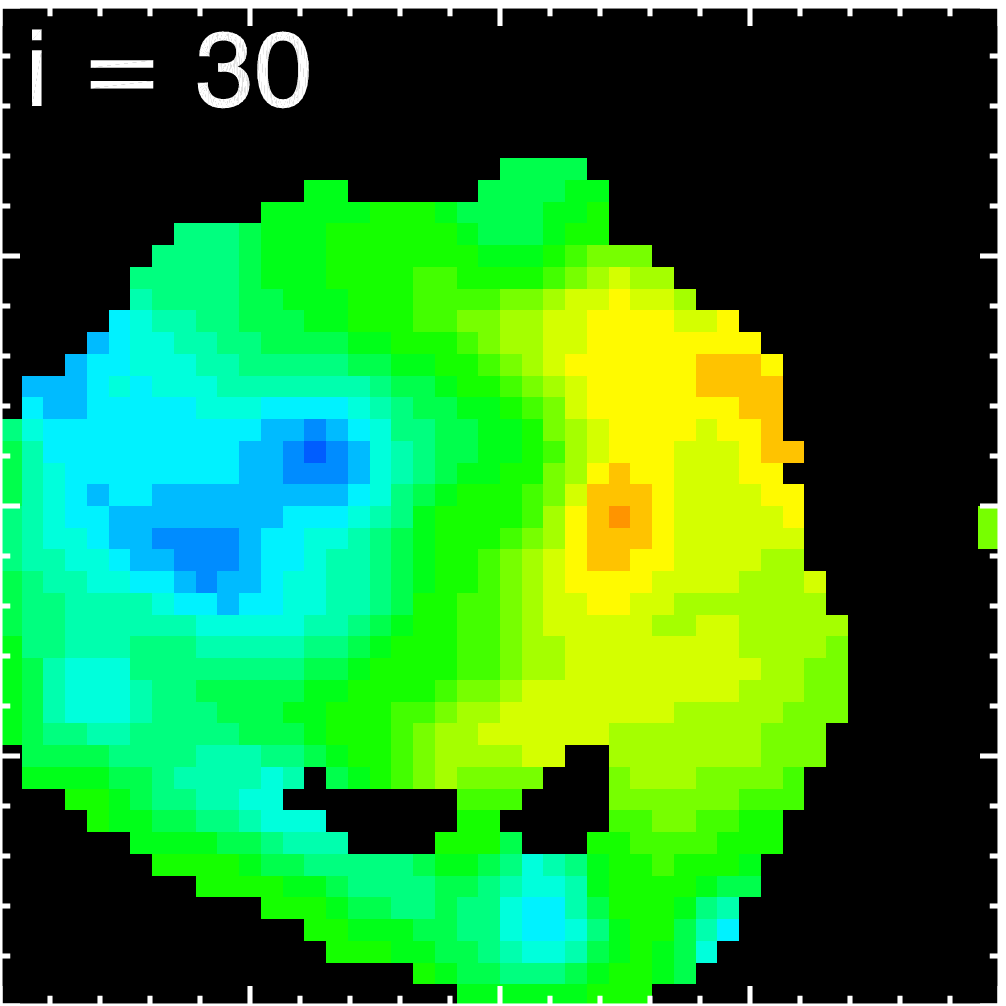}
\includegraphics[scale=0.3]{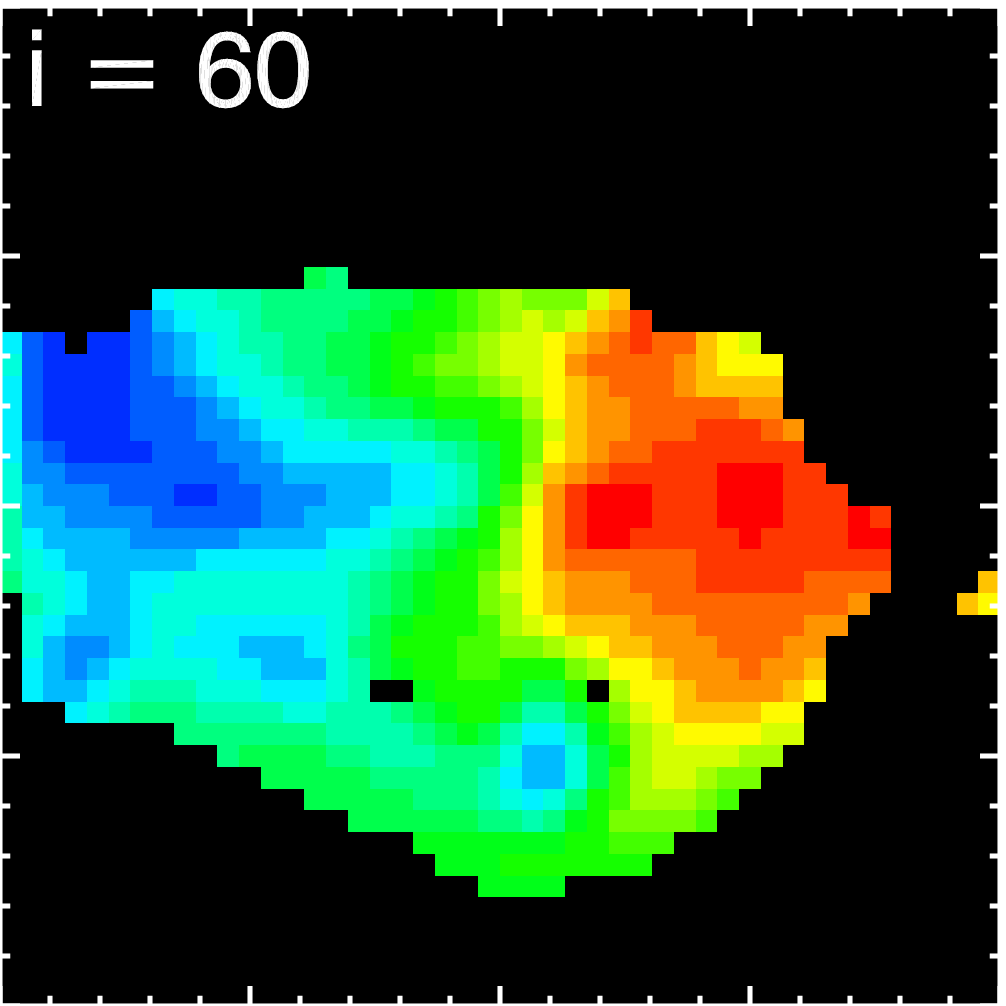}
\includegraphics[scale=0.3]{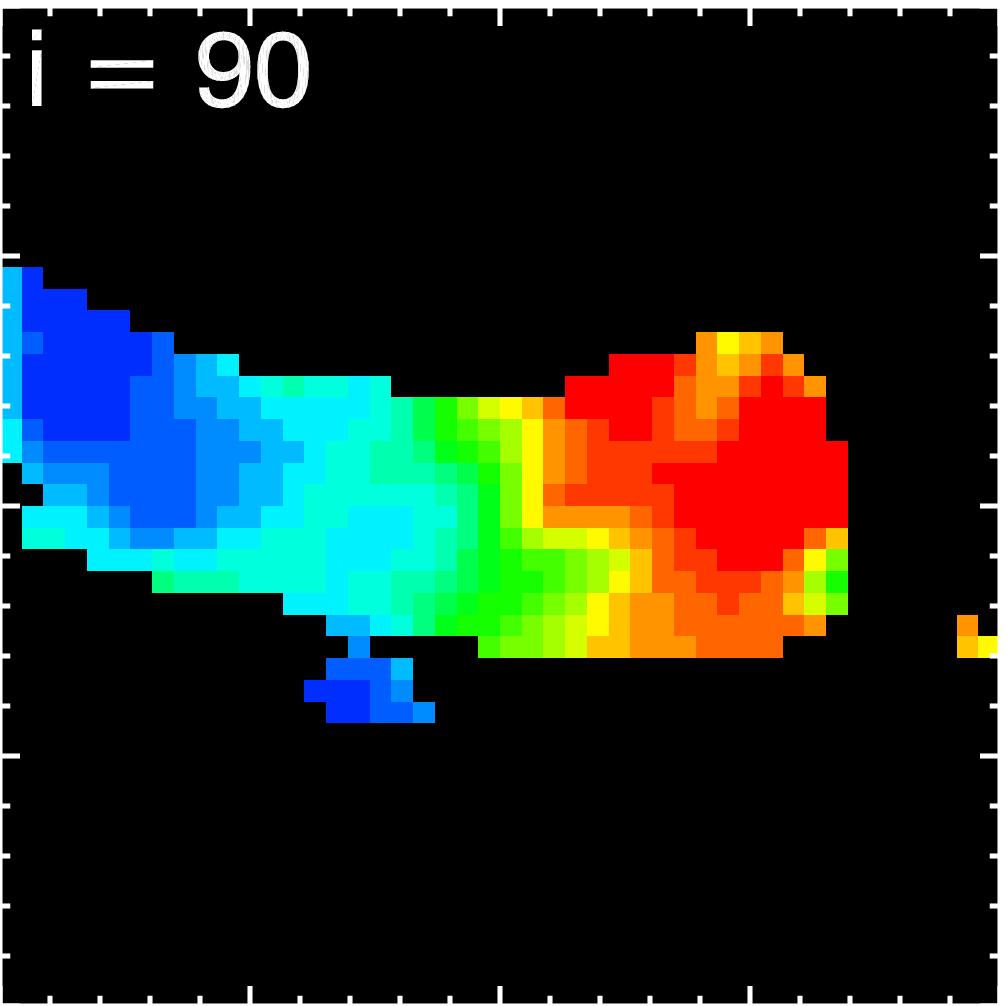}
\includegraphics[scale=0.4]{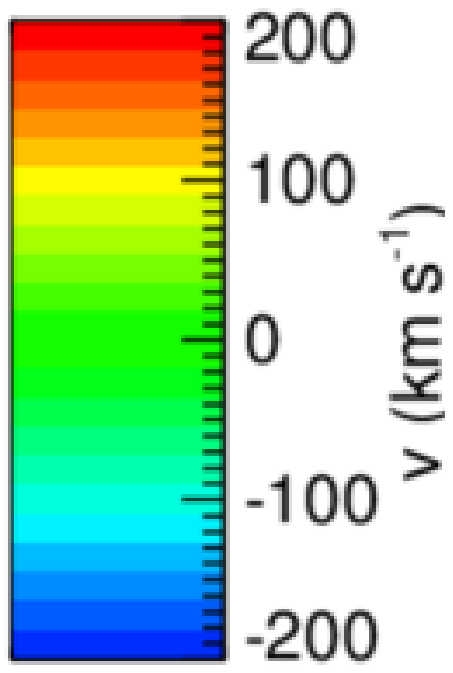}

\includegraphics[scale=0.3]{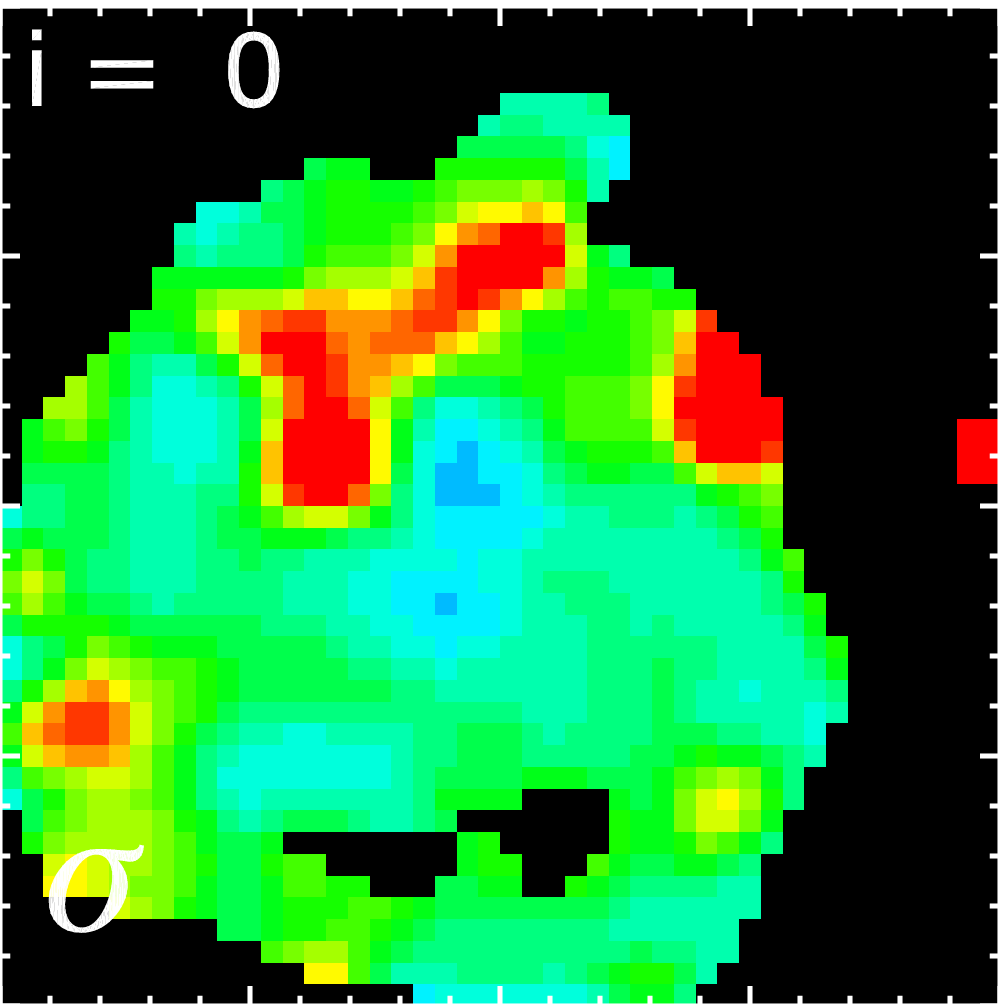}
\includegraphics[scale=0.3]{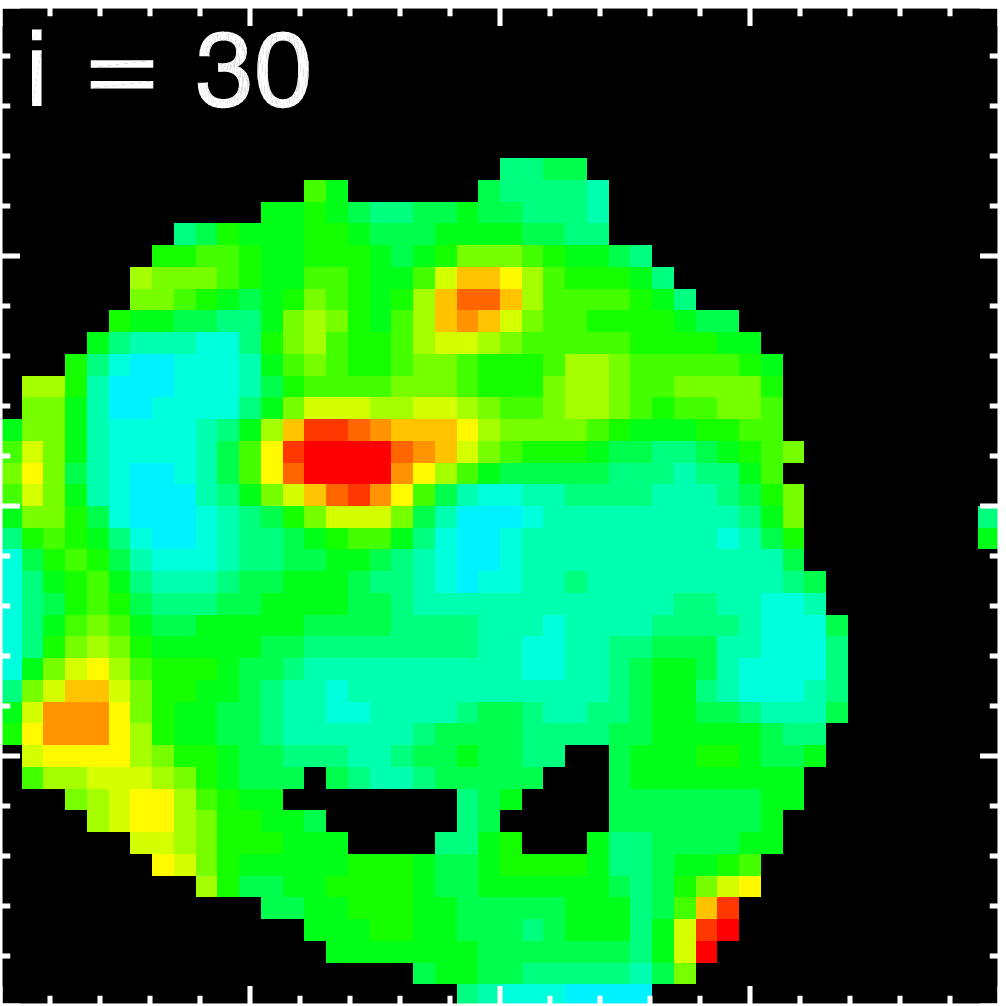}
\includegraphics[scale=0.3]{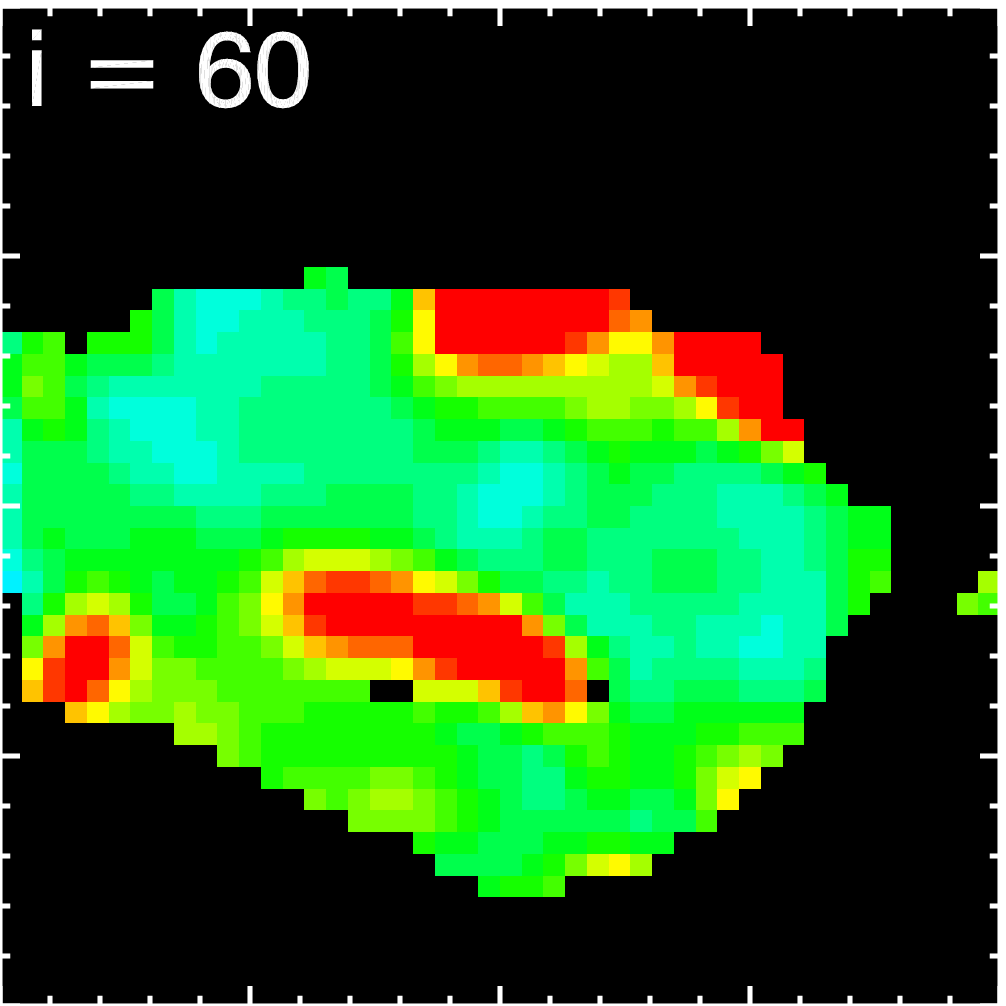}
\includegraphics[scale=0.3]{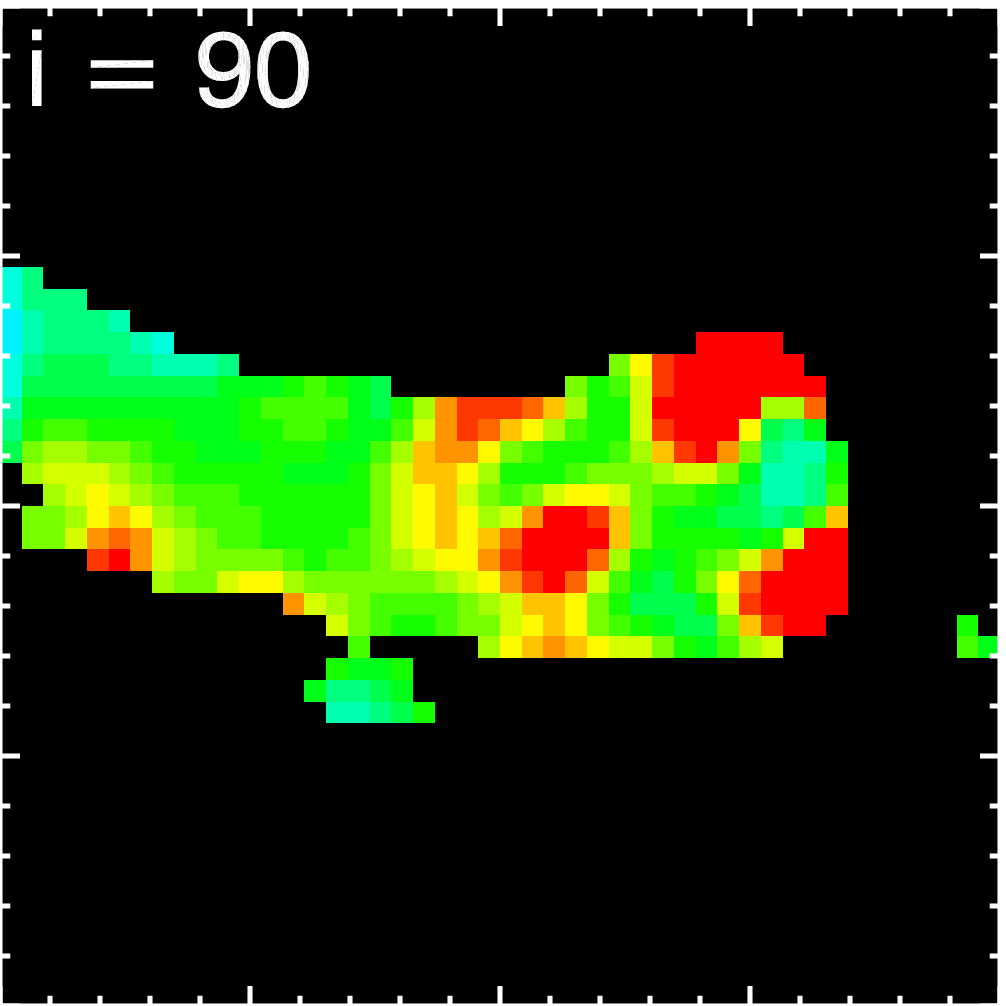}
\includegraphics[scale=0.4]{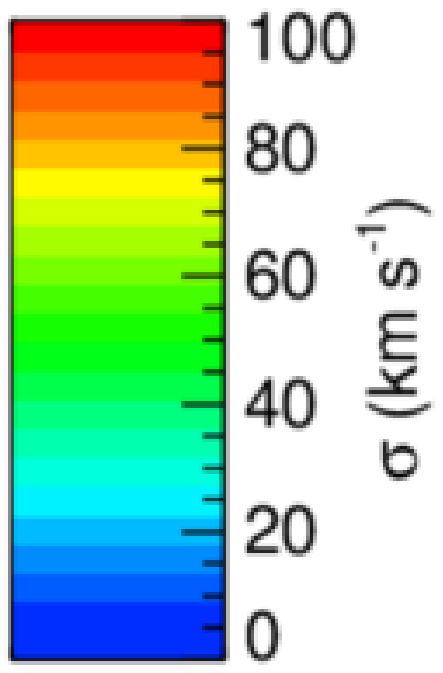}

\end{center}
\caption{Same as Figure~\ref{fig:mock54} for galaxy g222.}
\label{fig:mock222}
\end{figure*}

We begin by considering in detail the morphology of the two
galaxies that were specifically chosen for resimulation,
g54 and g222, and therefore lie near the center of the zoomed region.
Recall that g54 was chosen to be somewhat more isolated and have a fairly 
quiet merger history, while g222 lived in a denser region with a more
violent merger history.

Figure~\ref{fig:mock54} shows two-dimensional projected views of
galaxy g54 at four different inclination angles.  The top three
rows show the stellar surface density, gas surface density, and SFR
density distributions resulting from our fiducial simulation including
momentum-driven winds.  The three bottom rows show mock H$\alpha$
line intensity, line-of-sight velocity, and velocity dispersion
maps of galaxy g54 that mimic the resolution of current near-infrared
integral field spectroscopic observations with SINFONI.  Simulated
SFRs have been converted to H$\alpha$ luminosity using $L_{{\rm H}\alpha}
[{\rm erg\,s^{-1}}] = 2.3 \times 10^{41} {\rm SFR}$ [\Msun\,yr$^{-1}$]
\citep[from][corrected for a \citet{cha03} IMF]{ken98b}.  Then,
mock H$\alpha$ line intensity maps have been obtained by (1)
placing the simulated galaxy at $z = 2$, converting its physical
size to its apparent angular size, (2) convolving the obtained
H$\alpha$ flux map with a 0.17\arcsec~beam, and (3) matching
the pixel size to those of typical observations (0.05\arcsec\,pix$^{-1}$).
Mock H$\alpha$ flux-weighted line-of-sight velocity and velocity
dispersion maps have been obtained by degrading the spatial resolution
in a similar way.

With stellar mass $\sim 2.5 \times 10^{10}$\,\Msun, gas fraction
$f_{\rm gas} \approx 0.48$, and SFR $\sim 13.6$\,\Msun\,yr$^{-1}$,
galaxy g54 has a prominent two-arm spiral structure extending up
to $\sim 10$\,kpc scales that could be observable even at high
inclination angles ($i \leq 60^{\circ}$).  Star-forming gas and
stars generally trace the same structure, though the stars show a
prominent bulge that is absent in the gaseous component.  Mock
H$\alpha$ line intensity maps show a somewhat compact central nucleus
with radius $\sim 1$\,kpc and a factor $\sim 2$ increased brightness
relative to the spiral arms (depending on inclination angle).
Observations using star formation tracers and rest-frame optical
observations should, thus, infer similar spiral morphologies.  The
velocity maps illustrate the effects of the inclination angle on the
observed kinematic properties of galaxies.  A very smooth velocity
gradient along the morphological major axis can be clearly identified
for this galaxy at a wide range of inclination angles ($i \geq
30^{\circ}$).  Interestingly, galaxy g54 shows a characteristic
``spider diagram" pattern in the velocity iso-contours at intermediate
inclination angles ($i \approx 30^{\circ}-60^{\circ}$), as expected
for inclined rotating disks.  For nearly edge-on observations,
line-of-sight projected peak-to-peak velocities reach values
 $> 400$\,km\,s$^{-1}$, consistent with the circular velocity
$v_{\rm circ} \approx 230$\,km\,s$^{-1}$ calculated at the outer
edge of the disk ($R \sim 10$\,kpc).  With a relatively flat velocity
dispersion map, this galaxy would be identified as a large, thick,
rotation-dominated disk.  These properties are in general quite
similar to, though somewhat scaled down from BX442, the $z=2.18$
grand design spiral observed by \citet{law12}.  BX442 has a measured
stellar mass $\sim 6 \times 10^{10}$\,\Msun, H$\alpha$-derived SFR
$\sim 45$\,\Msun\,yr$^{-1}$, inclination-corrected circular velocity
$v_{\rm circ} \approx 234$\,km\,s$^{-1}$, and very similar size and
morphology compared to our simulated galaxy g54.

Figure~\ref{fig:mock222} shows similar two-dimensional maps for
galaxy g222, a marginally higher mass galaxy that has undergone
more frequent interactions and mergers.  Indeed, there is an infalling
galaxy along the lower spiral arm, which is prominent in the stellar
distribution but much less evident in the gas, possibly because its
gas has been stripped during infall.  The stellar surface density
distribution shows a high concentration of stars at the center of
g222 and possibly a weak, extended spiral structure.  The gas and
SFR distributions reveal high levels of turbulence on this galaxy.
The morphology of g222 appears highly disturbed in H$\alpha$ emission
and it does not seem to trace the smooth stellar component as in
the case of g54.  The velocity maps show signs of ordered rotation
in the underlying large-scale gas distribution at high inclination
angles ($i \geq 60^{\circ}$), but the inferred velocity gradients
are significantly disturbed.  The velocity dispersion map reveals
an irregular gas clumpy structure with significant turbulent motions.
The H$\alpha$ shows up to several clumps, depending on the viewing
angles, but these clumps are not apparent in the stellar distribution,
which suggests these are short-lived gaseous clumps as found in similar
simulations by~\citet{gen12b}, and as inferred in observations by
\citet{wuy12}; we will examine clump properties in more detail in
future work.  Overall, this simulated galaxy shares various characteristic
morphological and kinematic properties with typical clumpy disk galaxies observed at 
$z \sim 2$~\citep{for09,for11,genz11,swi12a,swi12b}.

Note, however, that star formation in our simulated galaxies appears to be in general more centrally concentrated than observed $z \sim 2$ disk galaxies (Figure~\ref{fig:gasima}), 
for which a central H$\alpha$ peak is sometimes not present, particularly for less massive galaxies.  
This might be partially explained by obscuration effects, which we will quantify in future work in order to make a more detailed comparison between observed H$\alpha$ line intensity maps and the intrinsic SFR surface density distribution of simulated galaxies.  
One numerical issue is that the pressurized ISM resulting from the star formation prescription \citep{spr03a} may inhibit the formation of off-center clumps by gravitational instability, especially in galaxies undergoing a more quiescent evolution.  
Despite this, the two example galaxies presented here---with stellar masses $M_{\rm star} > 10^{10}$\,\Msun~that are actually comparable to commonly observed $z \sim 2$ systems---illustrate the wide diversity in morphologies predicted in these simulations, as well
as its dependence on tracer (gas, stars, H$\alpha$).  Disk structures
are usually present, but can range from quiescent ``grand-design"
spirals to turbulent and clumpy disks.

\subsection{Comparison to SINS and SHiZELS}

\begin{figure*}
\begin{center}
\includegraphics[scale=0.5]{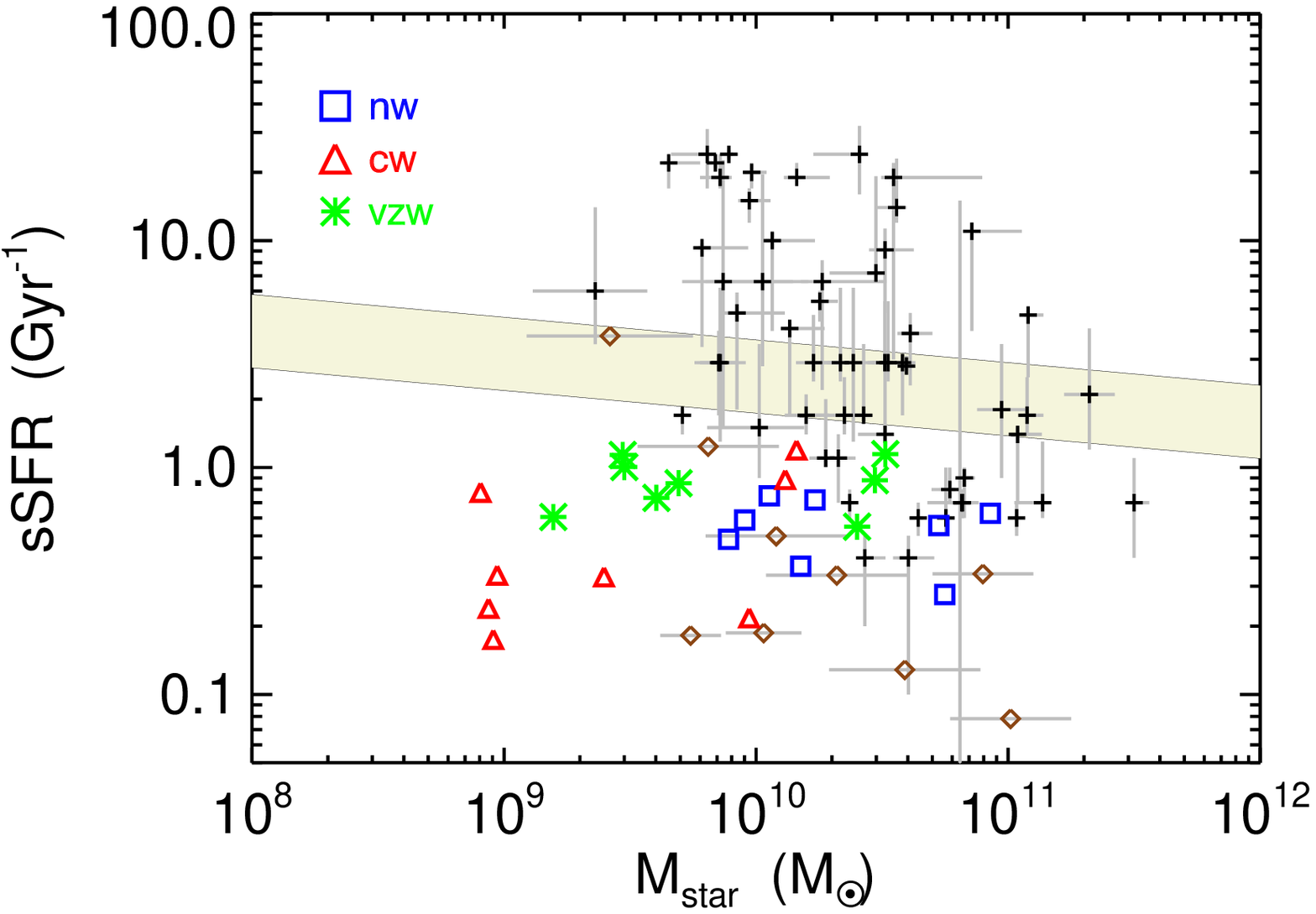}
\includegraphics[scale=0.5]{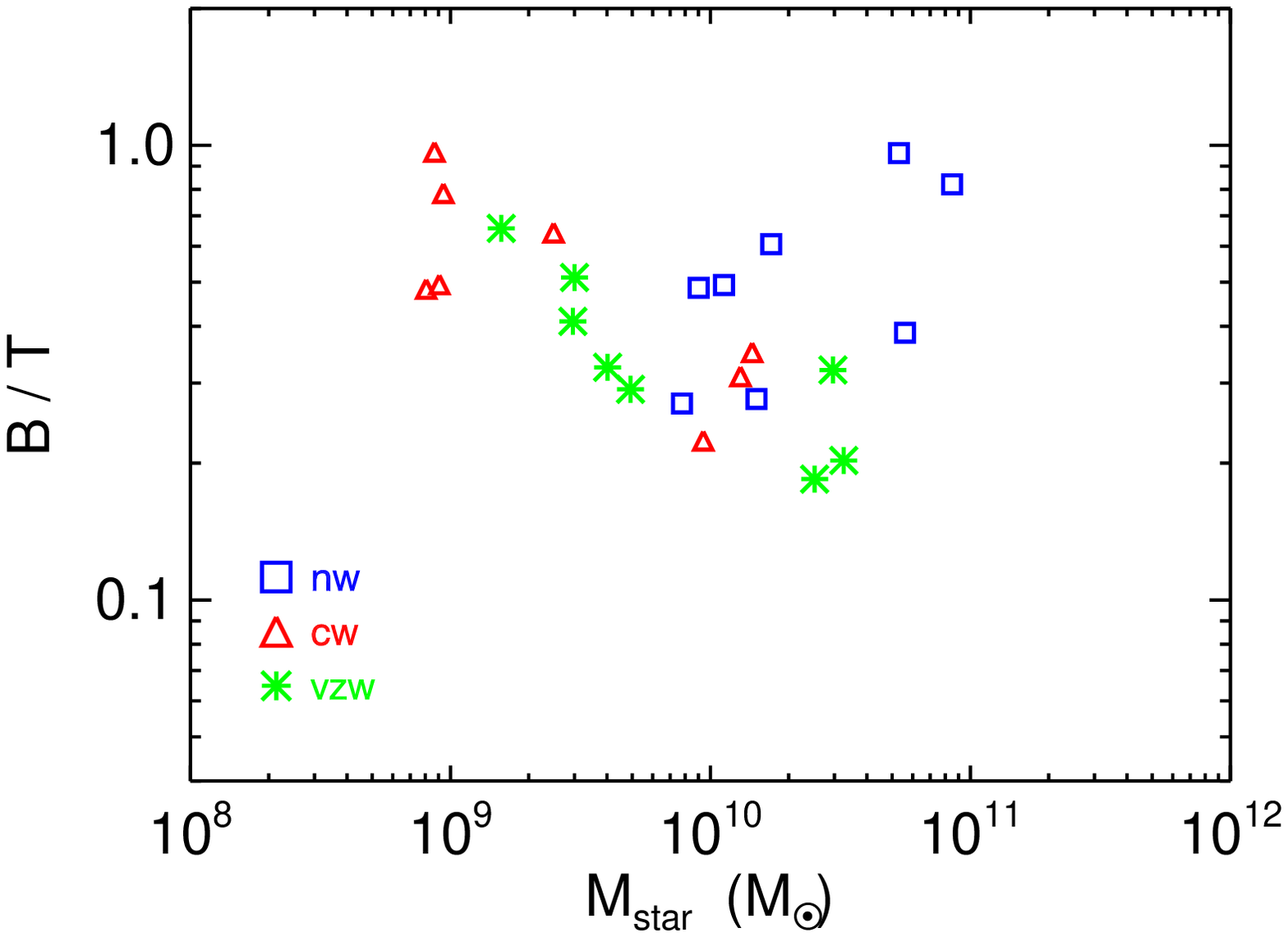}
\includegraphics[scale=0.5]{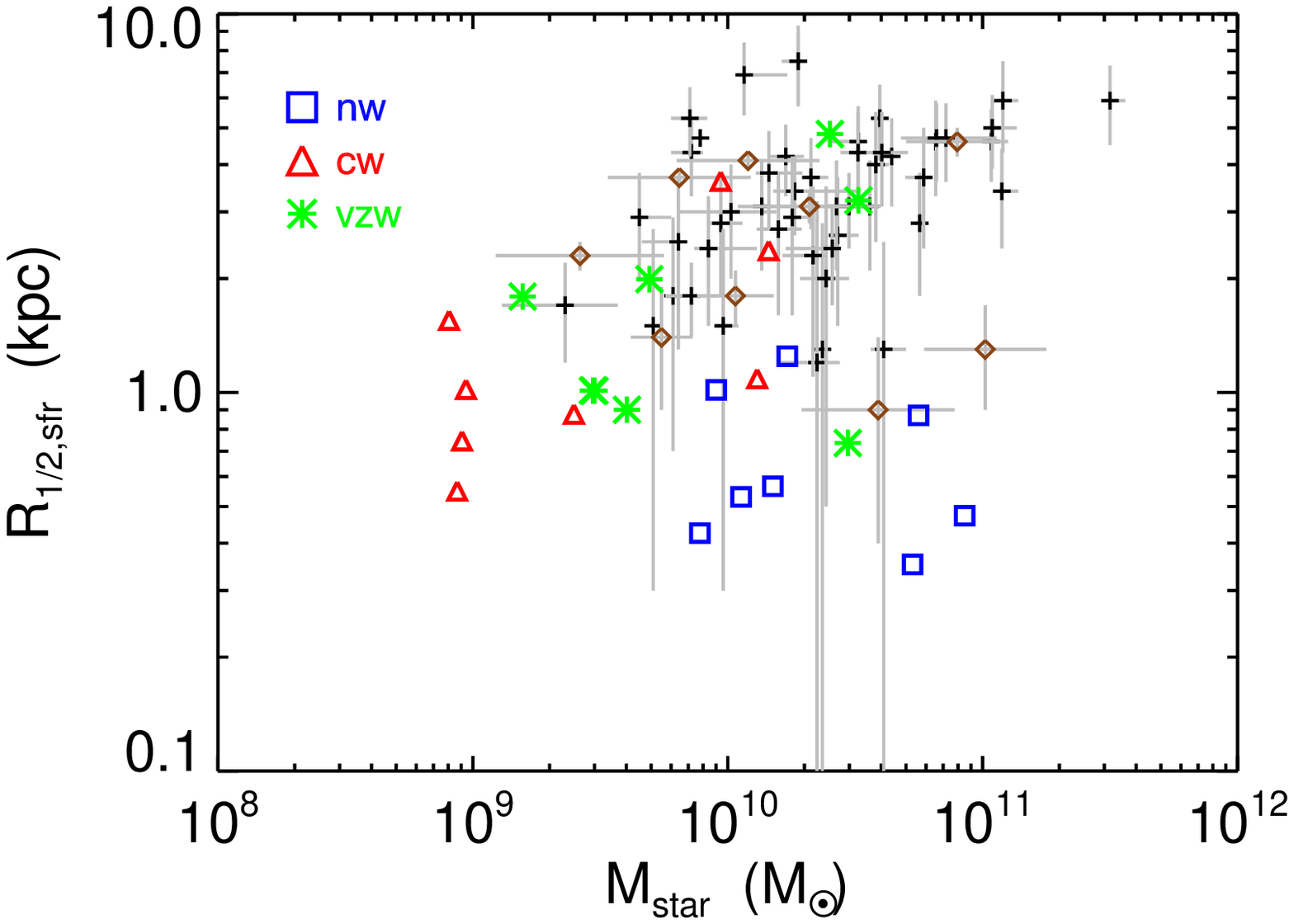}
\includegraphics[scale=0.5]{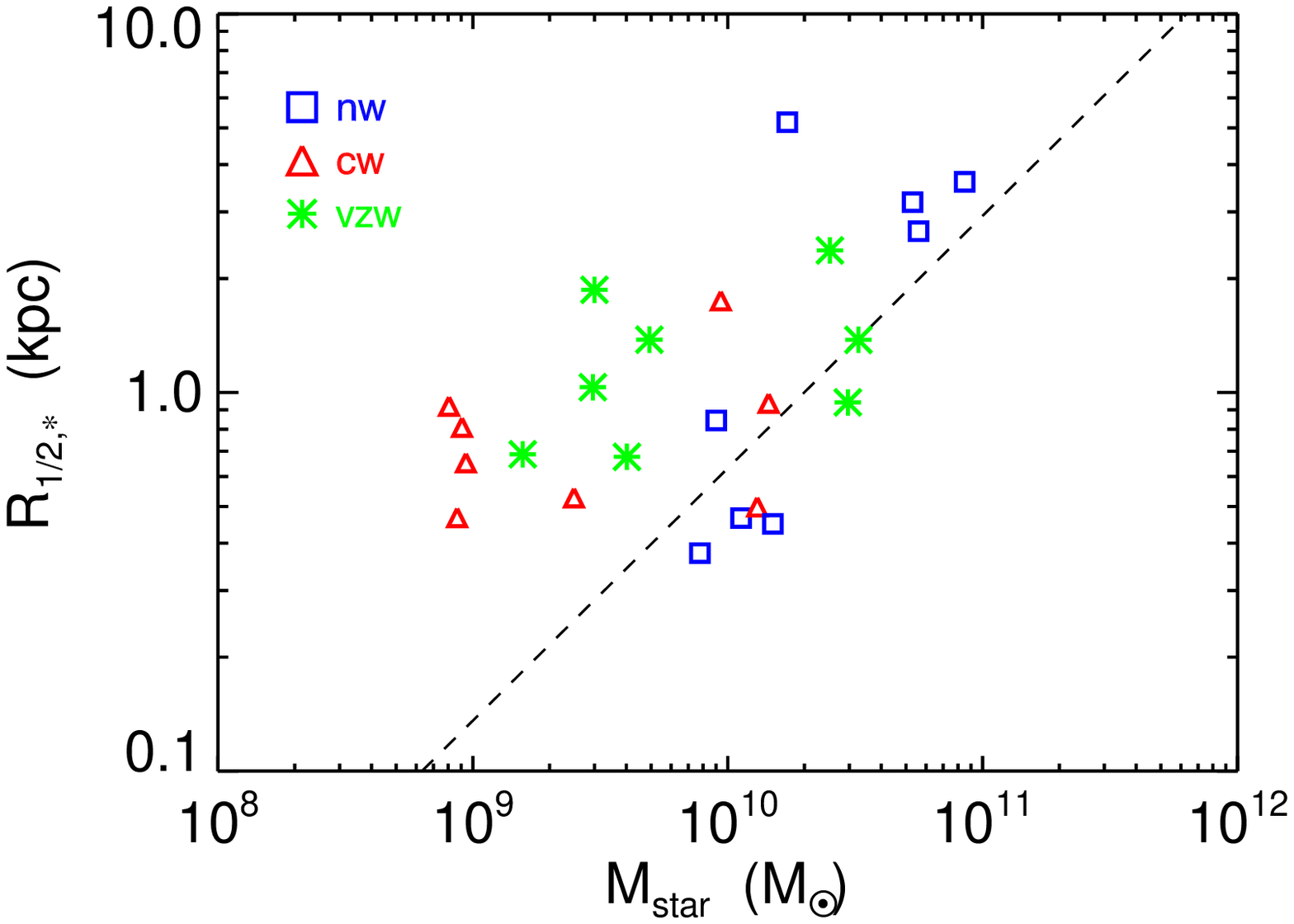}
\includegraphics[scale=0.5]{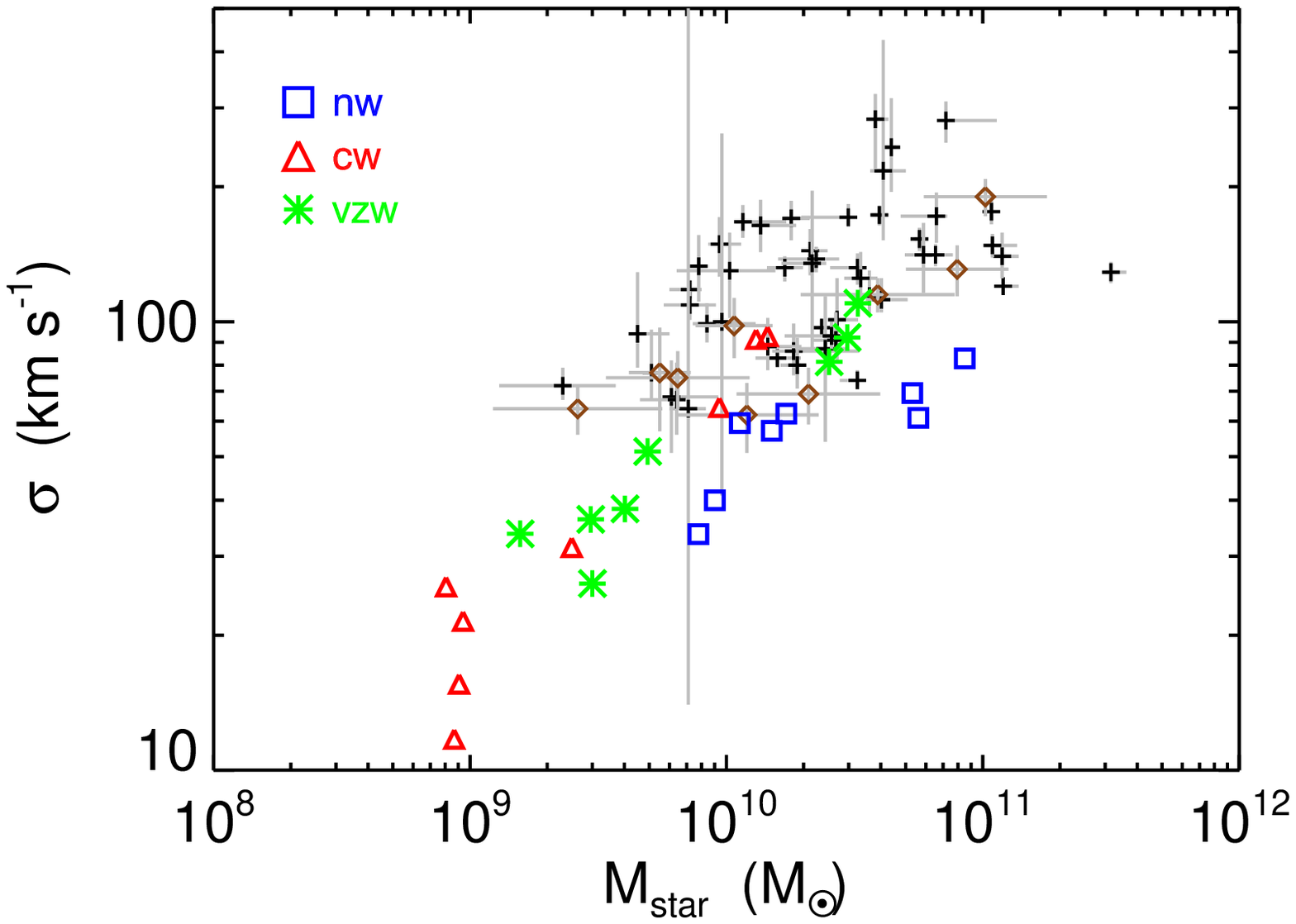}
\includegraphics[scale=0.5]{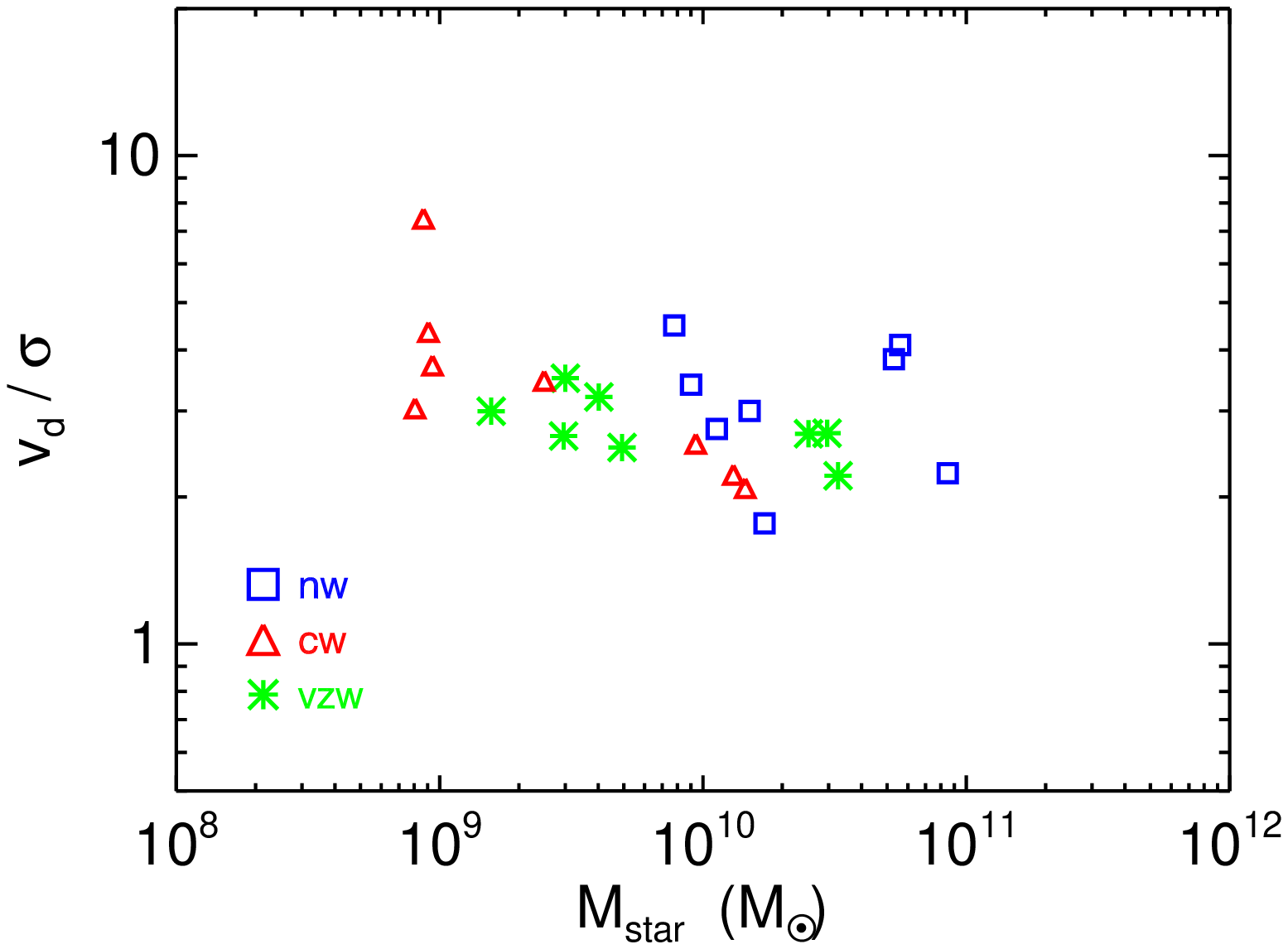}
\end{center}
\caption{Structural and dynamical quantities calculated from our sample of simulated galaxies at $z=2$, for our momentum-driven wind (vzw; green stars), constant wind (cw; red triangles), and no wind (nw; blue squares) simulations.  Black crosses with grey error bars in each panel represent the $z \sim 2$ star-forming galaxies of the SINS survey \citep{for09}, while the brown diamonds correspond to a sample of galaxies from the SHiZELS survey \citep{swi12b}.  
{\it Top left}: specific star formation rate (sSFR) as a function of stellar mass ($M_*$).  The shaded beige band shows the $z = 2$ $M_{\star}$--sSFR correlation of \citet{dad07} with a scatter (semi-interquartile range) of 0.16 dex in sSFRs.
{\it Top right}: stellar bulge mass fraction as a function of $M_*$.
{\it Middle left}: radius enclosing half of the total star formation rate as a function of $M_*$.
{\it Middle right}: radius enclosing half of the total stellar mass as a function of $M_*$.  The dashed line shows the nominal threshold stellar surface density $M_*/R_{\rm half,*}^{1.5}=10^{10.3}\,M_\odot$\,kpc$^{-1.5}$ for compact galaxies from~\citet{barro13}.
{\it Bottom left}: velocity dispersion $\sigma$ of the gas component of simulated galaxies as a function of $M_*$. 
{\it Bottom right}: disk peak rotational velocity $v_{\rm d}$ divided by $\sigma$ of the star-forming gas, as a function of $M_*$ for all simulated galaxies.}
\label{fig:sins}
\end{figure*}

We now conduct a more quantitative comparison to the
structural and dynamical properties of $z\sim2$ galaxies inferred
from near-infrared integral field spectroscopic observations obtained
with SINFONI at the Very Large Telescope.  In particular, we focus on the ``H$\alpha$
sample" of the SINS survey presented in \citet{for09}, consisting
of 62 rest-frame UV/optically selected star-forming galaxies at $z
\approx 1.5$--2.5, and the SHiZELS survey \citep{swi12b}, consisting
of nine H$\alpha$-selected galaxies at $z \approx 0.84$--2.23 drawn
from the HiZELS near-infrared narrow-band survey \citep{gea08}.  In
Figure~\ref{fig:sins}, we show some key structural and dynamical
quantities versus stellar mass, for our simulated galaxies color-coded
by wind model, as well as for the SINS (black crosses) and SHiZELS
(brown diamonds) samples.

The upper left panel of Figure~\ref{fig:sins} shows the specific SFR as a
function of stellar mass for our sample of galaxies at $z = 2$.
Our simulations naturally predict the existence of a correlation
between galactic SFRs and stellar masses, $M_{\star} \propto $ SFR,
fairly insensitive to stellar feedback models, due to the dominance
of smooth and steady cold accretion \citep[e.g.,][]{fin06,dav08,dav11a}.
As expected from models, a tight $M_{\star}$--SFR relation has been
observed out to $z \sim 2$, with a roughly constant slope close to
unity \citep[e.g.,][]{dad07,elb07,noe07}.  However, the observed
normalization of this relations increases more rapidly with redshift
from $z=0\rightarrow 2$ that current cosmological simulations
predict, resulting in too low predicted specific SFRs at $z\sim
2$~\citep{dad07,dav08}.  Our zoom simulations follow this trend---Figure~\ref{fig:sins} shows that they lie below the $M_{\star}$--SFR
correlation of \citet{dad07} by a factor $\times 2$--3 in specific
SFRs (for the vzw model), and by somewhat higher factors for the
other wind models.  The discrepancy is slightly worse when compared
to the SINS galaxies, although the H$\alpha$ selection could play some
role in preferentially picking out high-SFR objects \citep{for09}.  
On the other hand, the difference between our simulations and the SHiZELS galaxies
appears to be not as large, but note that their typical redshifts correspond to the low-$z$ end of the redshift distribution of the SINS sample ($z \sim 1.5$), below the final redshift reached by our simulated galaxies.  
This discrepancy between models and
data has been much debated; for instance recent {\it Herschel}
results suggest that H$\alpha$-inferred SFRs may slightly overestimate
the true (i.e., far-infrared bolometric) SFR by some non-trivial
factor~\citep{nordon10}.  Other ideas for explaining this discrepancy
invoke variations in the stellar initial mass function~\citep{nar12}
or modifying the star formation law to account for metal-dependent H$_2$ formation \citep{kru12}.
We will not pursue this here, except to note that our zoom simulations
do not alleviate this discrepancy known from our larger cosmological
runs.

Simulations with momentum-driven winds result in higher specific
SFRs compared to simulations with no winds at $z = 2$, more in
agreement with observations.  In the vzw model, the mass loading
factor scales as $\eta \sim 1/\sigma$ and outflows tend to suppress
early star formation while providing high gas fractions to maintain
comparatively high SFRs at later ($z \sim 2$) times, as seen in
Figures~\ref{fig:evol} and~\ref{fig:cSFE}.  This suggests that even
stronger outflows might be needed in simulations in order to match
the $z \sim 2$ $M_{\star}$--SFR relation \citep{dav08}.  This is
not trivial, however, since outflows must not be too strong lest
they fail to produce enough early metals to enrich the IGM
\citep{opp06,opp08} and enough photons to reionize the
universe~\citep[e.g.,][]{fin12}.

The middle left panel of Figure~\ref{fig:sins} shows the $z = 2$ size--mass
relation for all simulated galaxies.  The effective radius $R_{\rm
1/2,sfr}$ is defined to enclose half of the total SFR and it is
calculated as a two-dimensional projected radius averaged over 100
random viewing angles.  Here we see that there is a clear separation
of galaxies in the $M_{\star}$--$R_{\rm 1/2,sfr}$ plane for the
different wind models.  For a given stellar mass, simulations with
no winds produce galaxies with a very compact distribution of gas,
with most of the star formation happening within $<1$\,kpc from
their centers.  Simulations with constant winds populate the low
mass and small size region of the plot, and only the more massive
galaxies in the sample are able to maintain gas disks with sizes
$R_{\rm 1/2,sfr}> 1$\,kpc.  Momentum-driven winds produce more
extended galaxies with sizes of several kpc, along with a minority
of more compact galaxies.

Outflows affect sizes by preferentially ejecting the star-forming
gas from the central regions and having it re-accrete over larger
scales~\citep{brook12}, resulting in large, extended star-forming
disks that are more in agreement with SINS and SHiZELS data.
Nonetheless, it is interesting that compact galaxies can also occur
in the vzw case; galaxy g2403 is one such compact galaxy (see
Figure~\ref{fig:prf}).  From our visualizations\footnote{See {\tt
http://www.physics.arizona.edu/\~{}angles/movies/}}, it appears
that these more compact systems generally arise during or shortly
after a significant merger event \citep{wuy10,bour11}, but we will quantify
this more rigorously in the future.

The diversity of sizes is also reflected in the half-mass radii of
the stars $R_{\rm 1/2,*}$ (Figure~\ref{fig:sins}, middle right).
Generally, the sizes are comparable to the half-SFR radius, but in
some cases $R_{\rm 1/2,*}$ can be $<1$~kpc.  This may have interesting
consequences for the progenitors of so-called compact ellipticals,
i.e., early-type galaxies at $z\sim 2$ that show very high stellar
densities.  We show as the dashed line a nominal threshold stellar
surface density of $M_*/R_{\rm 1/2,*}^{1.5}=10^{10.3}$\,\Msun\,kpc$^{-1.5}$
for compact galaxies from~\citet[see their Figure~1]{barro13}.  The
radius used for these observations is actually the $H$-band half-light
radius, but this should be fairly comparable to the stellar half-mass
radius.  Even with winds, a small fraction of our galaxies lie above
this surface density threshold (i.e., below the line).  This suggests
that our simulations do, with some frequency, produce galaxies that
have sufficient stellar densities to be the progenitors of compact
ellipticals.  These simulations have no mechanism for quenching
star formation, and hence our compact galaxies are still star-forming;
it is an additional constraint on models to form compact passive
systems that whatever feedback mechanism is responsible for quenching,
it operates when the galaxy is in a dense (likely post-merger)
state.

Galactic outflows seem to affect the bulge mass fraction of galaxies,
as shown in Figure~\ref{fig:sins}, upper right panel.  To estimate
the stellar bulge mass fraction, we perform a simple bulge-disk
kinematic decomposition for the stellar content of all simulated
galaxies.  We calculate the azimuthal velocity $v_{\rm \phi}$ of
each star particle with respect to the direction of the total angular
momentum of the galaxy as for our rotation curves, and estimate the
bulge mass as double the mass of particles moving with $v_{\rm \phi}
< 0$.  Though this is not analogous to observational determinations
of the bulge mass, it accurately characterizes the mass in the
spheroidal component of our simulated systems.  With winds, the
bulge fraction decreases with increasing stellar mass, suggesting
that galaxies start out small and dispersion-dominated, and move
toward being more ordered disks.  The opposite trend is seen for
our no wind simulations, tracking more the canonical behavior that
galaxies start out as disks and then merge together to form more
dispersion-dominated systems~\citep[e.g.,][]{whi91}.  SINS observations
suggest that smaller galaxies tend to be more dispersion-dominated,
qualitatively favoring our wind simulations, although a more careful
comparison that accurately mimics how bulge-to-disk ratios are
measured in data are needed for a more definitive result.

The lower left panel of Figure~\ref{fig:sins} shows the gas velocity dispersion,
$\sigma$, as a function of stellar mass at $z = 2$.  Here $\sigma$
is calculated as the spatially integrated (SFR-weighted) one-dimensional
velocity dispersion calculated within $R_{\rm 1/2,sfr}$ and averaged
over 100 random orientations.  Despite being extended and rotationally
supported disks, most of our galaxies are characterized by high
velocity dispersions ($> 30$\,km\,s$^{-1}$).  This suggests that
turbulent motions are significant even in rotationally supported
$z \sim 2$ galaxies, as inferred from observations
\citep{law09,for09,wri09,swi12b} and reported in previous simulations
\citep[e.g.,][]{ceve10,gen12b}.  
Our simulated galaxies are characterized by high $\sigma$ values but still lower relative to observed $z \sim 2$ galaxies with similar stellar masses, suggesting that turbulent motions are not fully resolved at scales comparable to the spatial resolution of our simulations.  Indeed, the self-regulated multi-phase model for star-forming gas is meant to capture the turbulent pressure arising from the continuous formation and disruption of gas clouds at a sub-grid level \citep{spr03a}.  For our simulated galaxies, the effective sound speed of the star-forming gas may reach values $\sim 150$\,km\,s$^{-1}$ \citep{ang13}, well above the resolved large-scale turbulent velocities.

For a given stellar mass, simulations
including outflows result in galaxies with higher gas velocity
dispersion, in better agreement with observations.  
Note that since outflowing gas is hydrodynamically decoupled as it is ejected from
the ISM, the outflows themselves are not directly injecting turbulence.
Instead, this increased turbulence relative to simulations without
winds likely owes to the higher gas and disk fractions, implying
lower Toomre $Q$ parameters for similar mass galaxies \citep{toomre64} and hence increased gravitational fragmentation, and possibly to the injection of energy due to the recycling of the outflowing gas back into the galaxies, in qualitative agreement with the analytic equilibrium model of \citet{gen12a}.

The lower right panel of Figure~\ref{fig:sins} shows the disk rotation velocity,
$v_{\rm d}$ divided by velocity dispersion $\sigma$, as a function
of stellar mass.  Here $v_{\rm d}$ is taken as the peak azimuthal
velocity ($v_\phi$) from the gas rotation curves obtained in
Section~\ref{sec:rot}.  
Direct comparisons
with data are not straightforward since rotation is computed from
a variety of means in the data, and we have not tried to mimic this
in detail; hence we have not plotted data here.  
Nonetheless, the momentum-driven wind model yields a
typical $v_{\rm d}/\sigma \sim 3$, which is in broad agreement with
the turbulent high-$z$ disks seen in SINS and SHiZELS.  
Constant and no wind models result in slightly higher $v_{\rm d}/\sigma$ on average,
though still within the range of the data.  There is little trend with mass, much less than
for the stellar bulge fraction, so even though the stellar components of higher mass galaxies are
diskier (for simulations with outflows), their gas content is not more rotationally supported with increasing stellar mass.  Note that the stellar component is less rotationally supported than the gas in all cases (cooling of shock-heated gas may dissipate turbulent energy) and, therefore, the gas component can be rotationally supported ($v > \sigma$) even for stellar bulge-dominated galaxies.  Galaxy g2743 represents an extreme example for the simulation with constant winds, with its stellar bulge fraction close to unity and still a rotationally supported gas disk (see Figures~\ref{fig:gasima} and~\ref{fig:starima}).

Overall, the sizes and dynamical properties of simulated
disks at $z=2$ are in fair agreement with observations from the
latest integral field unit studies of high-$z$ star-forming galaxies, particularly
in the case of momentum-driven winds.  
This occurs despite some 
overly simple assumptions in the modeling, such as decoupling of
wind material escaping the disk, and the usual concerns about the
ability of SPH to suppress viscosity and resolve dynamical
instabilities~\citep[e.g.,][]{agertz07}.   
Indeed, the properties of simulated galaxies are more strongly dependent on feedback models relative to the details of the specific hydrodynamic technique \citep[e.g.,][]{hop13b}.
This suggests that our simulations capture the dominant processes for
establishing the structural properties of galaxies and provide a plausible model for the
formation of disks during this cosmic epoch.

\subsection{Evolution of Physical Properties}

\begin{figure*}
\begin{center}
\includegraphics[scale=0.5]{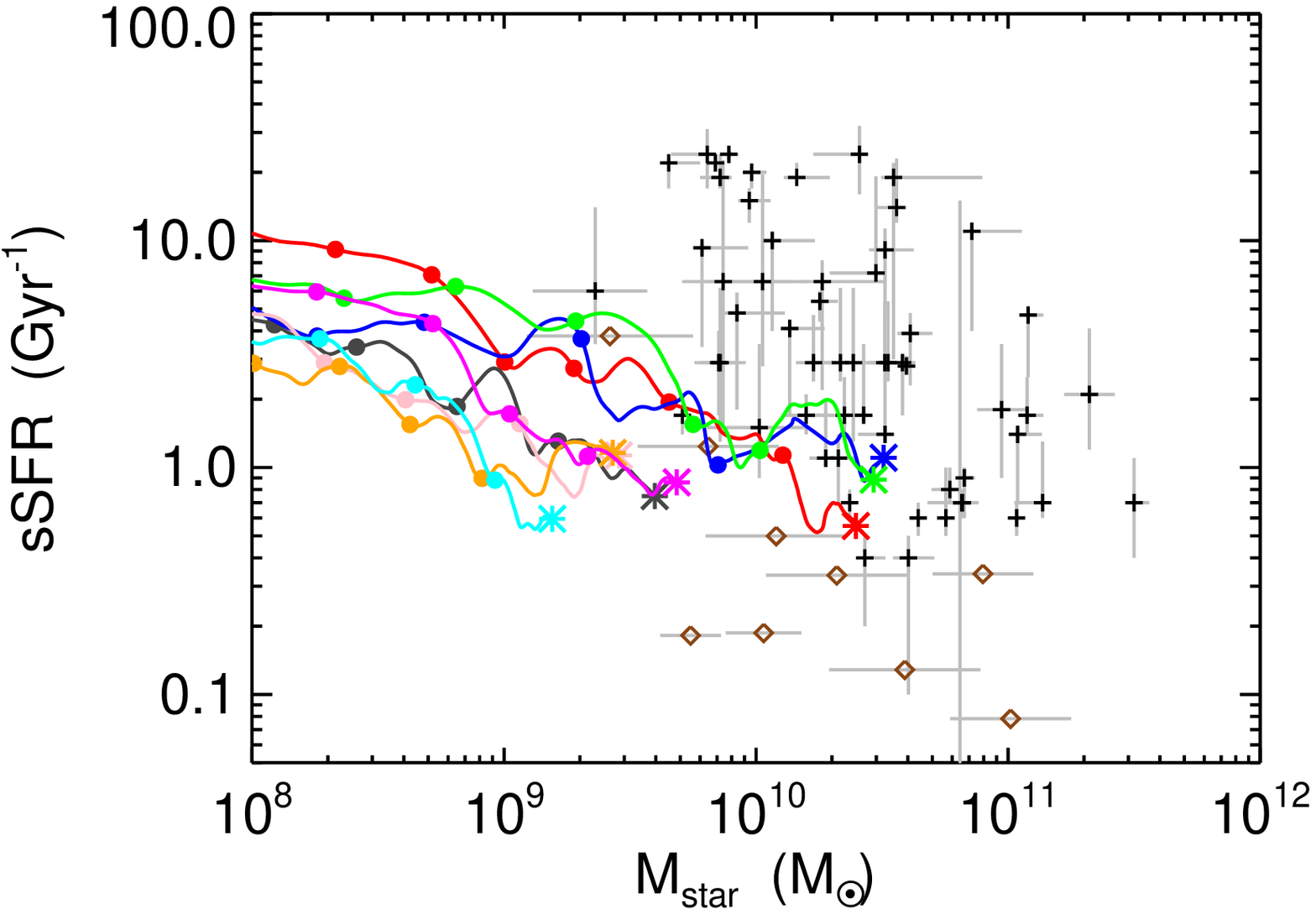}
\includegraphics[scale=0.5]{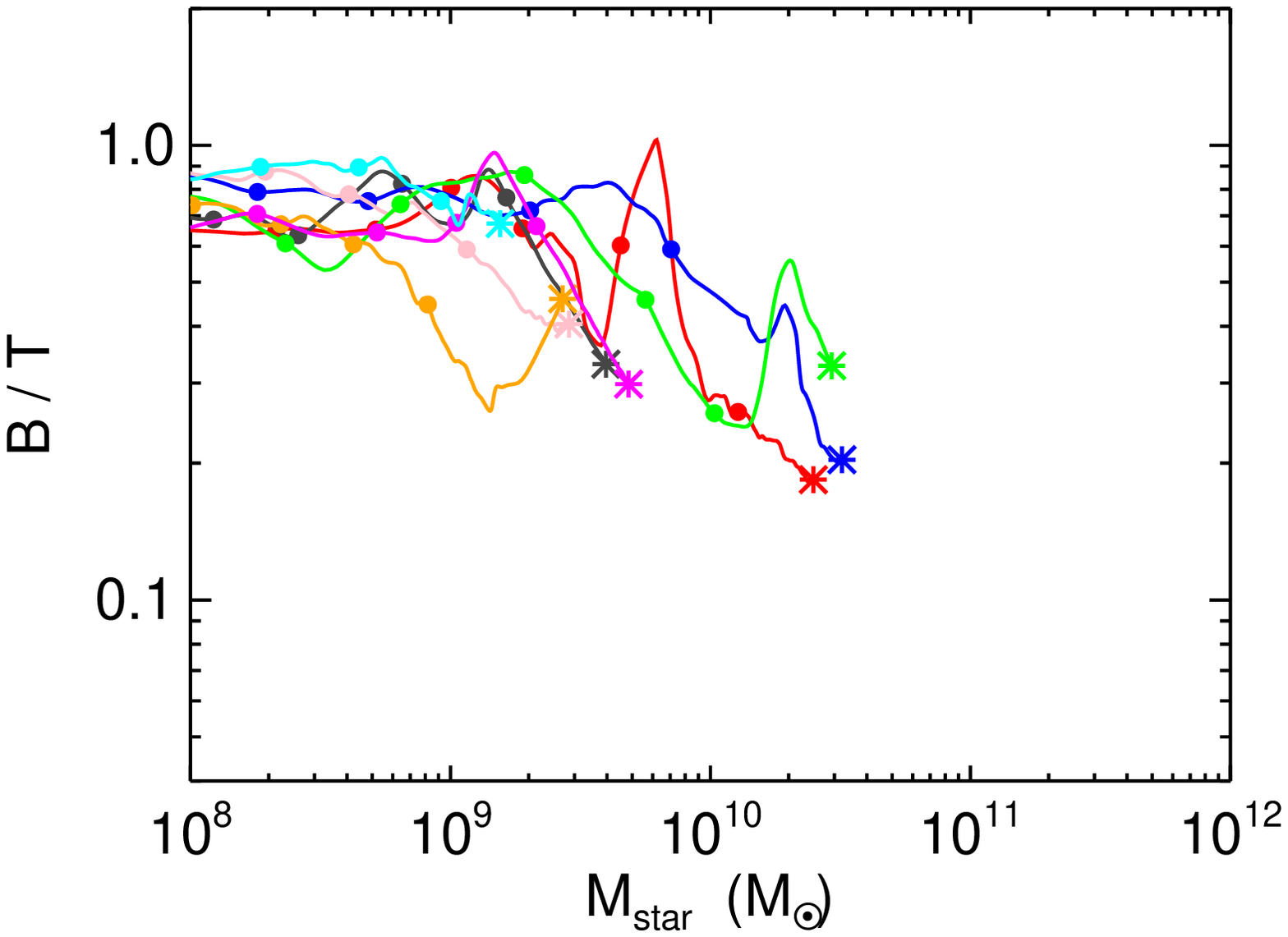}
\includegraphics[scale=0.5]{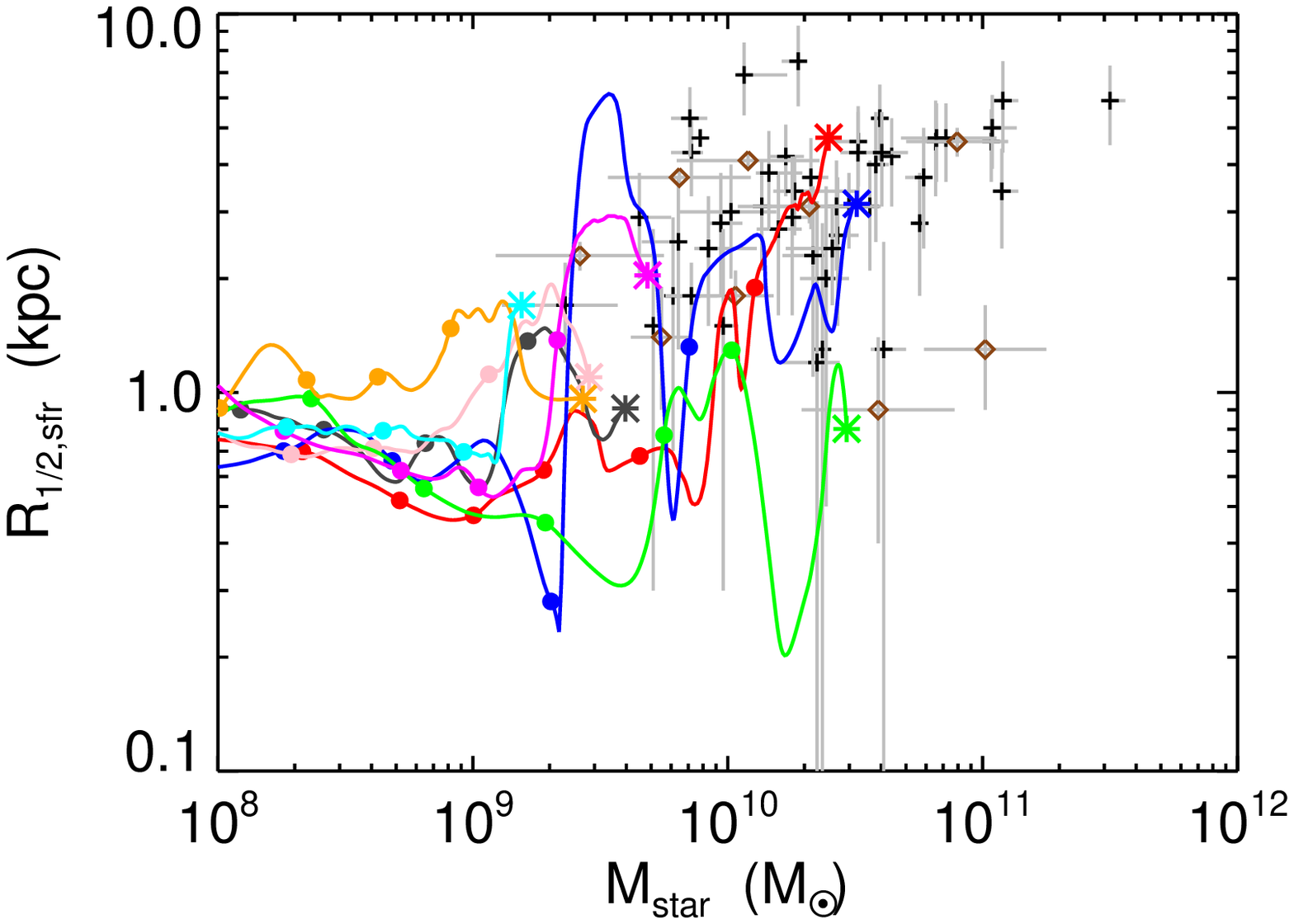}
\includegraphics[scale=0.5]{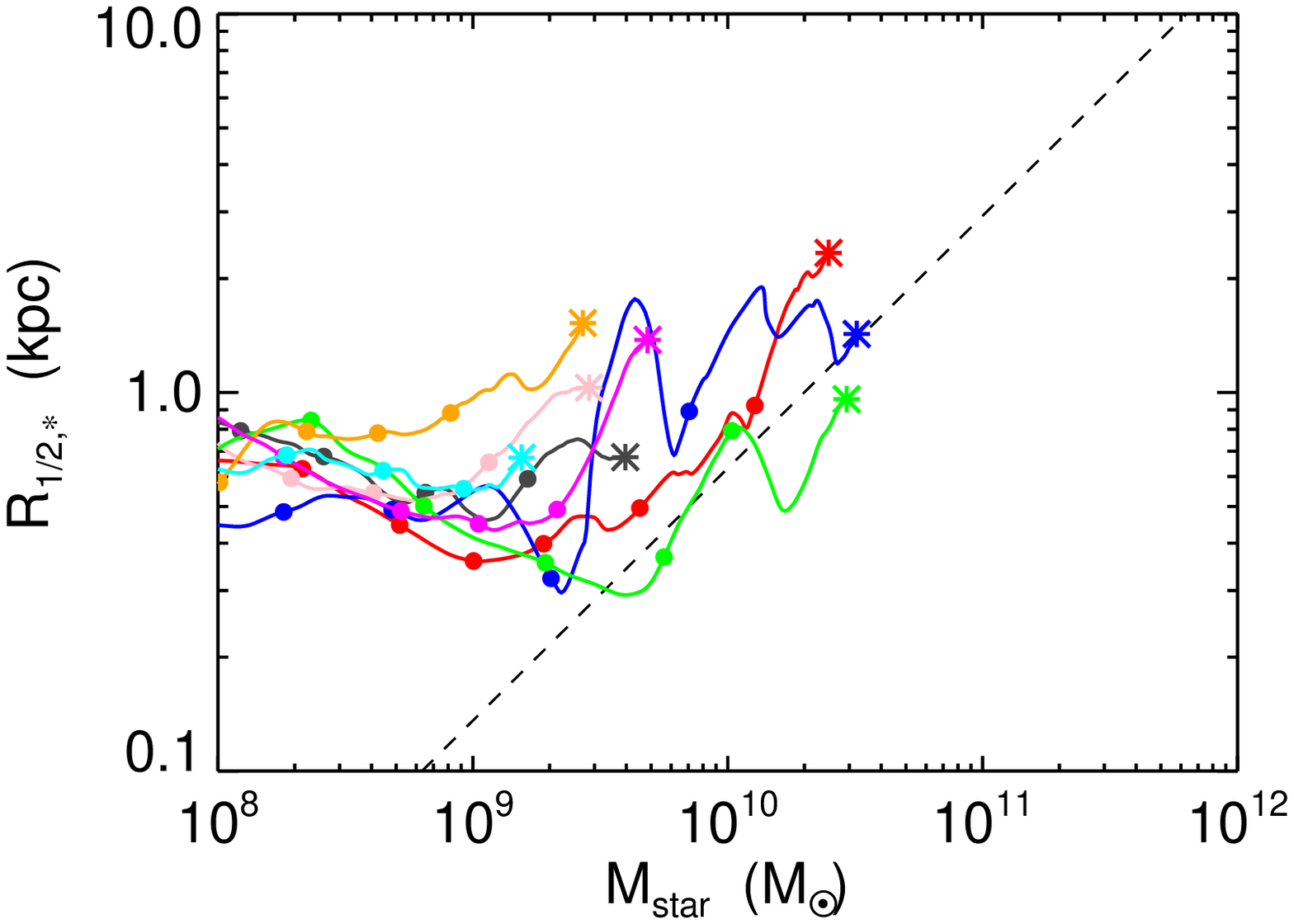}
\includegraphics[scale=0.5]{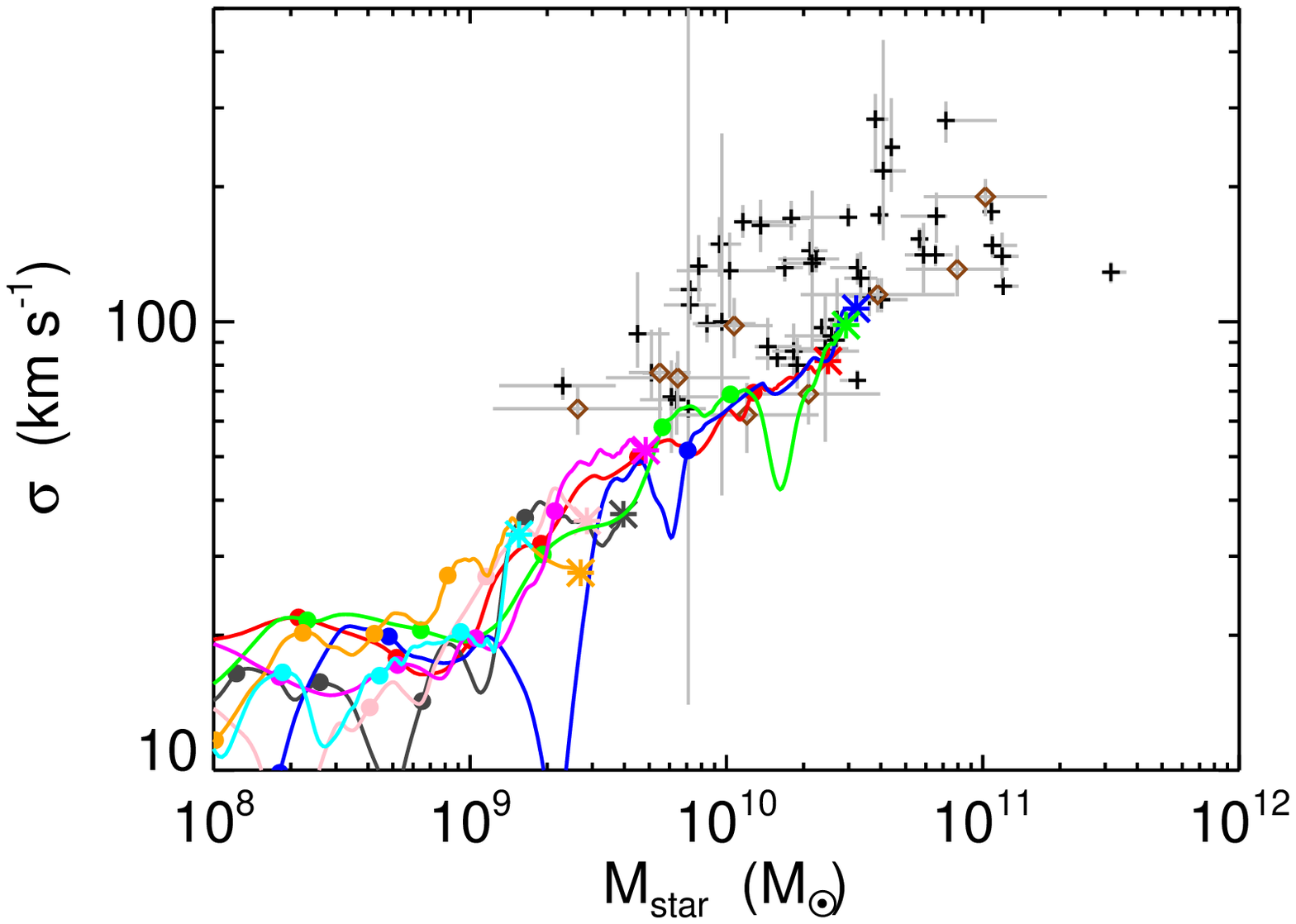}
\includegraphics[scale=0.5]{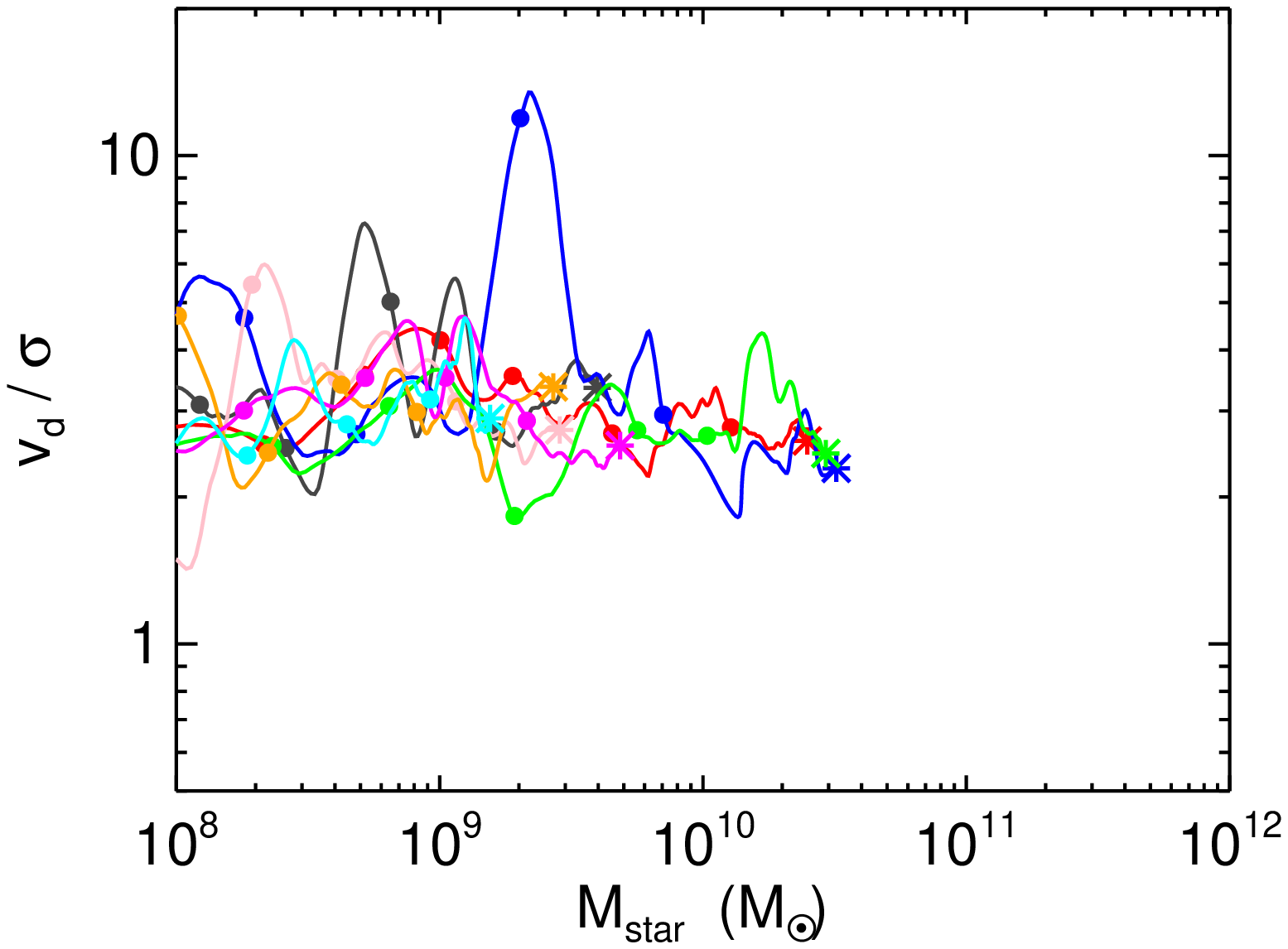}
\end{center}
\caption{Evolution of structural and dynamical quantities calculated from our sample of simulated galaxies from $z=8\rightarrow 2$. 
Different color tracks show each of our eight momentum-driven wind run 
galaxies, with filled circles indicated at integer redshifts (stars show the location of galaxies at the end of the simulation at $z = 2$).  Other lines and symbols are as in Figure~\ref{fig:sins}.
{\it Top left}: specific star formation rate (sSFR) as a function of stellar mass ($M_*$).  
{\it Top right}: stellar bulge mass fraction as a function of $M_*$.
{\it Middle left}: radius enclosing half of the total star formation rate as a function of $M_*$.
{\it Middle right}: radius enclosing half of the total stellar mass as a function of $M_*$. 
{\it Bottom left}: velocity dispersion $\sigma$ of the gas component of simulated galaxies as a function of $M_*$. 
{\it Bottom right}: disk peak rotational velocity $v_{\rm d}$ divided by $\sigma$ of the star-forming gas, as a function of $M_*$ for all simulated galaxies.
The time evolution of all physical quantities has been averaged over time intervals of $\sim 150$\,Myr.
}
\label{fig:sins_evol}
\end{figure*}

Using our plausible simulated population of $z = 2$ disk galaxies, we
now examine the evolution of physical properties from $z = 8 \rightarrow 2$.
We focus here on the momentum-driven wind model, since from a variety of wider constraints
it is our favored model of the three presented here~\citep[see,
e.g.,][]{dav11a,dav11b}.  However, many of the results are broadly
similar for our other wind models.  We show the evolutionary tracks 
for our eight galaxies in various colors in Figure~\ref{fig:sins_evol}. 
Since we are interested on the evolution of galaxies on cosmological time-scales, 
all physical quantities have been averaged over time intervals of $\sim 150$\,Myr.
All tracks go from left to right (i.e., lower to higher mass),
with individual unit redshifts indicated by the points along the
tracks starting at $z=8$.  The same data as shown in Figure~\ref{fig:sins}
is reproduced here, but note that this is for observed $z \approx 2$
galaxies and hence it is shown here for reference.

The top left panel of Figure~\ref{fig:sins_evol} shows the evolutionary tracks
of simulated galaxies in the main sequence ($M_*$--sSFR) plane.  
All galaxies show a similar evolution of decreasing specific SFR as 
they increase their stellar mass, consistent with a roughly
linear SFR to $M_*$ relation with the overall normalization decreasing
with redshift.  
Despite the increase in stellar mass, mergers cause a temporary enhancement of specific SFRs relative to the dominant decreasing trend.  These variations in specific SFR are apparent in the evolutionary tracks of our simulated galaxies even after averaging over time intervals of $\sim 150$\,Myr.  For the three most massive galaxies---g222 (blue), g2403 (green), and g54 (red)---major mergers can be identified in Figure~\ref{fig:evol} as abrupt changes in their stellar mass, and connected to the effects on the evolutionary tracks in Figure~\ref{fig:sins_evol} by direct comparison to the location of the points indicating integer redshifts.

At $z = 2$, galaxy sizes scale with stellar mass for simulations
with winds, in agreement with observations (Figure~\ref{fig:sins}).
Middle panels in Figure~\ref{fig:sins_evol} show that galaxies tend
to increase in size with time, as expected, but their evolutionary
tracks exhibit significant variation between galaxies, resulting
in a large scatter in the size--mass diagram at any given redshift.
Large variations of the half-SFR radius, $R_{\rm 1/2,sfr}$, tend to occur
in redshift intervals during which a major galaxy merger is taking place.
This suggests that major mergers, despite representing only a fraction of the total
mass growth in galaxies \citep[e.g.,][]{ker05,rodighiero11}, have
a significant impact on galaxy sizes, as it appears from our visualizations.
The evolution of the stellar half-mass radius, $R_{\rm 1/2,*}$, 
roughly follows that of the star-forming gas and is also
imprinted by significant variations occurring during galaxy mergers.
Interestingly, our most compact galaxy at $z = 2$ (g2403; green evolutionary track) 
was also in the compact regime at $z\sim 4$, both cases 
preceded by a major merger (see Figure~\ref{fig:evol}, middle panel).
This suggests that major mergers may drive galaxies toward the region 
of the size--mass diagram populated by compact ellipticals.
At the earliest epochs, the radius of galaxies does not change much,
and may in fact become smaller with time.  This partly reflects
our star formation criterion of $n_H>0.13$~cm$^{-3}$ that results 
in star formation occurring over a wide area when the universe is
very dense; it is therefore unlikely to be a robust prediction.

The upper right panel of Figure~\ref{fig:sins_evol} shows the stellar bulge
fraction evolution.  Early on, most galaxies are bulge-dominated,
as a result of high merger rates and large gas reservoirs that 
keep the mass distribution disordered.
As time proceeds, the combined effects of smooth gas accreting into
galaxies with higher specific angular momentum and galactic outflows
removing preferentially low angular momentum gas from their centers,
cause galaxies to increase their sizes and reduce their stellar
bulge fractions on average \citep{gover09,gover10,brook11}.  
It is interesting that this process only kicks in around $z\sim 4$
to start producing disk-dominated systems, at least
for the range of galaxy masses considered here.  
Our simulations, thus, predict that $z\sim 4$ was the 
beginning of the epoch of disk formation for massive galaxies.  
By $z\sim 2$, this results in a trend of decreasing bulge fraction with increasing
stellar mass in this wind model.  We note that this trend is opposite to
simple expectations from classic hierarchical galaxy formation
models~\citep[e.g.,][]{whi91}, in which disks form first and then
merge later to give rise to more dispersion-dominated systems.

The bottom left panel of Figure~\ref{fig:sins_evol} shows how simulated galaxies
evolve with redshift in the $M_*$--$\sigma$ plane.  Galaxies tend
to evolve along the correlation of $\sigma$ and $M_*$ observed at $z\sim 2$,
extrapolated to lower masses.
Disk peak rotation velocities follow a similar trend with
stellar mass, giving rise to a nearly constant ratio $v_{\rm d}/\sigma
\approx 3$ for all simulated galaxies independent of redshift, as
shown in Figure~\ref{fig:sins_evol}, bottom right panel.  Hence
while the predicted $v_{\rm d}/\sigma$ values are similar to that observed at $z=2$, 
it remains to be seen whether these same simulations can produce very thin disks
as seen at $z=0$.  In order to do so, something must alter the
current evolution of $v_{\rm d}/\sigma$, perhaps as a result of the
dropping accretion rate or else the dropping outflow rate.
Note that the simulated velocity dispersions shown here have been averaged over many random directions, in analogy with the spatially integrated $\sigma$ values uncorrected for average background velocity gradients inferred for the SINS galaxies \citep{for09}. 
Contributions from disk rotation velocities may thus overestimate $\sigma$ values and underestimate the inferred rotational support of galaxies.
For simulated galaxies, we can eliminate the contributions from ordered rotation simply by calculating the velocity dispersion along the line of sight perpendicular to the plane of the disk, $\sigma_z$.  Interestingly, while we find $\sigma_z < \sigma$ in most cases, as expected, $\sigma_z$ is also correlated with the stellar mass of galaxies.

Overall, the redshift evolution of these properties shows a 
consistent buildup of size and velocity dispersion with time (and mass), 
although mergers can particularly
impact the inferred sizes substantially over short time periods.
The large scatter in observed sizes may, therefore, reflect the
short-term merger history of galaxies even more reliably than instantaneous
SFRs or bulge-to-disk ratios.
Galaxies evolve from being bulge-dominated when small to disk-dominated
when larger, with disks becoming prominent only at $z\la 4$.
Despite this, galaxies are always rotationally supported even at early
times when bulge-dominated, suggesting that the present-day association
between small bulge-to-disk ratio and large rotational support does not necessarily
apply to high-redshift galaxies.

\section{Resolution Convergence}\label{sec:num}

\begin{figure*}
\begin{center}
\includegraphics[scale=0.5]{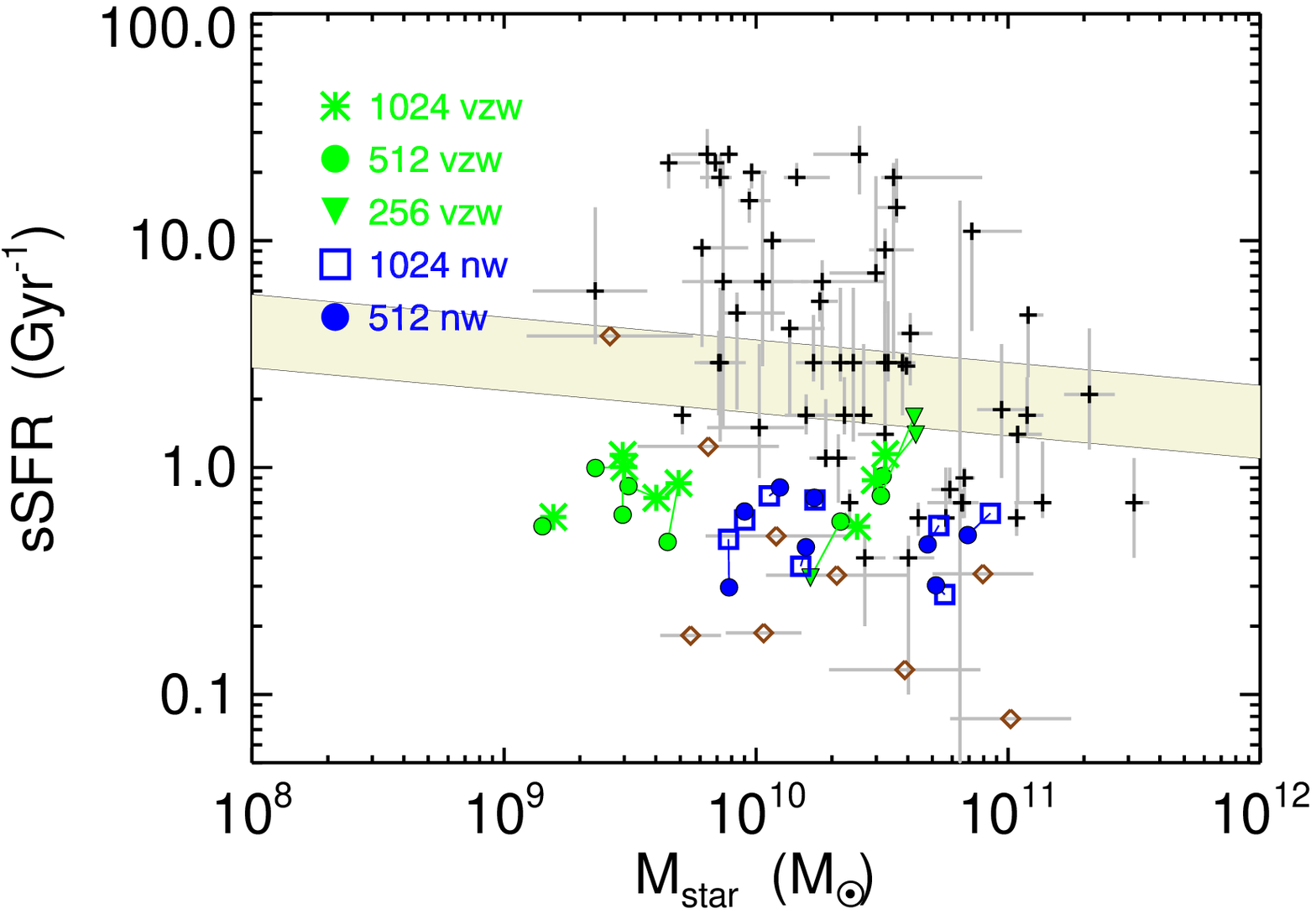}
\includegraphics[scale=0.5]{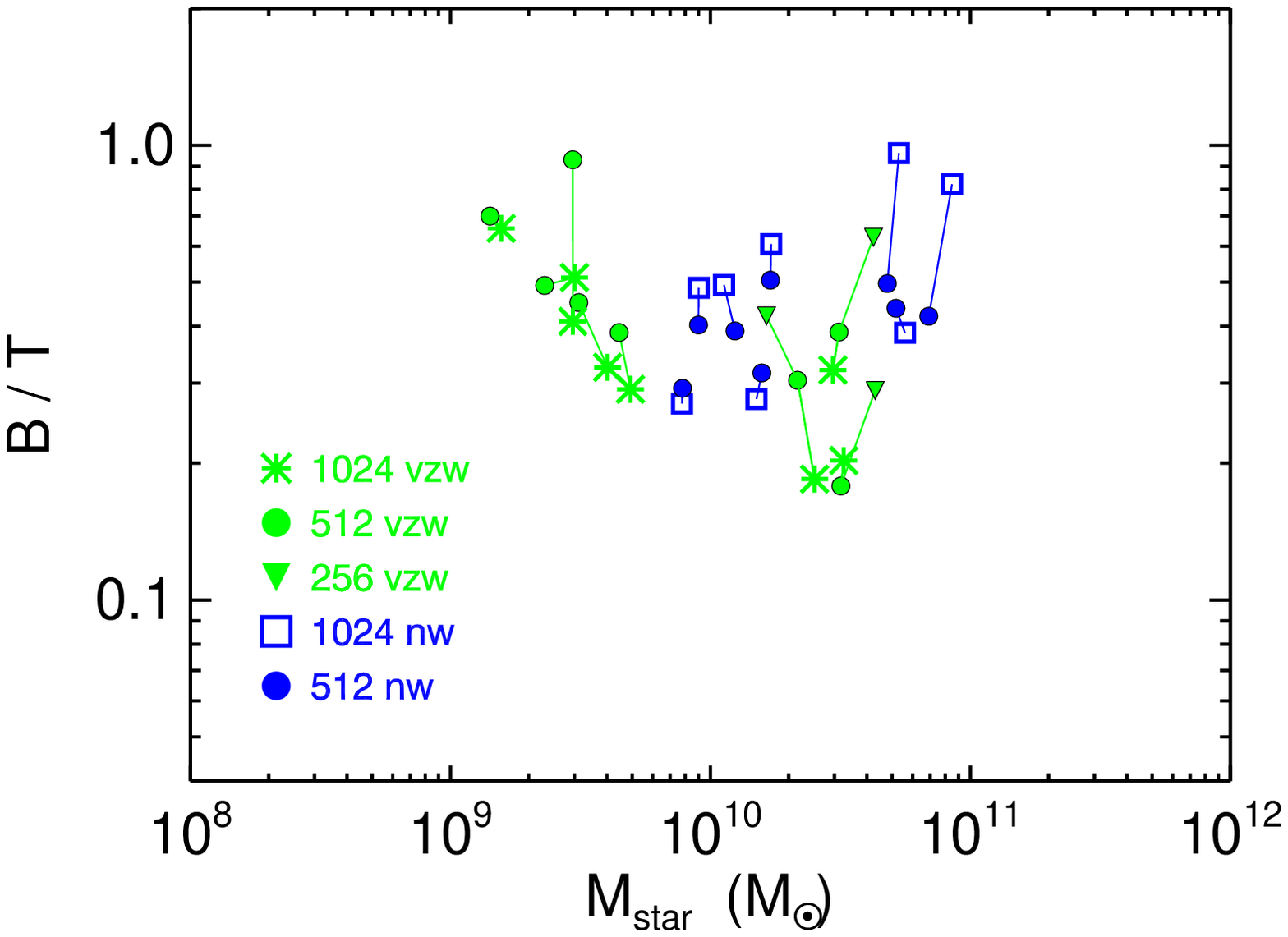}
\includegraphics[scale=0.5]{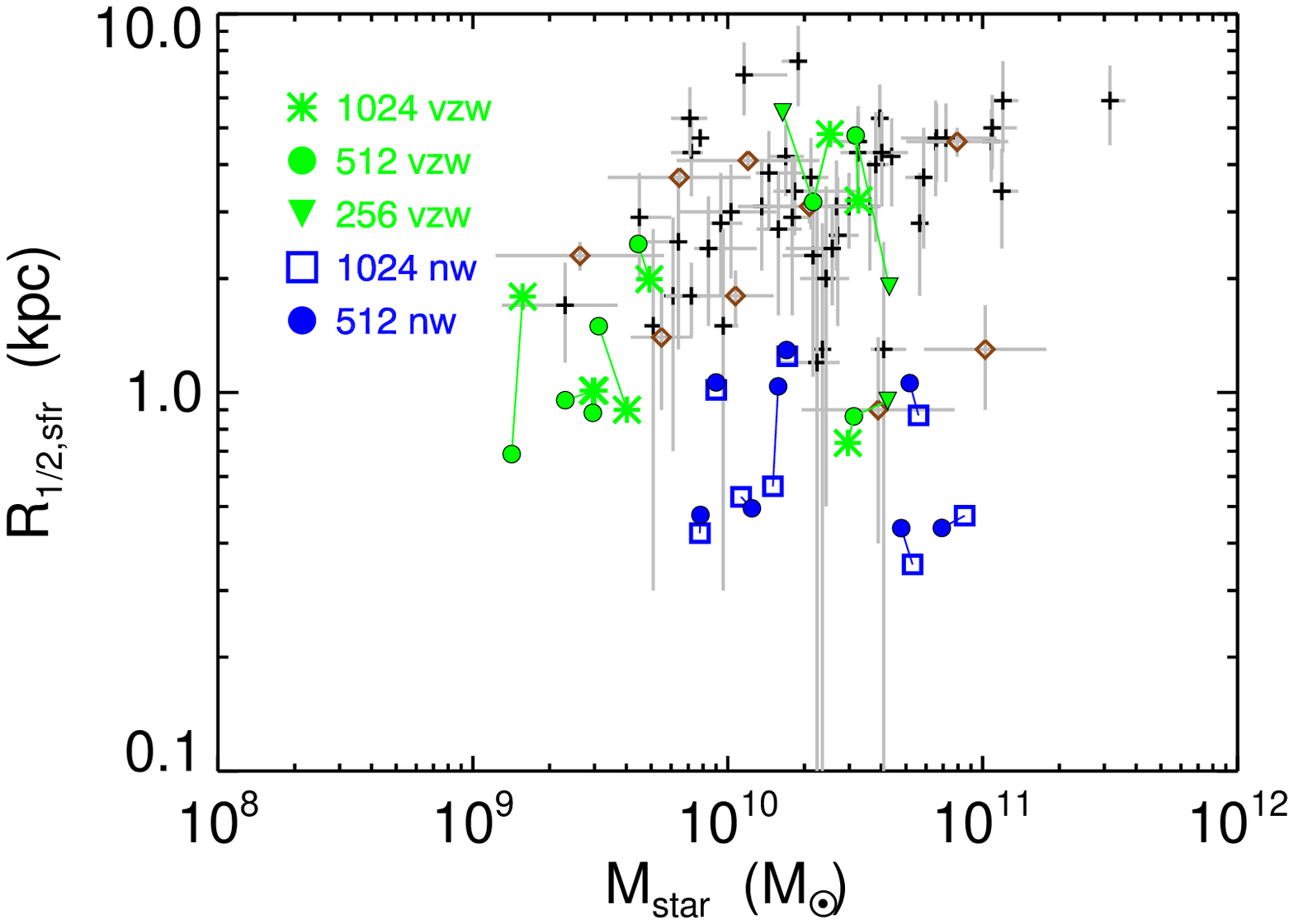}
\includegraphics[scale=0.5]{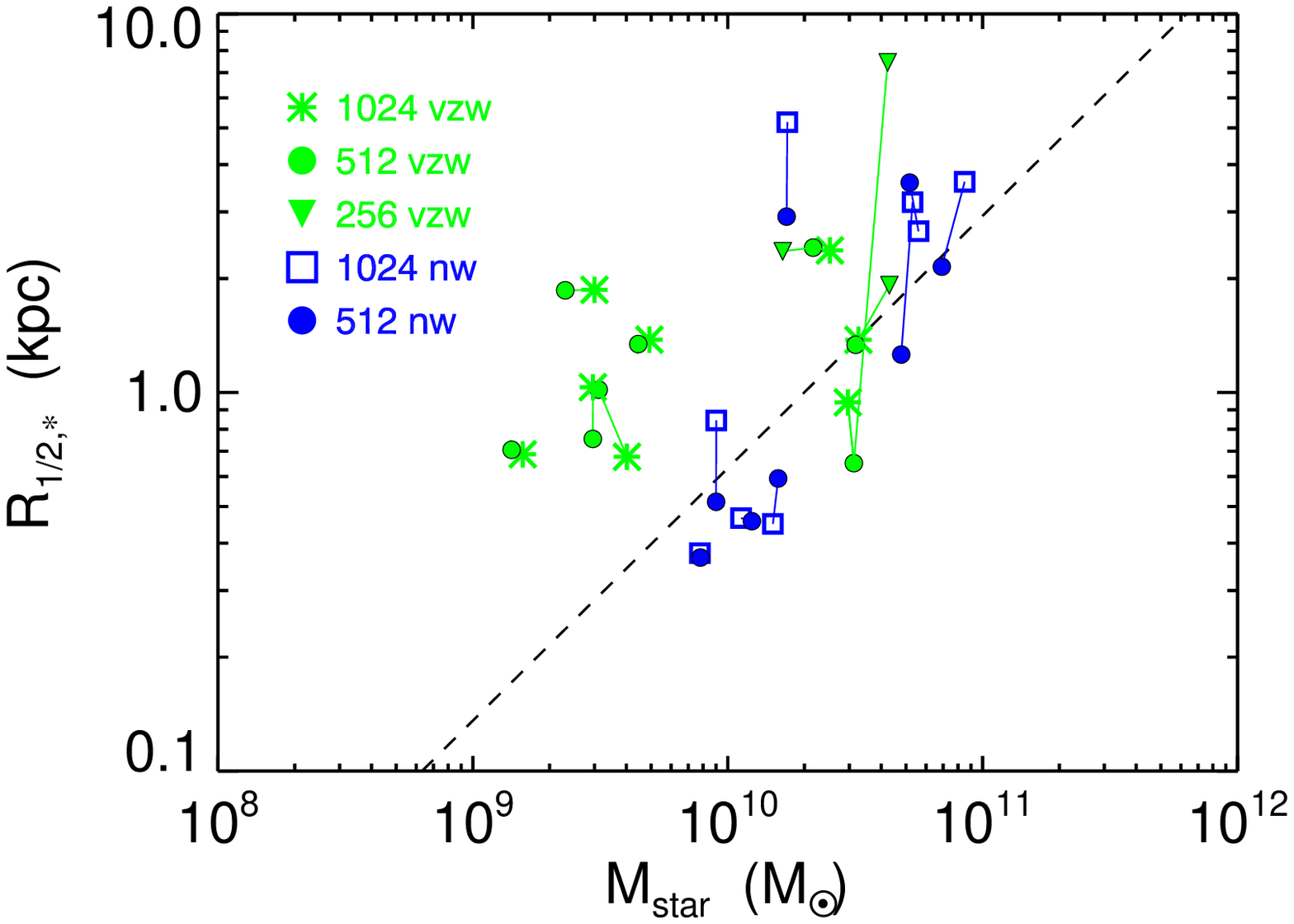}
\includegraphics[scale=0.5]{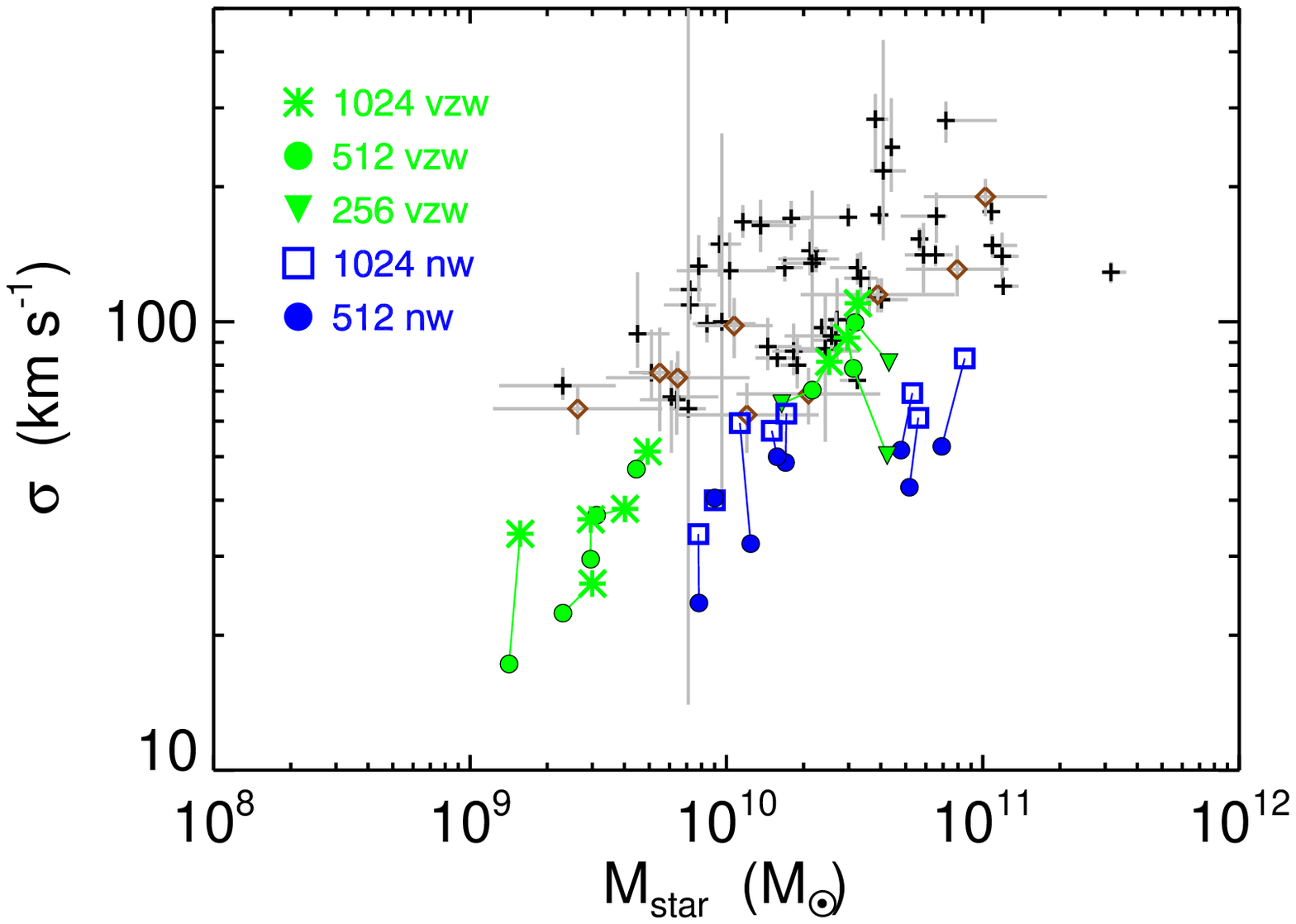}
\includegraphics[scale=0.5]{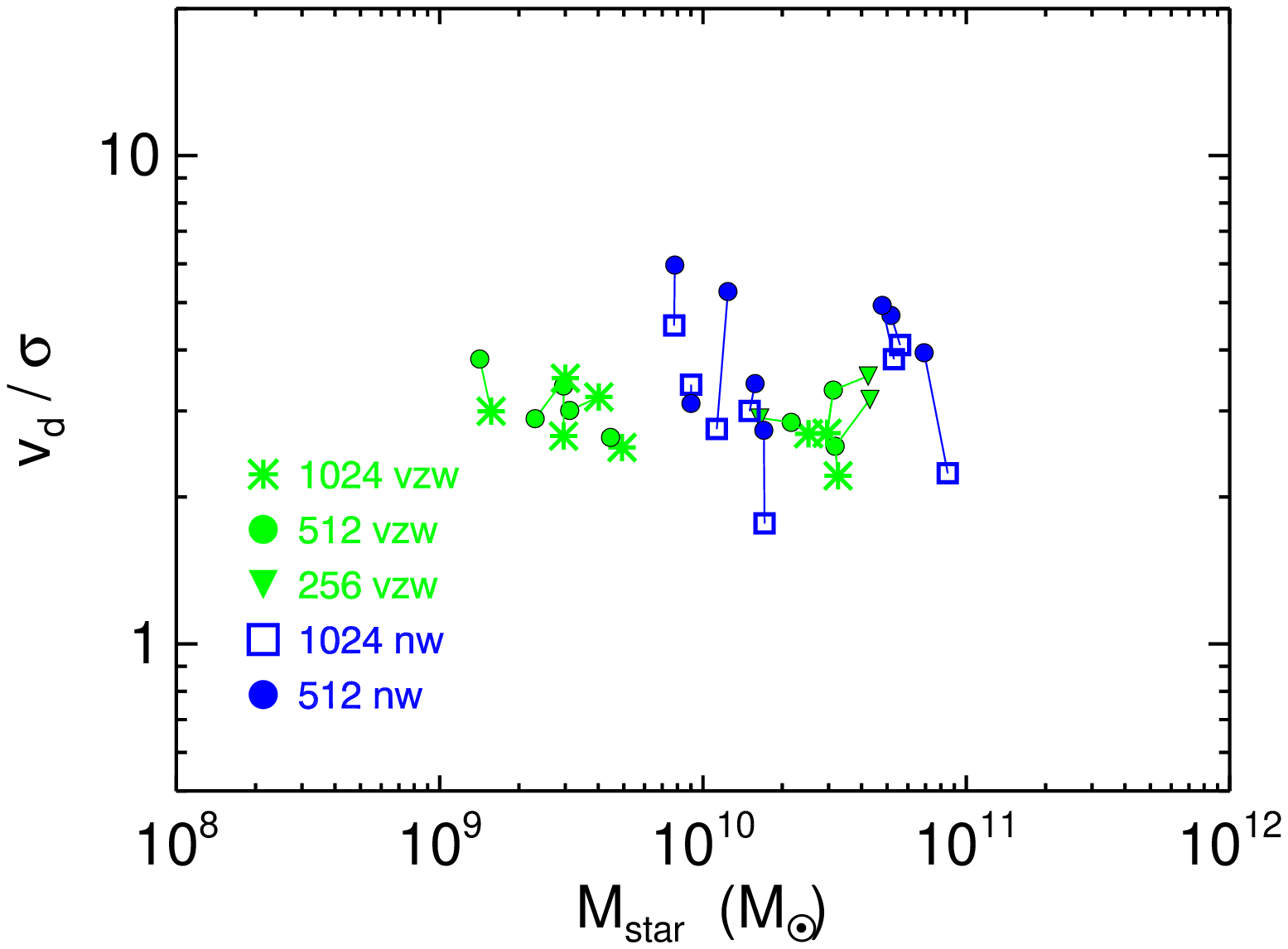}
\end{center}
\caption{Resolution convergence of physical quantities at $z=2$ for our momentum-driven wind (vzw; green) and no wind (nw; blue) simulations, comparing our effective $1024^3$ runs (green stars and blue squares for vzw and nw respectively) with effective $512^3$ runs (green and blue filled circles for vzw and nw respectively) of the same initial conditions.  Green and blue solid lines connect individual galaxies corresponding to the higher resolution and lower resolution simulations for the vzw and nw simulations respectively.  
The green triangles show the physical properties of the three most massive galaxies as obtained from the original large-scale 256$^{3}$ simulation with momentum-driven winds. 
Other lines and symbols are as in Figure~\ref{fig:sins}.
{\it Top left}: specific star formation rate (sSFR) as a function of stellar mass ($M_*$).  
{\it Top right}: stellar bulge mass fraction as a function of $M_*$. 
{\it Middle left}: radius enclosing half of the total star formation rate as a function of $M_*$.
{\it Middle right}: radius enclosing half of the total stellar mass as a function of $M_*$. 
{\it Bottom left}: velocity dispersion $\sigma$ of the gas component of simulated galaxies as a function of $M_*$. 
{\it Bottom right}: disk peak rotational velocity $v_{\rm d}$ divided by $\sigma$ of the star-forming gas, as a function of $M_*$ for all simulated galaxies.
}
\label{fig:sins_res}
\end{figure*}

A key computational issue we face in this study is that the spatial resolution of the observations we are
comparing our results against is comparable to the numerical resolution of our simulations. 
Our nominal $\sim 220$\,pc resolution at $z = 2$ corresponds to the equivalent Plummer force softening length; thus, the scale at which we compute exact gravitational forces is 2.8 times this length, or $\sim 600$\,pc.  Furthermore, SPH techniques may suffer from large viscous transport of angular momentum.  Indeed,  the moving mesh code Arepo \citep{spr10} is seen to produce larger-scale disks than {\sc Gadget}~\citep{vog12,torr12} from identical initial conditions.  Other recent code comparisons suggest that differences in sub-grid physics may have an even stronger impact on the properties of simulated galaxies relative to differences in hydrodynamic techniques  \citep{sca12,hop13,hop13b}.  It is, therefore, important to carry out a basic test of resolution convergence of our key results.

We run simulations with a factor of two lower spatial resolution and
a factor of eight lower mass resolution, equivalent to $512^3$
resolution in a $[24\,\hmpc]^3$ box, from otherwise identical initial
conditions, and identify the galaxies that correspond to our high-resolution galaxy sample.  
In addition, we extend our resolution study by using the original large-scale cosmological simulation with momentum-driven winds and 256$^{3}$ resolution.  The unique identification of galaxies corresponding to our highest resolution galaxy sample is increasingly difficult as the resolution decreases and galaxy trajectories begin to differ.  Nonetheless, we have been able to identify and calculate the properties of our three most massive galaxies, extending the resolution study over 64$\times$ in mass and 4$\times$ in spatial scale.

Figure~\ref{fig:sins_res} shows the
effects of numerical resolution on key structural and dynamical
quantities of galaxies versus stellar mass for our momentum-driven
winds and no wind simulations at $z = 2$ (as in Figure~\ref{fig:sins}).
Here, solid lines (green and blue for vzw and nw respectively)
connect individual galaxies corresponding to the higher and lower
resolution simulations to help identify any systematic trends.
We note that a galaxy-by-galaxy comparison represents a strong
numerical convergence test given that even small deviations of
orbital parameters in galaxy mergers may result in rather different
structural and kinematic properties of remnant galaxies at a given
time.  Global properties of galaxies are thus expected to exhibit
better numerical convergence.  Indeed, Figure~\ref{fig:sins_res}
shows that total stellar masses and specific SFRs of galaxies are
both well converged for the vzw and nw simulations.
The average ratios $\langle {\rm sSFR}_{512} / {\rm sSFR}_{1024} \rangle$ and $\langle M_{*,512} / M_{*,1024} \rangle$ are consistent with unity within the 1$\sigma$ dispersion when comparing the effective $512^3$ and $1024^3$ simulations.

The size of galaxies is somewhat more sensitive to resolution, as shown in
Figure~\ref{fig:sins_res}, middle panels.  For individual galaxies,
we find that the half-SFR ($R_{\rm 1/2,sfr}$) and half-mass ($R_{\rm
1/2,*}$) radii may differ by up to a factor of $\sim 2$ for the
effective $1024^3$, $512^3$, and $256^3$ simulations at $z = 2$.  Such sensitivity
might be expected given that radii track recent merger activity,
which can vary substantially owing to the chaotic nature of orbits
within hierarchically growing halos.  
Indeed, the most compact galaxy in the sample is in the early stages of a major merger with a gas poor galaxy at $z = 2$ in our $256^3$ simulation, resulting in a strong variation in stellar effective radius without significantly affecting $R_{\rm 1/2,sfr}$. 
Despite this, we find no systematic trend in the whole sample
with resolution or wind model---the average size ratio $\langle R_{512} / R_{1024} \rangle$ is consistent with no resolution dependence for both the stellar and gas distributions.
Our lower resolution simulations confirm that outflows generally
produce more extended star-forming disk galaxies but that may also
result in compact galaxies (similar to simulations with no winds)
with some frequency.

Stellar bulge mass fractions are also somewhat
sensitive to resolution, as shown in the upper right panel of Figure~\ref{fig:sins_res}.  
In the vzw case, bulge fractions are generally higher at low 
resolution.  This is expected if the lack of gravitational resolution results in increased random motions and the overall reduction of ordered rotation due to a comparatively shallower gravitational potential.
Such a trend is not as apparent in the no wind case,
though in two cases the bulge fractions are substantially different.
Overall, our general result that bulge fractions are
higher for similar mass galaxies in simulations with no winds seems unaffected by
resolution.

We find a systematic trend for gas velocity dispersion ($\sigma$)
to decrease with resolution (Figure~\ref{fig:sins_res}, lower left),
though this is significantly more evident for simulations without 
winds.  This suggests that our
effective $1024^3$ simulations may not have reached numerical
convergence and that higher resolution simulations with outflows
could result in slightly higher $\sigma$ values for a given stellar
mass, more in agreement with observations.  Despite this, gas
recycling for simulations with outflows seems to result in increasing
$\sigma$ values and, therefore, turbulence in galaxies regardless
of numerical resolution.  Figure~\ref{fig:sins_res}, lower right,
shows that the ratio $v_{\rm d}/\sigma$ is reasonably well converged
for the vzw model but increases systematically for nw simulations
with lower resolution due to the overall decrease in $\sigma$.

In short, our main results are generally though not optimally resolution converged.
Higher resolution seems to result in lower stellar bulge fractions and higher gas velocity dispersions, while we find no systematic trend for galaxy sizes with resolution.  However, the resolution convergence exhibited by the wind simulations is better than that for no winds.  Interestingly, this trend is also seen for global mass and SFR functions~\citep{dav11a}.  By increasing
sizes and having a physically motivated driver of wind recycling that
sets the velocity dispersion, winds actually seem to help resolution
convergence somewhat.  Nonetheless, this convergence experiment only
spanned a factor of 4 in spatial scale and 64 in mass (for our three most massive galaxies), and hence
simulations with greater dynamic range will be needed to fully
assess the numerical robustness of these results.

\section{Summary and Conclusions}\label{sec:con}

Powerful galactic outflows are ubiquitous in high-redshift galaxies
and likely play a central role in the evolution of galaxies and the
IGM.  Here, we have presented high-resolution cosmological zoom
simulations that follow the evolution of a sample of eight central
galaxies down to $z = 2$, focusing on the impact of strong outflows
on their morphologies, kinematics, and star formation properties.
Our main results can be summarized as follows:

\begin{itemize}

\item Despite the limited sample of eight galaxies presented here, our
simulated systems span a wide range of morphological characteristics
at $z = 2$.  Disk structures are prevalent but can range from very
compact gas and stellar distributions, to extended quiescent
``grand-design" spirals, to turbulent and clumpy disks.  Inferred
morphologies can depend on the observed tracer.

\item Simulations with no winds produce rapidly rising SFRs that
result in higher stellar masses, higher metallicities, and lower
gas fractions for all galaxies at all times relative to simulations
with galactic outflows.  Momentum-driven winds cause an effective
delay in star formation by ejecting significant amounts of gas from
small, early galaxies and having it reaccreted at later times,
resulting in higher gas fractions and star formation histories more
in agreement with observations.  All wind models fail, however, in
reproducing the normalization of the observed $z = 2$ $M_{\star}$--SFR
relation, though the late-time recycling by the momentum-driven wind
model comes the closest; this suggests that even stronger 
outflows at early times and/or small masses may be required.

\item Galactic outflows affect the amount and distribution of metals
in galaxies by regulating star formation, ejecting metals into the
surrounding gas preferentially from their centers, and the recycling of enriched
gas back into galaxies over larger scales.  This results in lower
metallicities and less steep metallicity gradients relative to
simulations with no winds.  The resulting central metallicities
are somewhat super-solar.  Examples of inverted metallicity gradients 
are uncommon among our galaxy sample.

\item Winds have a significant impact on the structural properties
of simulated galaxies.  No wind simulations generally produce more
compact galaxies with higher stellar surface densities, higher
stellar mass bulge fractions, and most of the star-forming gas
concentrated within the inner kiloparsec.  Galactic winds usually
yield more extended disks and tend to reduce bulge fractions by
preferentially removing low angular momentum gas from their centers.
Nonetheless, simulations with winds may produce, in some cases,
galaxies with stellar surface densities above the threshold for
compact ellipticals, usually occurring shortly after a major merger event.
Sizes in general are quite sensitive to 
the merger history, more so than star formation or bulge fraction,
and hence this may be the best way to assess the hierarchical 
buildup of galaxies.

\item Simulations with winds produce galaxies with more gradually rising
rotation curves compared to the more centrally peaked rotation
curves of galaxies without winds.  When calculated from the gas
rotation velocities rather than the enclosed mass, rotation curves
are more smoothly rising because of the increased dispersion support
in the central regions of galaxies.

\item Peak rotation velocities and velocity dispersions scale with
stellar mass for all wind models, in a manner broadly consistent
with observations.  The inferred ratios $v_{\rm
d}/\sigma$ are consistent with rotationally supported turbulent
disks at $z = 2$.  Gas recycling and the high gas fractions of
galaxies from simulations with outflows yield higher gas
velocity dispersions and more turbulent disks compared to no-wind
simulations.  Early small galaxies have both high bulge fraction
and are rotation-dominated, counter to typical trends among
local galaxies, and suggests that the standard intuition from today's 
{\it Hubble} sequence may not apply to high-$z$ galaxies.

\end{itemize}

Our simulations complement previous studies by analyzing the effects
of large-scale outflows on the internal structure and evolution of
individual $z = 2$ galaxies.  We employ the same outflow mechanisms
used by \citet{dav11a,dav11b} in non-zoom cosmological simulations,
including an observationally constrained prescription for momentum-driven
winds, and with no further tuning of model parameters.  Encouragingly,
simulations with momentum-driven winds, which are favored by a wide
range of observations and recent idealized galaxy simulations
\citep{hop12}, yield similar trends on the global properties of
galaxies when applied to simulations with a factor $\times 150$
increased in mass resolution, and produce galaxies at $z = 2$ with
structural and kinematic properties in broad agreement with
observations.  This provides a new and non-trivial test of 
hierarchical galaxy formation models.

It is interesting that even among our limited galaxy sample, we produce
a diversity of morphologies, from grand design spirals to clumpy
turbulent disks to potential progenitors of compact ellipticals.
This diversity is enhanced by feedback, and may in fact be
governed by it.  
While there is an overall consensus on the importance of feedback
associated with star formation in determining the morphologies of
galaxies and their evolution over cosmic time, major uncertainties
remain on the nature and relative significance of different feedback
processes.  The specific outcomes of particular feedback models
often depend on numerical resolution, the treatment of ISM physics,
the star formation prescription, and other implementation details
\citep{ceve10,gover10,kru12,sca12,chris12,hop12}.  

Generically, we find that increasing the strength of feedback results in the
reduction of star formation efficiencies and the increase in galaxy
sizes for a given stellar mass, as found in, e.g., \citet{sal10}.  Galactic outflows
also increase the degree of rotational support of galaxies by
preferentially removing low angular momentum gas from their centers
\citep{gover10,brook11,brook12}.  Disk galaxies may survive or even
be produced from high angular momentum mergers of gas-rich systems,
provided that pressurization from a multiphase ISM prevents
fragmentation and efficient conversion of gas into stars
\citep{rob04,rob06,hop09a,hop09b}.  Simulations with no thermally
pressurized ISM, however, tend to produce more turbulent and clumpy disks and
may result in rather different merger remnant morphologies \citep{bour11}.  
These illustrate how the structure and
morphology of high-redshift galaxies provide yet another important
test of feedback processes during the peak epoch of galaxy growth.

The progress of high performance computing has enabled these types
of cosmological zoom simulations to become more routine, but they still 
rely on sub-grid prescriptions and are subject to numerical uncertainties.
Advancing this field will rely on developing more robust models for
the small-scale physics in addition to achieving higher dynamic
range.  This work presents a first step toward developing plausible
models for the formation of galactic systems down to sub-kpc scales,
but much work lies ahead to push both to smaller scales and to
understand better the underlying physical processes governing high-redshift galaxy assembly.

\section*{Acknowledgements}

We thank D. Ceverino, C. Christensen, C. Martin, B. Robertson, and G. Yepes 
for useful discussions, and the anonymous referee for a thoughtful report that helped improve the paper.
D.A.-A. thanks B. Robertson for very useful assistance with data visualization.  
F.\"O. gratefully acknowledges support from the Radcliffe Institute for Advanced Study at Harvard University.
The simulations were run on the University of Arizona's 512-processor SGI Altix system and the
TACC Sun Constellation Cluster (Ranger) at The University of Texas,
Austin. This work used the Extreme Science and Engineering Discovery
Environment (XSEDE), which is supported by National Science Foundation
grant number OCI-1053575.  This work was supported by the National
Science Foundation under grant numbers AST-0907998 and AST-1108753.
Computing resources were obtained through grant number DMS-0619881
from the National Science Foundation.

\end{document}